# Public salience mobilizes nationwide responses to violence against women

Augusto Cerqua[1*] Marco Letta[1] Costanza Giannantoni[1] Gabriele Pinto[1]

[1] Department of Social Sciences and Economics, Sapienza University of Rome, Rome, Italy.

[*] Corresponding author. Email: augusto.cerqua@uniroma1.it

## Summary

Does public salience around femicides mobilize societal responses or trigger further harm through imitation? Using newly assembled case-level data on femicides in Italy from 2017 to 2024, we study the nationwide consequences of the most salient cases for help-seeking, police reporting, and lethal violence. We find that highly salient intimate partner femicides (IPFs) generate nationwide increases in help-seeking, and that an exceptionally salient IPF also leads to a marked rise in police reports of violence against women. Yet we find no evidence of an increase in femicides—a pattern inconsistent with copycat theories. Responses are strongest during periods of peak salience. No comparable mobilization follows a lower-salience IPF, a similarly salient femicide committed by family members, a highly salient killing of a woman not classified as a femicide, or a lower-salience falsification case. Together, these comparisons indicate that mobilization depends on salience intensity, the victim–perpetrator relationship, and the gendered nature of the violence, rather than reflecting a generic response to salient violent crime. These findings suggest that public salience can contribute to prevention strategies against gender-based violence without amplifying lethal violence.

Femicides, defined as the intentional killings of women or girls because of their gender, are the most extreme manifestation of gender-based violence and remain a persistent problem worldwide, particularly in the context of intimate partner violence. According to recent estimates by the United Nations [1,2], while overall homicides globally have started to decline after peaking in 2021, female homicides have not followed the same trend. Women's empowerment may, in some contexts, provoke backlash and retaliation [3–5], potentially increasing the risk of intimate partner violence. Reducing gender-based killings is therefore both a major societal challenge and a central concern for policymakers.

Addressing femicides and, more broadly, gender-based violence requires not only "hard" policy tools—such as legal sanctions, policing, and preventive interventions—but also broader social change. Reducing stigma and breaking the silence surrounding violence against women are particularly important for encouraging help-seeking and reporting. Underreporting remains pervasive, often driven by fear of retaliation, limited trust in institutions, and social norms that discourage victims from coming forward [6], leaving estimates of the prevalence of violence against women highly uncertain and often discordant [7–11]. To reduce underreporting, many countries rely on awareness campaigns designed to increase the salience of gender-based violence and encourage victims and their social networks to seek help and report abuse. Yet the effectiveness of such campaigns is far from guaranteed [12], and isolating the causal role of awareness campaigns, and of public salience more broadly, is challenging.

Highly salient femicides provide a distinct setting in which sharp, unanticipated increases in the public salience of gender-based violence occur independently of institutional efforts, allowing us to study whether heightened attention itself mobilizes responses to gender-based violence. We contribute to the literature on how public salience shapes behavior and societal responses. Previous work shows that highly publicized cases of gender-based violence can increase help-seeking and reporting [13], while broader media exposure can reduce the social acceptability of domestic violence [14]. Yet this evidence leaves three important questions open. First, existing studies are largely local, leaving unclear whether individual femicides that achieve nationwide salience can mobilize responses across an entire country. Second, increases in help-seeking and reporting may reflect either greater awareness and reduced underreporting or increased violence through imitative behavior, yet little is known about whether public salience affects lethal violence. Third, the mechanisms underlying these responses remain poorly understood.

We address these questions by studying highly salient femicides and their consequences for help-seeking, police reporting, and femicides nationwide, while exploiting variation across events and outcomes to investigate the mechanisms behind these effects. Jointly examining behavioral responses and lethal violence is critical because femicides are measured with far less error than help-seeking and reporting [15]. This also allows us to test whether public salience mobilizes help-seeking and, potentially, prevention, or instead triggers copycat violence through imitative behavior [16–18]. The latter possibility is frequently raised in debates over extensive media coverage of gender-based violence, whereby men exposed to coverage of a femicide may identify with or emulate the perpetrator and subsequently commit similar acts of lethal violence. Finally, comparisons across different types of salient violent events and outcomes allow us to investigate more directly the mechanisms underlying these responses, including the role of the victim–perpetrator relationship and the gendered nature of the violence.[1]

Our empirical analysis begins with a newly hand-curated dataset covering femicide cases recorded across Italy between 2017 and 2024. For each case, we systematically classify the victim–perpetrator relationship—distinguishing intimate partners, family members, and other perpetrators—and, within intimate partner femicides (IPFs), the circumstances and motives associated with the killing, including jealousy or possessiveness, caregiver burnout, and other motives. Rather than selecting prominent cases ad hoc, we begin by objectively measuring the public salience of each femicide in the dataset. For every victim, we quantify media attention across 144 national and local news outlets, and for the most salient cases, we complement these measures with data on social-media posts and user engagement (see Supplementary Information A for more details).

Figure 1, Panel **a**, shows the 15 most salient cases among those eligible for the analysis (see Methods). Three cases clearly stand out at the top of the salience ranking and constitute the main events analyzed in the paper: the IPFs of Giulia Cecchettin and Giulia Tramontano, and the femicide of Saman Abbas by family members. The first stands out by far as the most salient event in the dataset, generating an exceptional wave of media attention and public mobilization that culminated in large demonstrations across the country. Tramontano is the second-most salient IPF, whereas Abbas is the most salient femicide occurring outside an intimate-partner relationship, with a level of salience similar to that of Tramontano. This combination allows us to compare highly salient IPFs of different intensity while also examining a similarly salient femicide involving a different victim–perpetrator relationship.

[1] Our contribution is also methodological. Existing studies often rely on unaffected geographic areas to identify the effects of salient events, an approach that is feasible when public attention remains geographically concentrated. Highly salient femicides, as we show in this study, can attract nationwide attention, leaving no unaffected area that can serve as a control group. We therefore exploit the sharp and plausibly unanticipated timing of each event within a high-frequency interrupted time series framework. This approach allows us to estimate responses to multiple salience shocks within a common institutional and data environment and to compare them across events differing in salience and type. See Supplementary Information D for further details on the research design.

Panel **b** shows the full distribution of media salience, with complementary Facebook evidence in the inset.

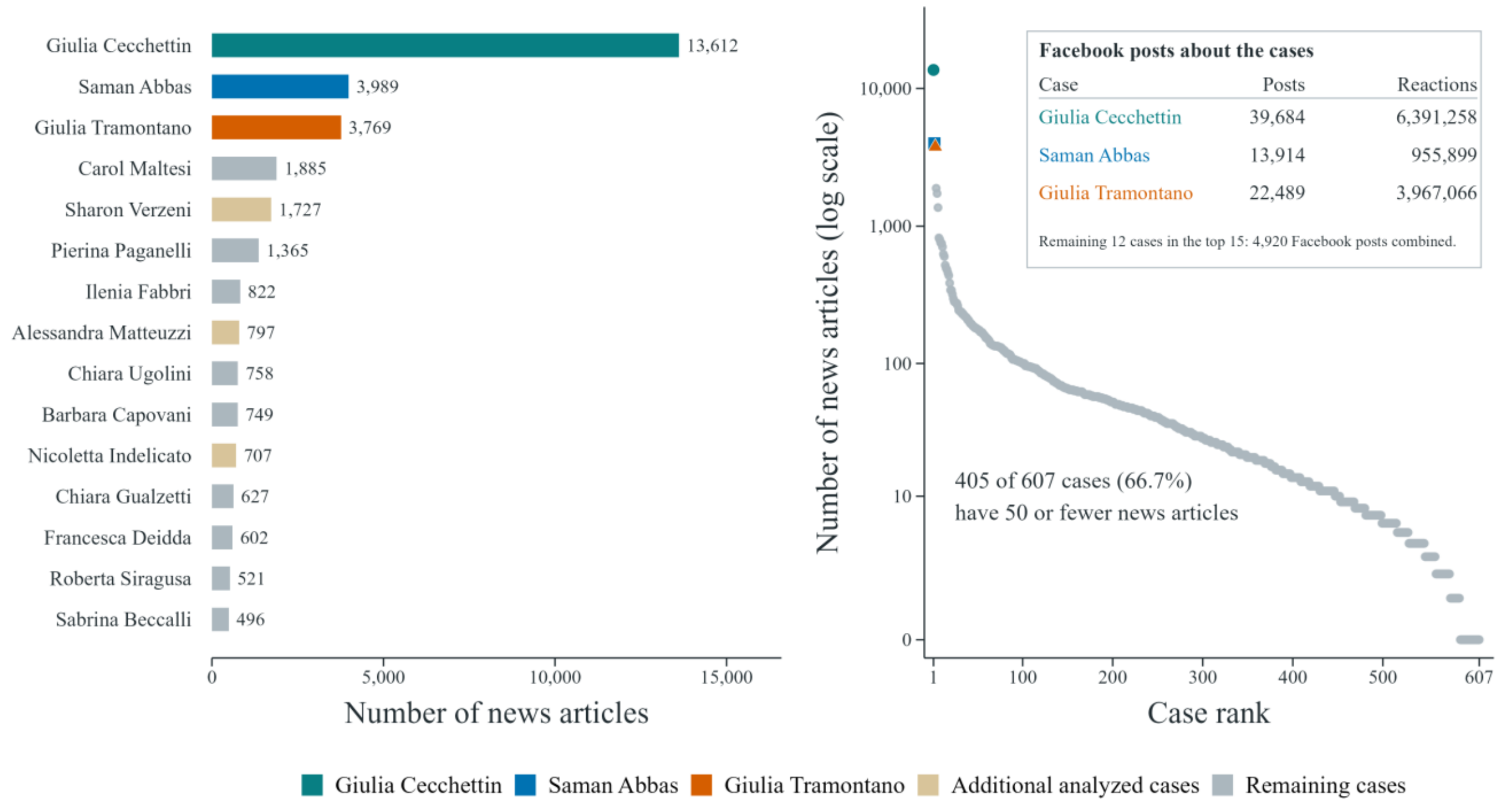

a) Media salience

b) Distribution of media salience

| Case | Classification | Victim-perpetrator relationship | Nationwide salience | Analytical role |
|---|---|---|---|---|
| Giulia Cecchettin | Femicide | IPF | Exceptional | Main case |
| Giulia Tramontano | Femicide | IPF | Very high | Main case |
| Saman Abbas | Femicide | Family-perpetrated | Very high | Main case |
| Sharon Verzeni | Killing of a woman, not classified as femicide | No prior relationship | High | Gender-related mechanism |
| Alessandra Matteuzzi | Femicide | IPF | Lower | Salience intensity comparison |
| Nicoletta Indelicato | Femicide | Non-IPF; multiple perpetrators | Lower | Falsification comparison |

c) Selected cases

**Fig. 1 | Public salience of femicides in Italy and selected cases. a,** The 14 femicide cases and one other killing of a woman receiving the highest number of news articles in Media Cloud. **b,** News article counts for all 607 femicide cases recorded between January 2019 and December 2024, ranked from highest to lowest. Cases occurring in 2017 and 2018 are excluded because our identification strategy requires at least 24 months of pre-treatment observations. The vertical axis is shown on a logarithmic scale, with tick labels reported in the original count units. Giulia Cecchettin, Saman Abbas and Giulia Tramontano, the three cases examined in the main empirical analysis, are highlighted in teal, blue and orange, respectively; together, they account for 34.2% of all articles attributed to individual femicide cases. Circles, squares and triangles identify these cases, respectively, in **b**. In **a**, muted gold highlights the three additional cases examined in the supplementary analyses: Sharon Verzeni, Alessandra Matteuzzi and Nicoletta Indelicato. The Verzeni case is included here solely to document its high media salience: it was considered in the initial screening of cases but excluded from the final femicide dataset following a case-by-case review, which classified it as a homicide rather than a femicide. The inset in **b** provides complementary evidence on public salience by reporting the number of Facebook posts concerning the main cases and the reactions to those posts. **c,** Classification of the six cases examined in the study according to the femicide classification, victim–perpetrator relationship, relative nationwide salience, and analytical role.

We complement these three main cases with three comparisons. To isolate the role of salience intensity, we select a substantially lower-salience IPF case (Alessandra Matteuzzi). We further select the most salient case among non-femicide killings of women, Sharon Verzeni, which allows us to examine whether salience itself outside the context of gender-based violence generates a similar response. Finally, we study a lower-salience, multiple-perpetrator femicide—Nicoletta Indelicato—as an additional falsification comparison. Panel **c** of Figure 1 summarizes the characteristics and analytical roles of the six selected cases. Together, the six selected cases provide variation in salience intensity, the victim–perpetrator relationship, and the gendered nature of the violence within a common institutional environment, allowing us to examine how each dimension shapes the response. Supplementary Information B provides detailed information on the background, timing, and aftermath of each case.

Exploiting the sharp and unanticipated timing of these events in a high-frequency interrupted time-series design, we estimate counterfactual post-event trajectories from pre-event dynamics. We find that the two highly salient IPFs generate substantial nationwide increases in help-seeking, with the strongest responses occurring during periods of peak public attention. Following the exceptionally salient IPF, we also document a marked nationwide increase in police reports of violence against women. By contrast, we find no evidence that either event increases total femicides, providing no support for copycat effects on lethal violence. If anything, point estimates following the exceptionally salient IPF case suggest a sizeable decline in the number of IPFs, despite the limited statistical precision inherent in rare outcomes. Administrative data on female homicides show a similar pattern, whereas male homicides do not. Subnational analyses further indicate that the increases in help-seeking and reporting are broadly distributed across the country rather than concentrated in the area where the events occurred.

The comparison cases help clarify when salience translates into mobilization. We find no comparable increase in help-seeking or police reporting following the lower-salience IPF or the falsification case. Nor do we observe the same positive mobilization following the highly salient family-perpetrated femicide or the salient killing of a woman not coded as femicide in our dataset. Together, these findings indicate that public salience can mobilize nationwide responses to violence against women without amplifying lethal violence, and that the response depends not only on the intensity of public salience, but also on the victim–perpetrator relationship and the gendered nature of the violence made salient.

## 2. Data and methods

### 2.1 Data

We assembled a new dataset by combining information from multiple sources, starting with the construction of a database of femicide cases in Italy that builds upon previous work [19]. This effort was necessary because no administrative dataset in Italy systematically identifies individual femicide cases on a case-by-case basis. The micro-level dataset covers all femicide cases occurring in Italy between 2017 and 2024 that could be documented from the available sources. We constructed it by digitizing information from reports produced by non-governmental organizations and journalistic sources, followed by an extensive case-by-case verification process (see Supplementary Information A for details).

Based on the relationship between the perpetrator and the victim, we classify femicides into three main categories: i) IPFs, ii) those committed by family members, and iii) other cases. As shown in Figure 2, Panel **a,** IPFs account for the majority of cases in our 2017–2024 sample (65.0% of all femicides), while 23.7% of femicides are committed by family members and 11.3% by friends, colleagues, clients, or strangers. We further divide IPFs into three subcategories to capture differences in motive: cases explicitly driven by jealousy or possessiveness; cases related to caregiver burnout; and cases associated with other motives, such as severe psychological disorders or acute economic distress.[2] Panel **b** shows that jealousy- or possessiveness-related cases constitute the majority of IPFs (60.5%), followed by other motives (23.0%) and caregiver burnout (16.5%). Jealousy- or possessiveness-related IPFs also constitute the category to which the Cecchettin and Tramontano cases belong and, descriptively, its incidence declined from an average of 40 cases per year in 2021–2023 to 32 cases in 2024, corresponding to a reduction of 20%. Panels **c** and **d** document an increase in helpline calls and reported gender-related crimes over time, with the upward trend particularly evident in 2023 and 2024.

[2] Caregiver burnout cases are particularly frequent among elderly couples.

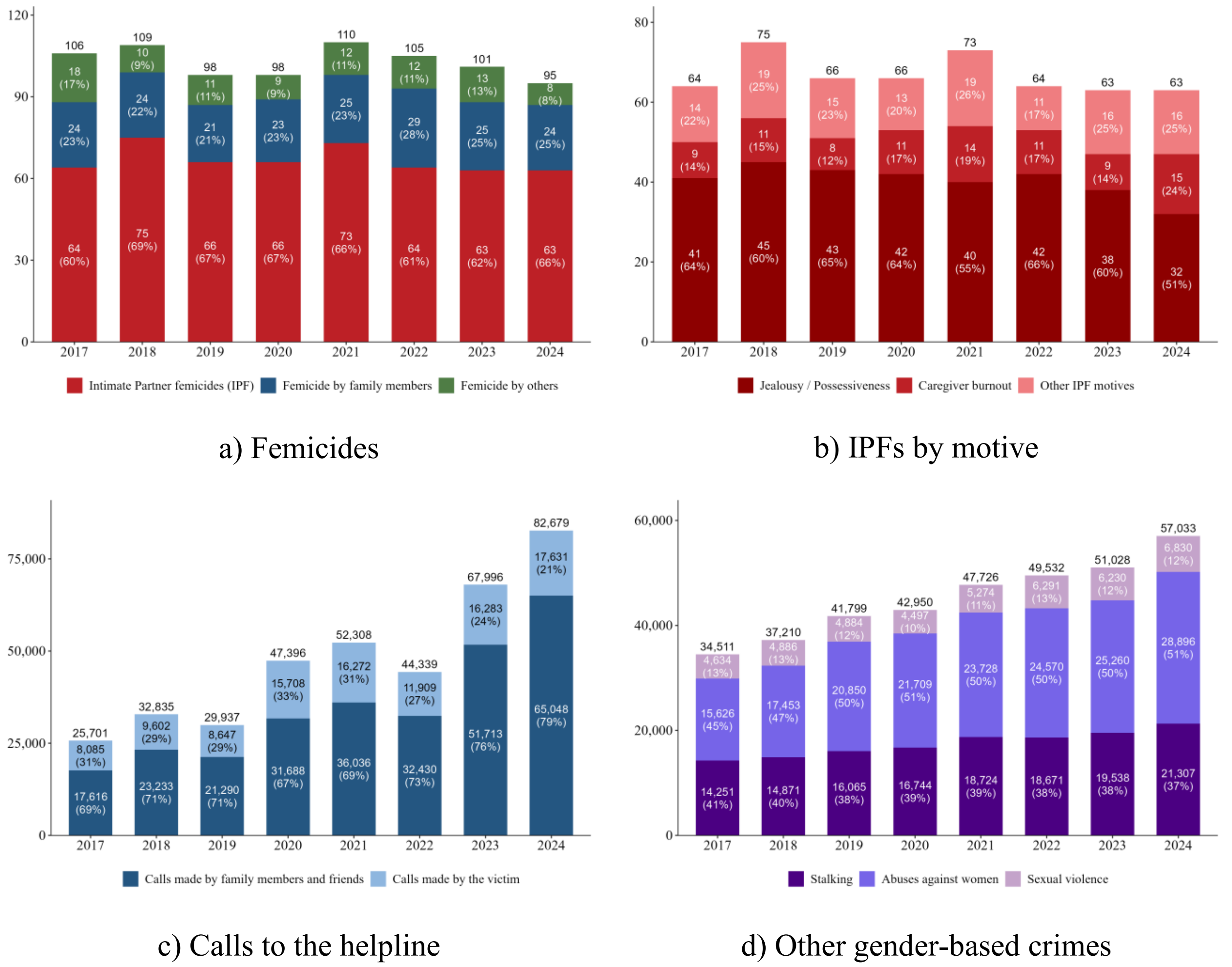

**Fig. 2 | Descriptive statistics.** The Figure shows the descriptive statistics of our data on femicides, help-seeking behavior, and other gender-related crimes. **a,** shows total femicides by the victim-killer relationship. **b,** identifies IPFs by their motive. **c,** distinguishes between calls to the national helpline made by victims and family or friends. **d,** shows the distribution of other gender-related crimes.

We complement these data with administrative records on gender-based violence from the central database of Italian Police (SDI - *Sistema di Indagine*). These records provide monthly information on gender-related crimes reported to the police, for the period 2017-2024. These data include stalking, sexual violence, and abuse against women. The SDI also records homicides, separately for male and female victims, and provides information on whether the perpetrator was an intimate partner of the victim. Finally, we use data from the Italian National Institute of Statistics (ISTAT) on calls to the national anti-violence helpline (1522), including information on whether calls were placed directly by victims or by relatives or friends. All these data can also be disaggregated at the sub-national level, distinguishing between Northern, Central, and Southern Italy, allowing us to test whether the effects of the salience shocks are localized near the site of the femicide or truly nationwide in scope.

For the main analysis, we consider six monthly outcomes spanning help-seeking, police reporting, and lethal violence: number of calls to the national 1522 anti-violence helpline; reports of abuse against women, stalking, and sexual violence; number of IPFs associated with jealousy or possessiveness; and total femicides, defined as the sum of jealousy- or possessiveness-related IPFs and all other femicides in our dataset. The first outcome captures help-seeking behavior, the next three capture violence reported to the police, and the final two capture lethal violence. As additional analyses, we disaggregate 1522 calls according to whether they were made directly by victims or by relatives and friends, allowing us to assess whether salience mobilizes only potential victims or also their broader social networks. To assess the robustness of the results on femicides, we additionally use administrative data on female homicides as an independently measured lethal-violence outcome. We also examine male homicides as a falsification outcome, allowing us to assess whether any post-event changes in homicide reflect broader homicide dynamics rather than responses specific to violence against women. Comparing these outcomes with those directly related to violence against women helps assess whether the observed mobilization reflects a gender-specific response rather than a broader reaction to salient violent crime.

## 2.2 Methods

All analyses use a common monthly sample beginning in 2017. Although the length of the pre-event period varies across cases depending on the timing of each femicide, this common starting point provides a sufficiently long history from which the models can learn underlying trends, seasonality, and other evolving features of the outcome series. This is particularly valuable because the three most salient cases occur relatively late in the sample. For each event, the model uses the full observed history up to the month preceding the femicide under analysis to construct the counterfactual path that would have occurred in its absence. The procedure is recursive: when analyzing a later event, the realized history following earlier salient cases is already part of the information used to construct the new counterfactual, while post-event predictions are generated sequentially from previous counterfactual values. The model therefore accounts for persistent responses to earlier events insofar as they are reflected in the pre-event history used to construct the baseline trajectory against which the later shock is evaluated. When post-treatment periods overlap, however, the effects of successive events cannot be separately attributed without additional assumptions. We therefore restrict event-specific interpretations to non-overlapping horizons. This issue is mainly relevant for Tramontano and Cecchettin, the two most salient IPFs: the main Tramontano analysis is limited to a six-month window ending before the Cecchettin shock, while at longer horizons we additionally report estimates that treat the two salient IPFs jointly rather than separately.

We use an interrupted time-series design [20–22], comparing observed post-event outcomes with the trajectory expected in the absence of the event. To estimate this counterfactual flexibly, we use TabPFN 3.0 [23,24]. Rather than imposing a fixed parametric trend or a specific autoregressive structure, TabPFN uses the pre-event evolution of the outcome together with lagged dynamics, seasonality, calendar effects, and information from related series to predict the path that would have been expected after the event. This is useful in our setting because the monthly series are relatively short and may exhibit nonlinear and changing dynamics.

We estimate both month-specific effects and cumulative effects over the post-event period and use the predictive distribution generated by TabPFN to characterize uncertainty around the counterfactual trajectory.[3] Supplementary Information E reports a broad set of robustness analyses and also repeats the main analysis using an established causal time-series method [25] based on a different modeling framework. The results are consistent with the main estimates and provide complementary inference under a different set of modeling assumptions.

Causal interpretation of the observed–counterfactual differences relies on the maintained assumptions that the timing of each event is unanticipated, that no contemporaneous shock independently generates the same post-event deviations and that, absent the event, the pre-treatment relationships remain informative over the post-treatment horizon.[4] Under these assumptions, the estimated differences can be interpreted as causal effects of the salience shocks on help-seeking, police reporting, and lethal violence. We assess the plausibility of these assumptions through pre-treatment out-of-sample predictive performance, comparisons across events and outcomes, falsification cases, robustness to the alternative causal time series method, and a detailed examination of potentially relevant contemporaneous shocks. Supplementary Information D provides full methodological details.

## 3. Results

Figures 3–5 report six-month post-event estimates for the three main cases, while Fig. 6 extends the horizon from the Tramontano occurrence through the end of 2024 to characterize the joint longer-run response to the two highly salient IPFs. The displayed series for Tramontano and Cecchettin begin in 2021 for readability, but all models are estimated using the common sample beginning in 2017. In each figure, the solid blue line shows observed outcomes, the dashed orange line pre-treatment one-step-ahead forecasts, and the dashed red line the recursively generated post-treatment counterfactual.

[3] These are pointwise, model-based predictive intervals around the counterfactual and are not formal causal confidence intervals for the estimated effects. See Supplementary Information D for more details.

[4] The exact timing of the femicides is plausibly exogenous to short-run dynamics in help-seeking, reporting, and lethal violence, and the events themselves could not reasonably have been anticipated. We account for recurring seasonal patterns, including the November period surrounding the International Day for the Elimination of Violence against Women, and examine potentially relevant contemporaneous events in Supplementary Information D.

Shaded bands denote 95% predictive intervals around the counterfactual.

**Highly salient IPF.** The highly salient IPF of Giulia Tramontano generated a clear increase in help-seeking (Fig. 3). Calls to the national 1522 helpline moved above their counterfactual trajectory soon after the femicide and remained elevated throughout much of the six-month window, yielding a cumulative increase of 29.48% over the six-month post-event window. By contrast, police reports of abuse against women, stalking, and sexual violence show no comparable departure from their counterfactual paths. Nor do we find evidence of an increase in total femicides or in jealousy- or possessiveness-related IPFs. The Tramontano case therefore provides the first indication that a highly salient IPF can mobilize the initial stage of help-seeking, increasing calls for support without producing a corresponding rise in police reports or lethal violence.

**Exceptionally salient IPF.** The response is substantially larger and broader following the exceptionally salient Cecchettin femicide (Fig. 4).[5] Helpline calls rise sharply, with the largest divergence occurring in the first months after the event and a cumulative increase of roughly one-half over the six-month window (+47.54% over the six-month post-event window). Police reporting also increases: reports of abuse against women remain persistently above their counterfactual trajectory (+22.84%), while stalking reports rise more moderately (+12.76%); reports of sexual violence show no comparable response (+7.56%), with none of the observed monthly outcomes exceeding the upper bound of the 95% predictive interval. Given the substantial underreporting of gender-based violence, these increases in police reports are more plausibly interpreted as previously hidden violence coming to the attention of the authorities than as additional violence induced by the salience shock. The concurrent absence of any increase in lethal violence further supports this interpretation (see Supplementary Information A for a fuller discussion). Total femicides lie predominantly below the estimated counterfactual, and jealousy- or possessiveness-related IPFs also decline on average, although these rare outcomes are estimated imprecisely. Taken together, these findings are inconsistent with the copycat hypothesis and instead point to heightened salience as a mechanism that brings otherwise hidden violence to light by prompting victims to recognize their own experiences, seek help, and report them, particularly when salience reaches exceptionally high levels.

**Highly salient non-IPF.** The highly salient family-perpetrated femicide of Saman Abbas provides a sharp contrast (Fig. 5). Despite a level of public salience comparable to Tramontano, we find no similarly coherent mobilization in helpline calls or police reporting, and no evidence of a systematic response in lethal violence. The contrast between Abbas and the two salient IPFs indicates that salience alone is not sufficient: the victim–perpetrator relationship, and the extent to which the event maps onto risks recognizable as intimate-partner violence, appear to shape whether public attention translates into help-seeking and reporting.

[5] Cecchettin was not simply the highest-ranked case in the media-based salience distribution, but an outlier across multiple independent indicators of public attention. Supplementary Information C complements the evidence in Figure 1 with descriptive evidence from online search, social media, and street mobilization. All three indicators show an unusually large and synchronized response following the femicide, including record-high search interest, intense social-media engagement, and exceptionally large demonstrations against gender-based violence.

**Joint longer-run response.** Because Cecchettin occurs approximately six months after Tramontano, effects at longer horizons cannot be separately attributed to the two events without additional assumptions. Figure 6 therefore treats the sequence of the two salient IPFs jointly, beginning from the Tramontano case. The longer window shows that the increase in helpline calls is not confined to an immediate short-lived spike: it intensifies sharply after Cecchettin and remains elevated through the end of the available series. Reports of abuse against women and stalking also move above their counterfactual trajectories after the Cecchettin shock, with the observed deviations falling outside the 95% predictive intervals in several post-event periods, whereas the lethal-violence outcomes provide no evidence of a sustained increase. The joint estimates thus reinforce the central pattern: repeated highly salient IPFs generate substantial and persistent mobilization without a corresponding amplification of lethal violence.

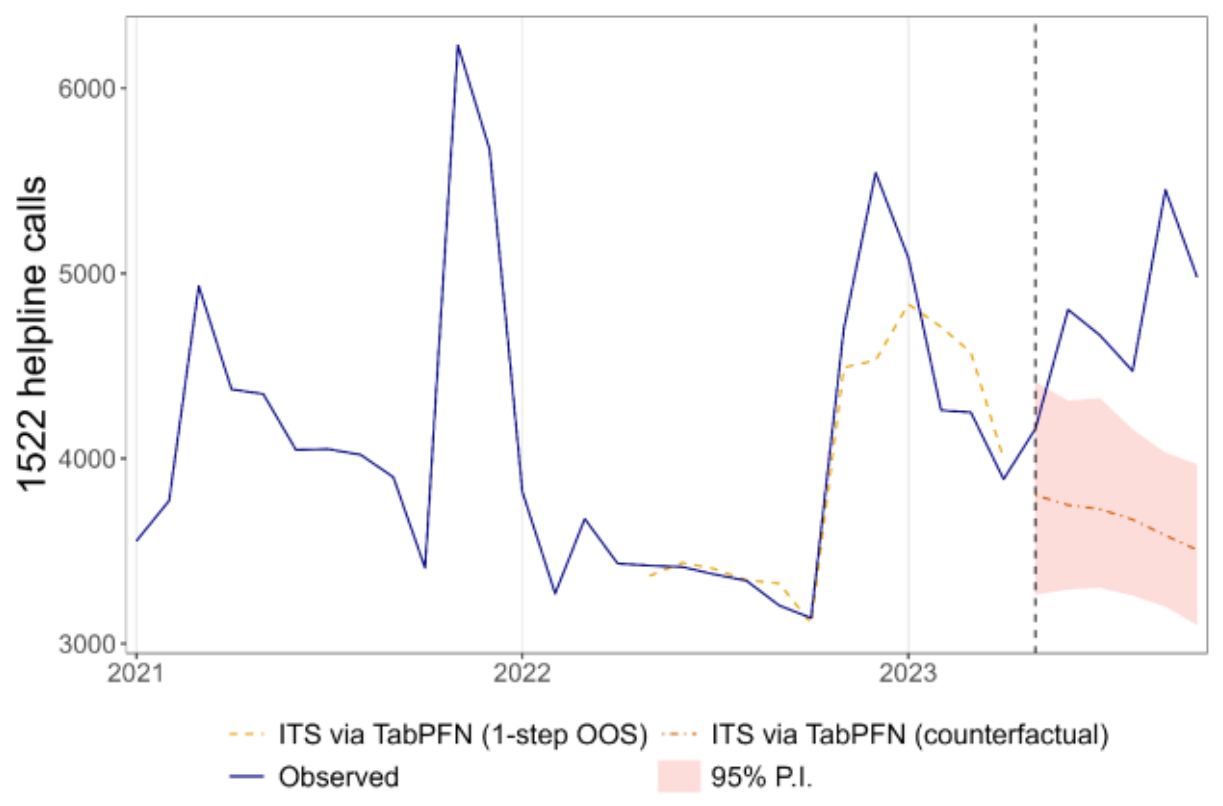

a) Calls to the helpline

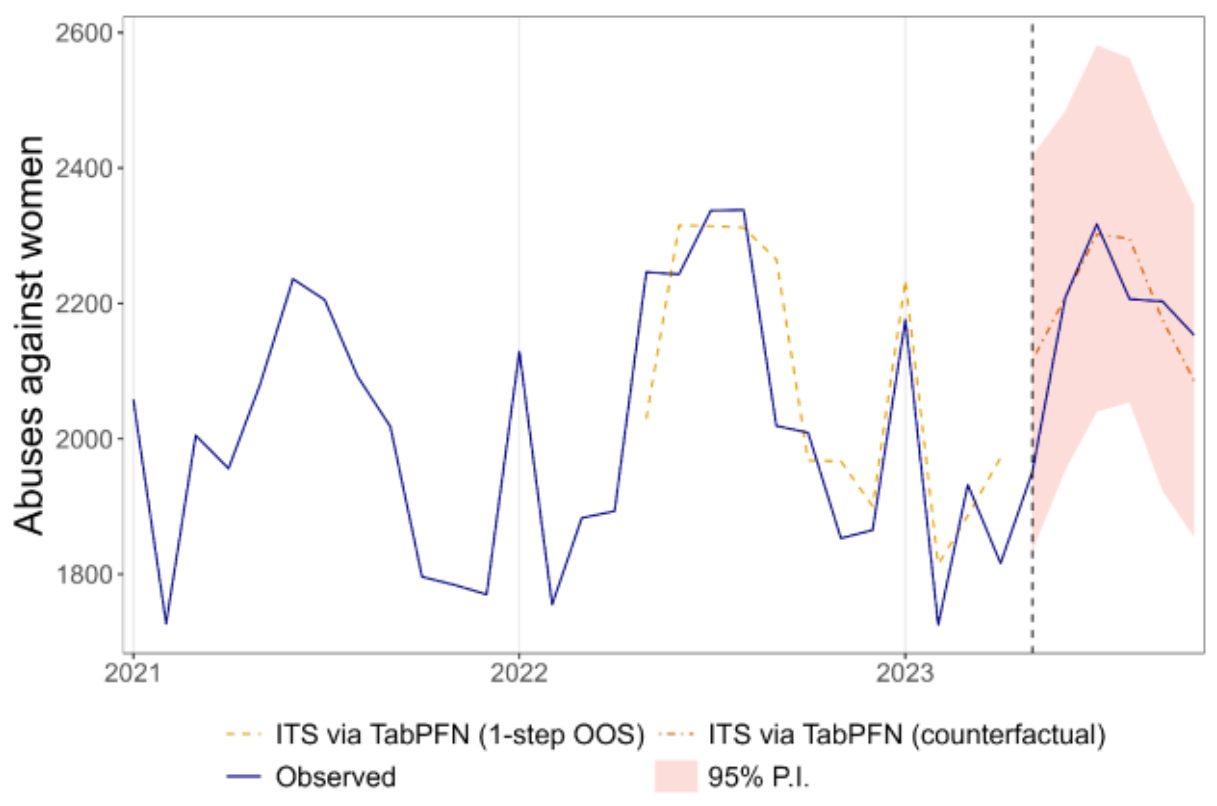

b) Abuses against women

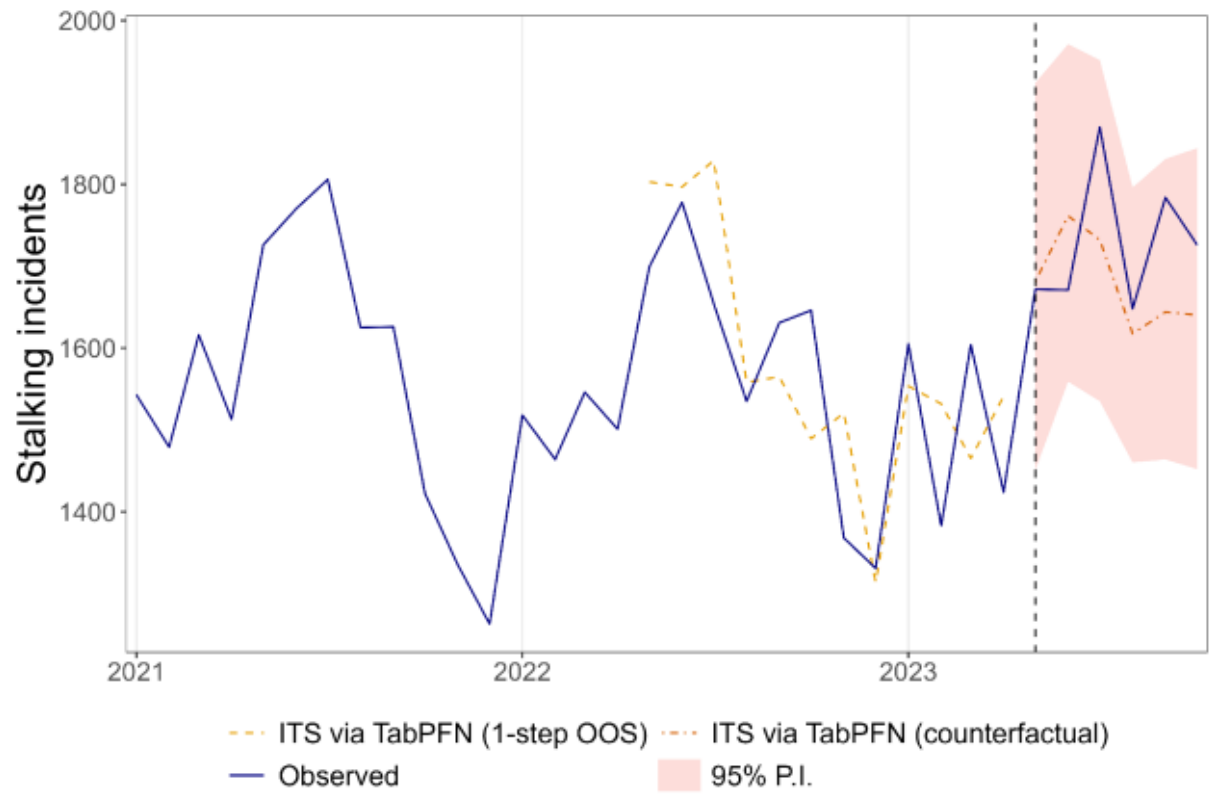

c) Stalking cases

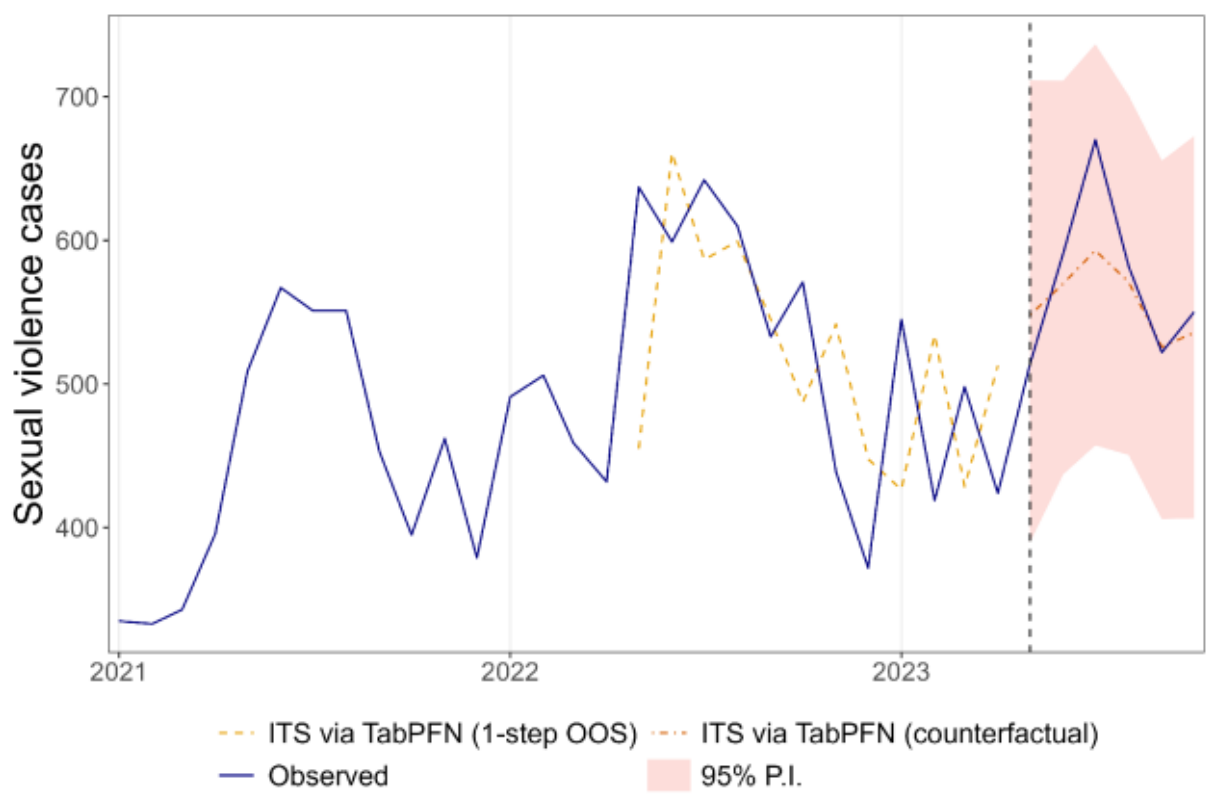

d) Sexual violence cases

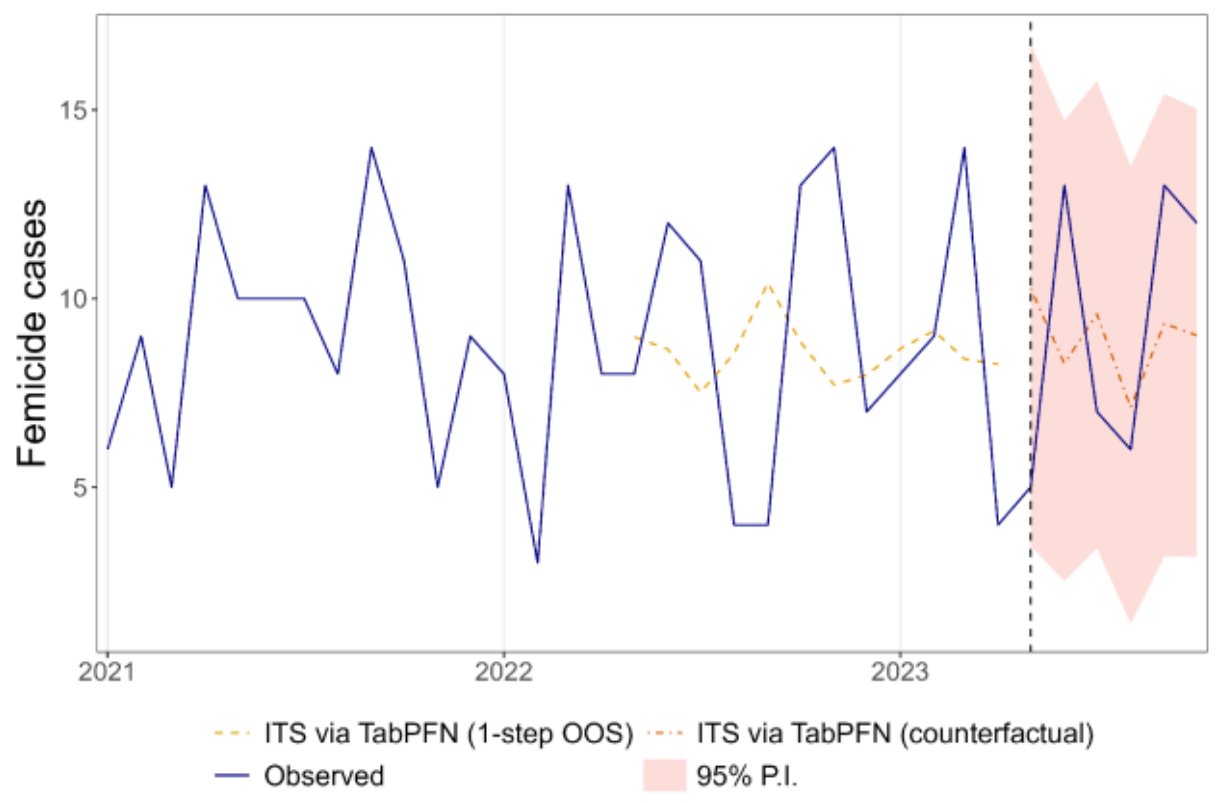

e) All femicides

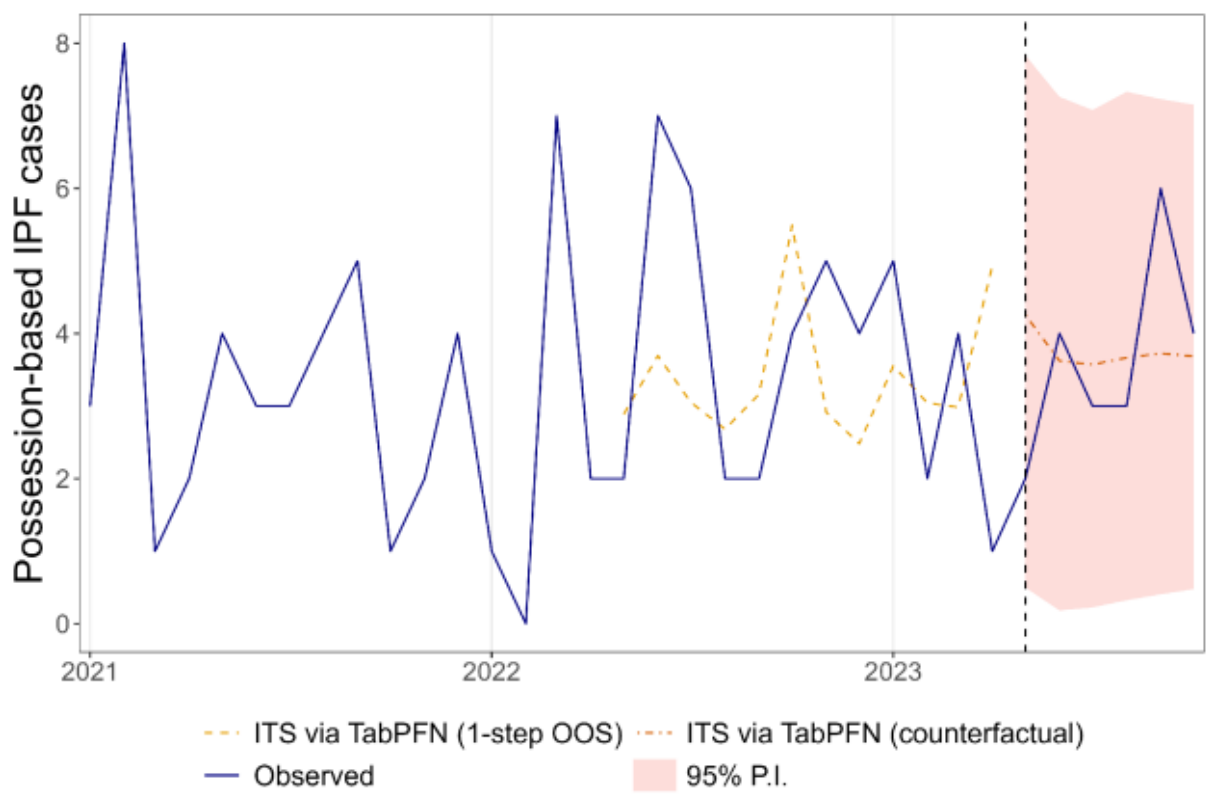

f) IPFs driven by jealousy or possessiveness

**Fig. 3 | Responses following the highly salient IPF of Giulia Tramontano. a,** Calls to the national 1522 helpline. **b,** Reports of abuse against women. **c,** Stalking. **d,** Sexual violence. **e,** Total femicides. **f,** IPFs associated with jealousy or possessiveness. The solid blue line shows observed monthly values; the dashed orange line shows pre-treatment one-step-ahead forecasts; the dashed red line shows the recursively generated post-treatment counterfactual. Shading denotes the 95% predictive interval around the counterfactual. The vertical dashed line marks the event month. See Supplementary Information D for estimation and uncertainty details.

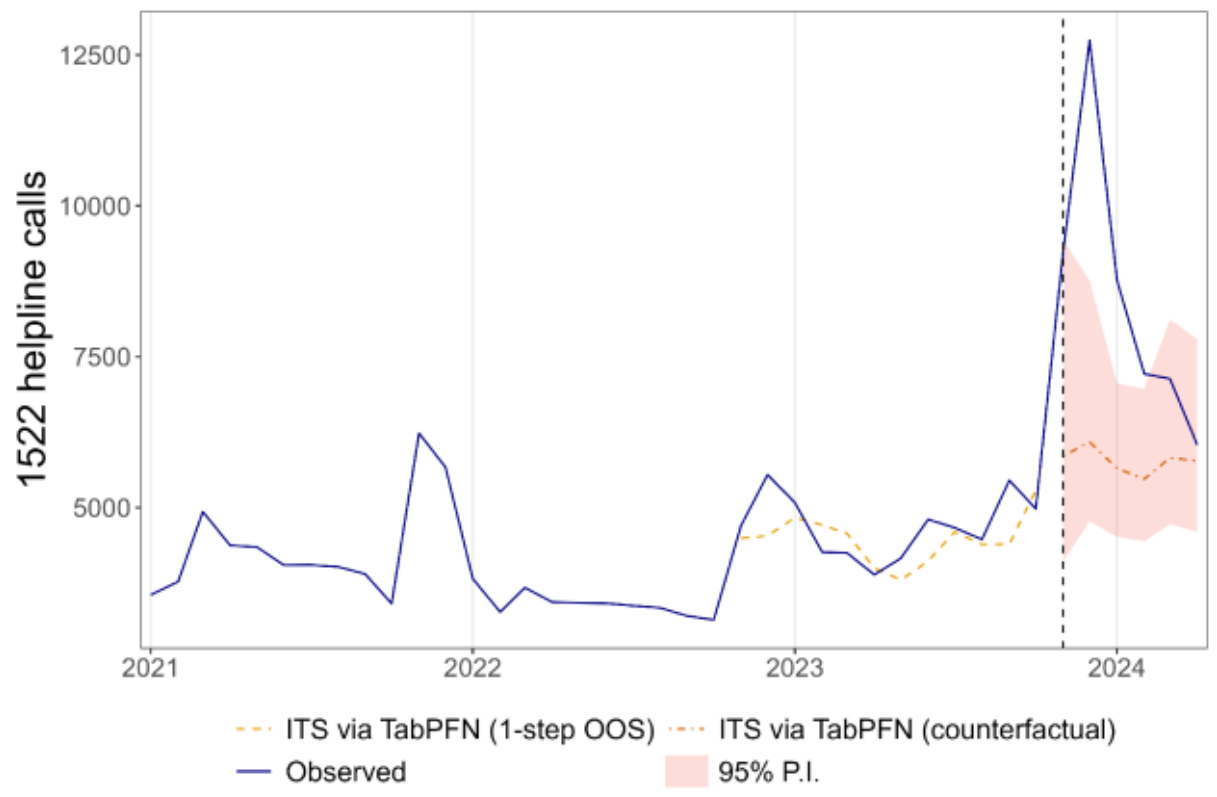

a) Calls to the helpline

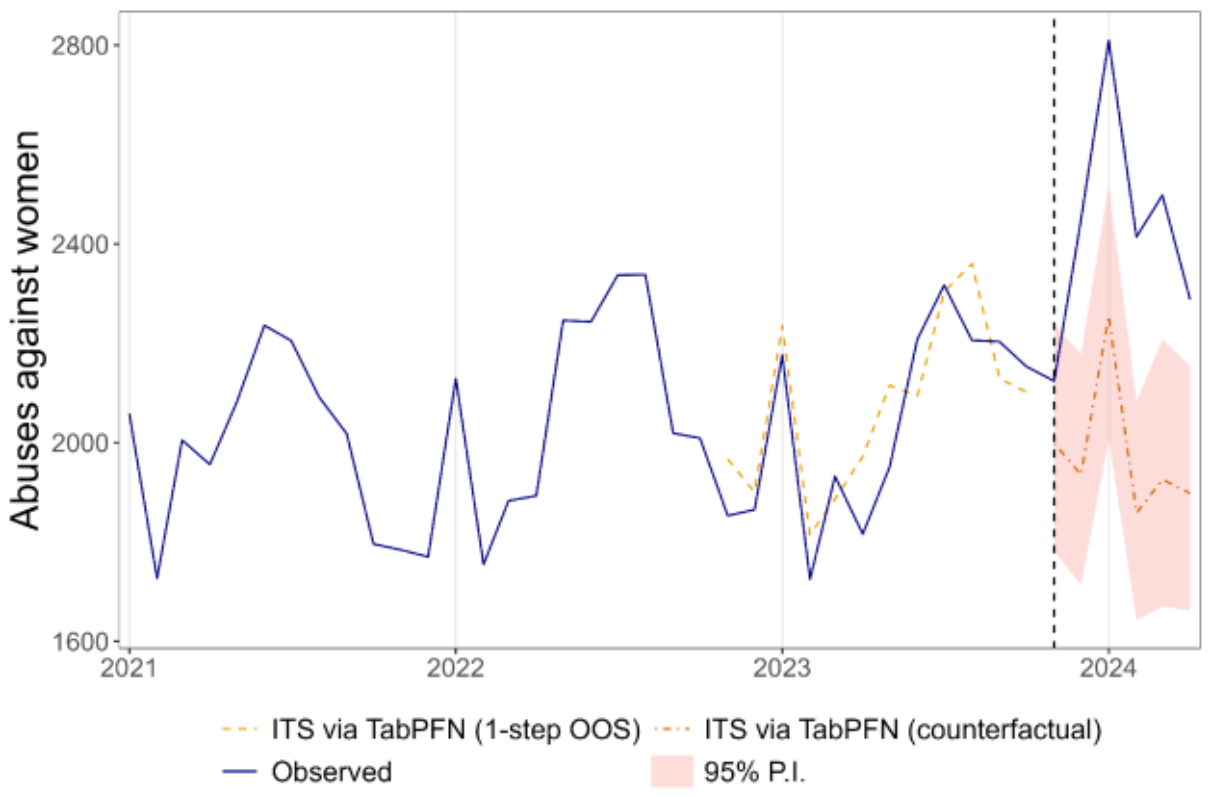

b) Abuses against women

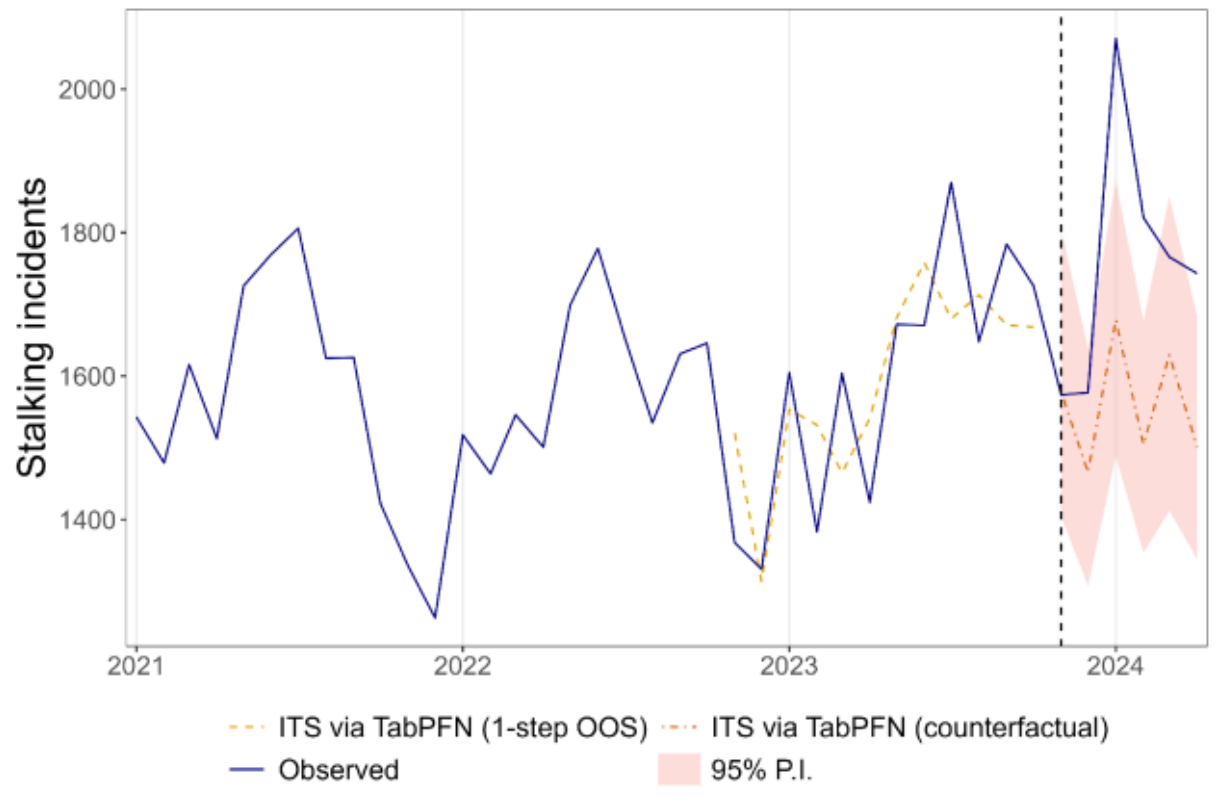

c) Stalking cases

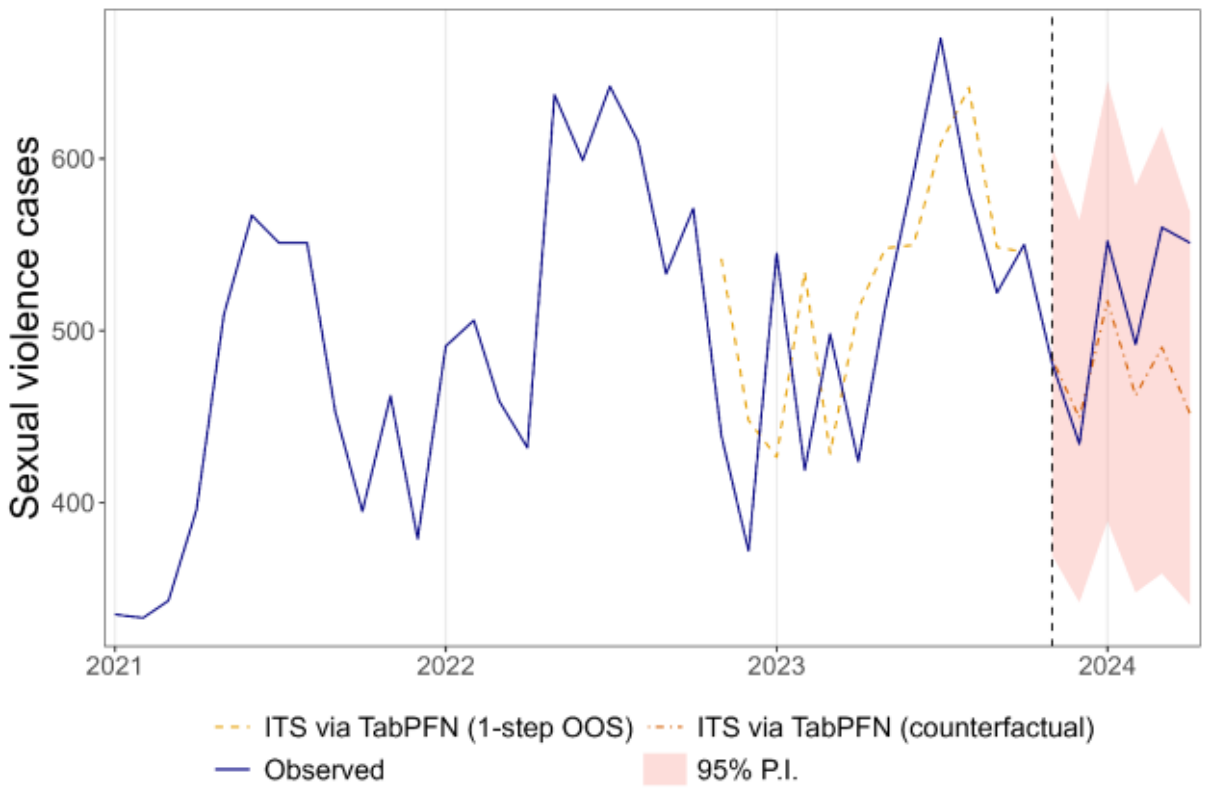

d) Sexual violence cases

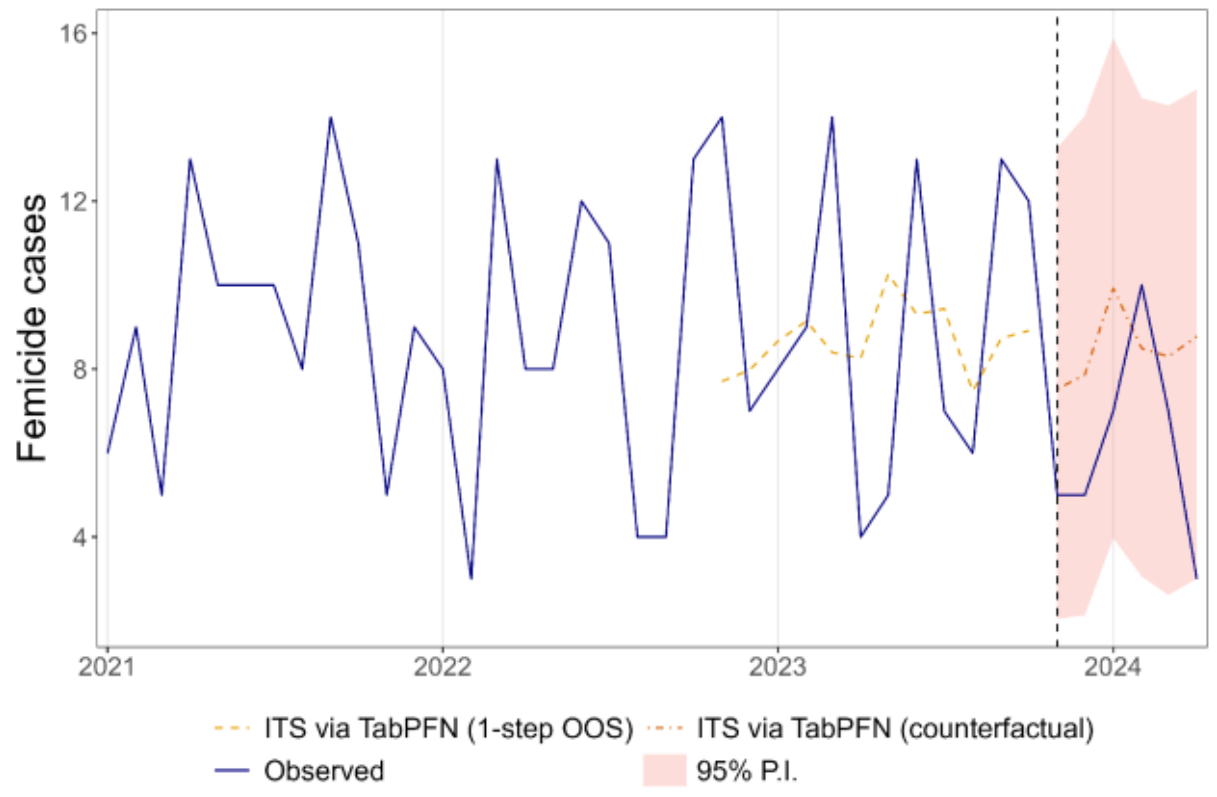

e) All femicides

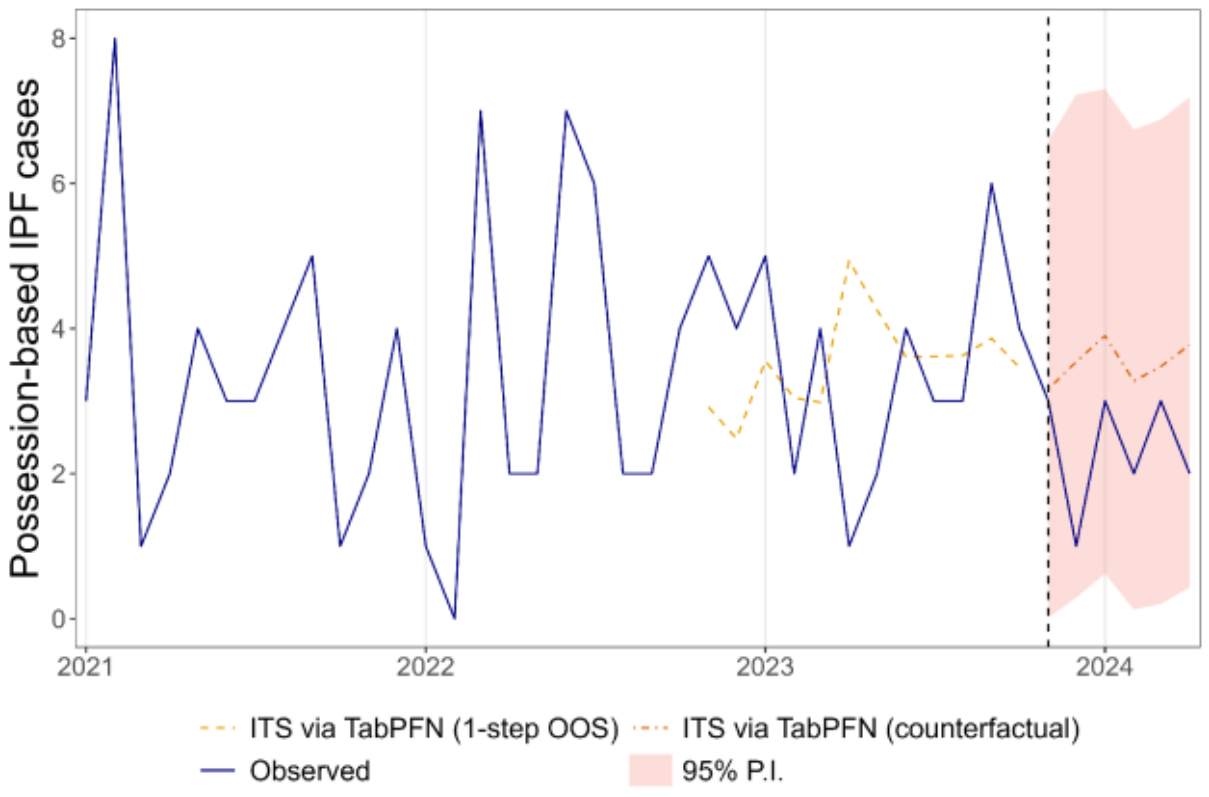

f) IPFs driven by jealousy or possessiveness

**Fig. 4 | Responses following the exceptionally salient IPF of Giulia Cecchettin. a,** Calls to the national 1522 helpline. **b,** Reports of abuse against women. **c,** Stalking. **d,** Sexual violence. **e,** Total femicides. **f,** IPFs associated with jealousy or possessiveness. The solid blue line shows observed monthly values; the dashed orange line shows pre-treatment one-step-ahead forecasts; the dashed red line shows the recursively generated post-treatment counterfactual. Shading denotes the 95% predictive interval around the counterfactual. The vertical dashed line marks the event month. See Supplementary Information D for estimation and uncertainty details.

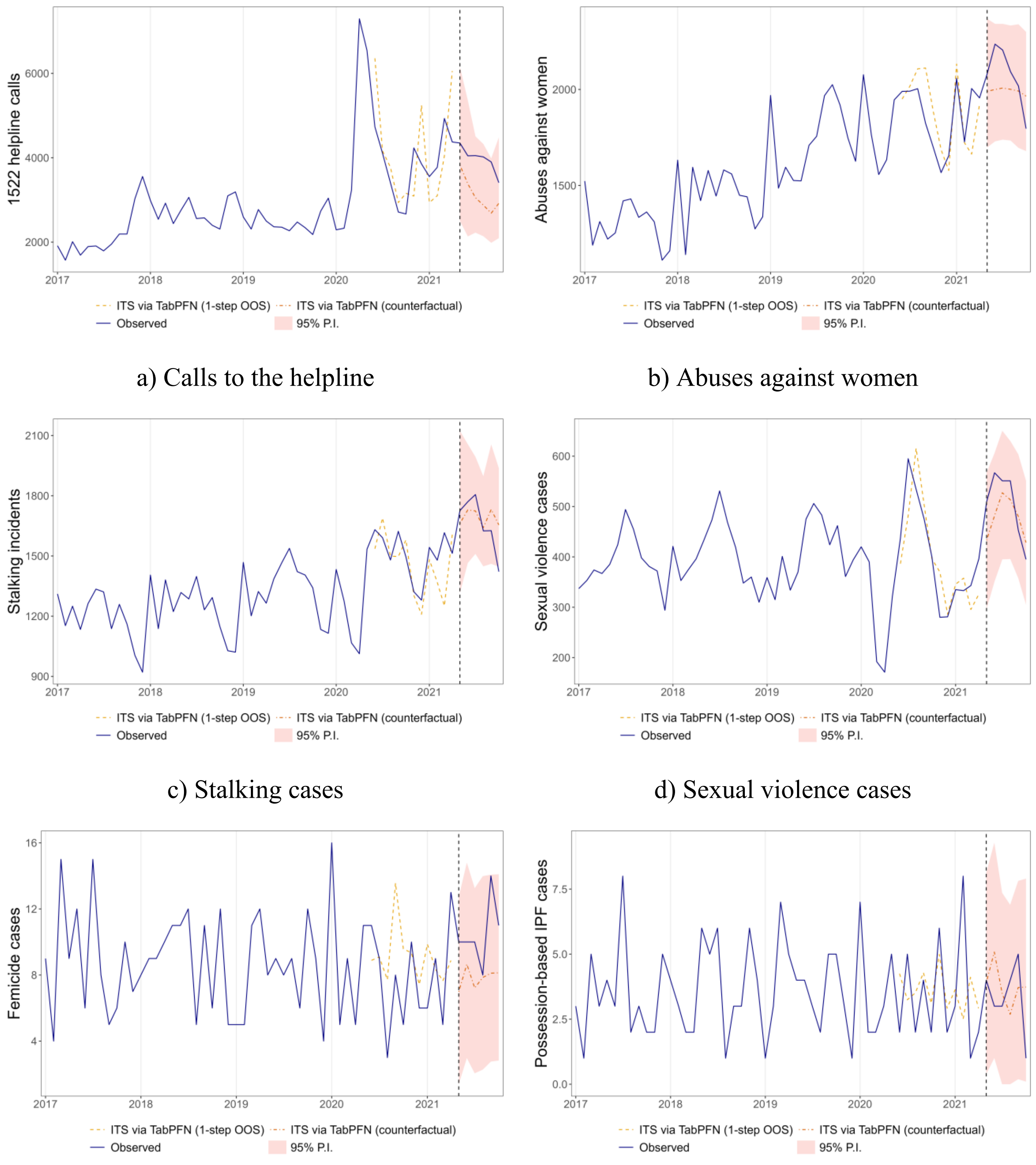

a) Calls to the helpline
b) Abuses against women
c) Stalking cases
d) Sexual violence cases
e) All femicides
f) IPFs driven by jealousy or possessiveness

**Fig. 5 | Responses following the highly salient non-IPF of Saman Abbas. a,** Calls to the national 1522 helpline. **b,** Reports of abuse against women. **c,** Stalking. **d,** Sexual violence. **e,** Total femicides. **f,** IPFs associated with jealousy or possessiveness. The solid blue line shows observed monthly values; the dashed orange line shows pre-treatment one-step-ahead forecasts; the dashed red line shows the recursively generated post-treatment counterfactual. Shading denotes the 95% predictive interval around the counterfactual. The vertical dashed line marks the event month. See Supplementary Information D for estimation and uncertainty details.

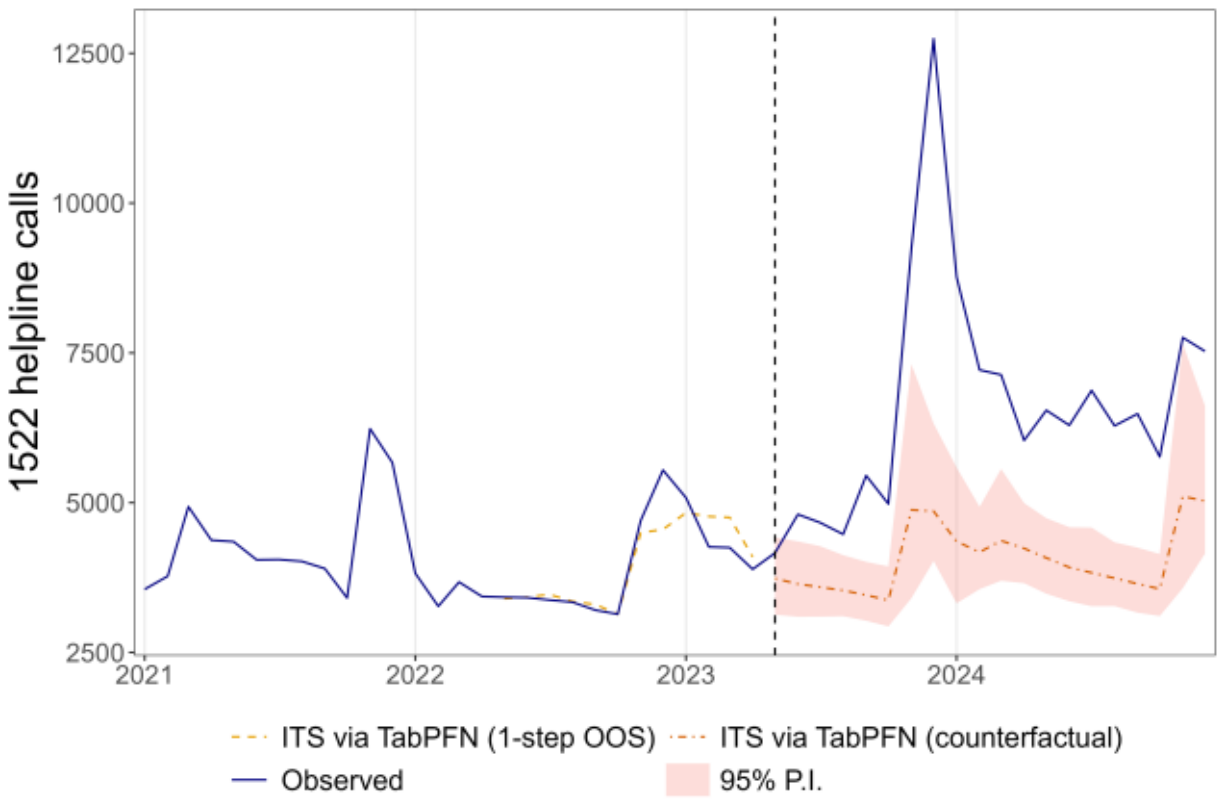

a) Calls to the helpline

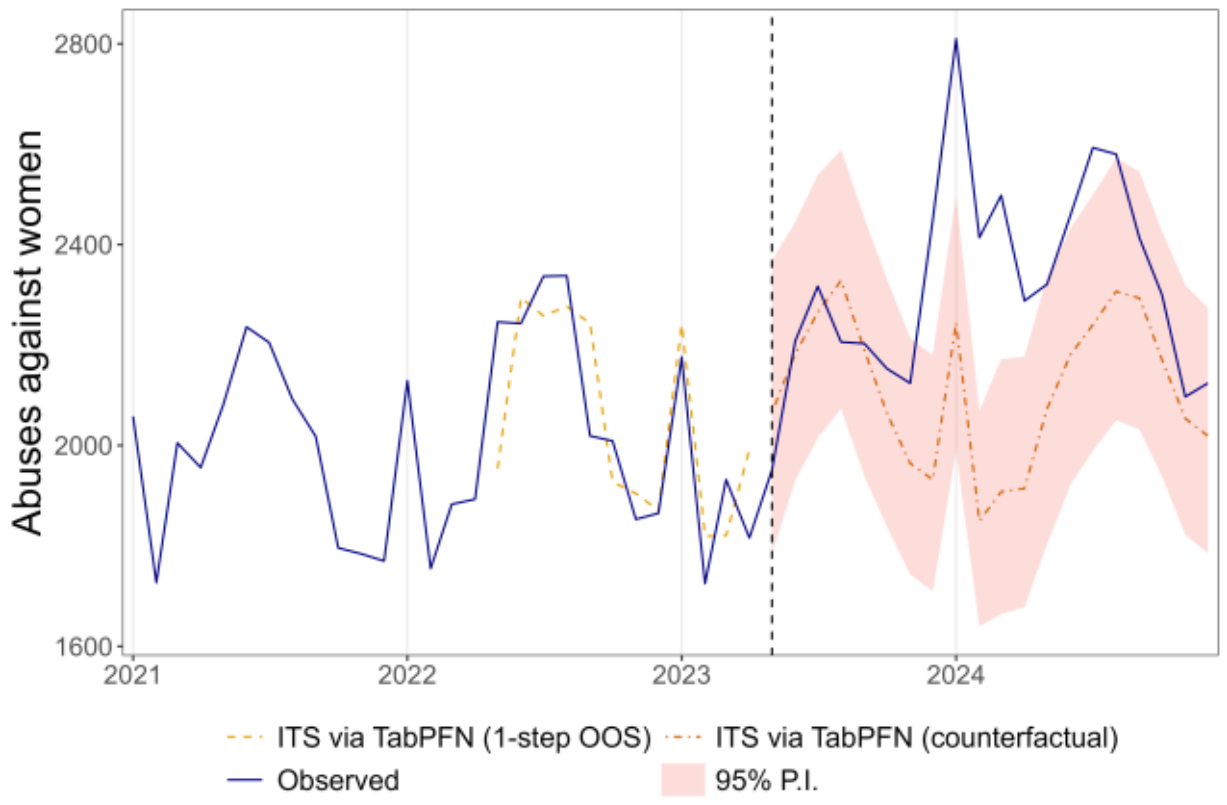

b) Abuses against women

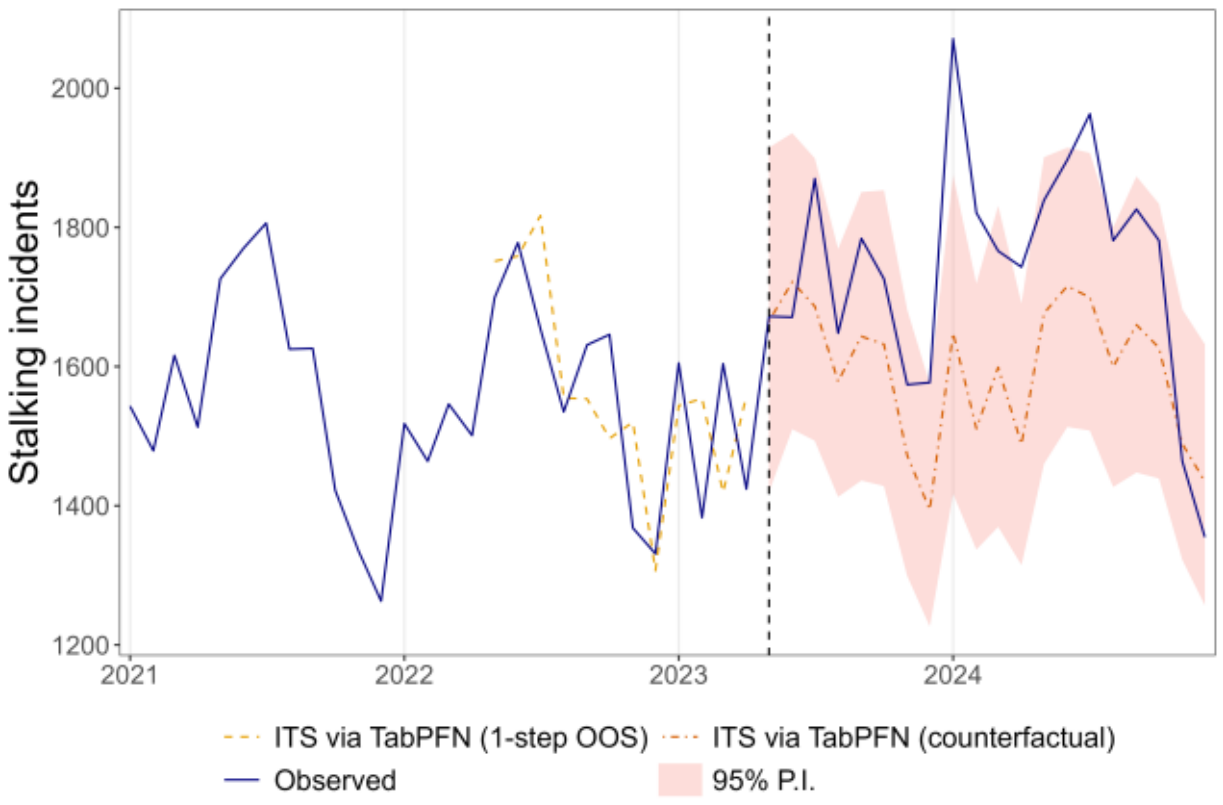

c) Stalking cases

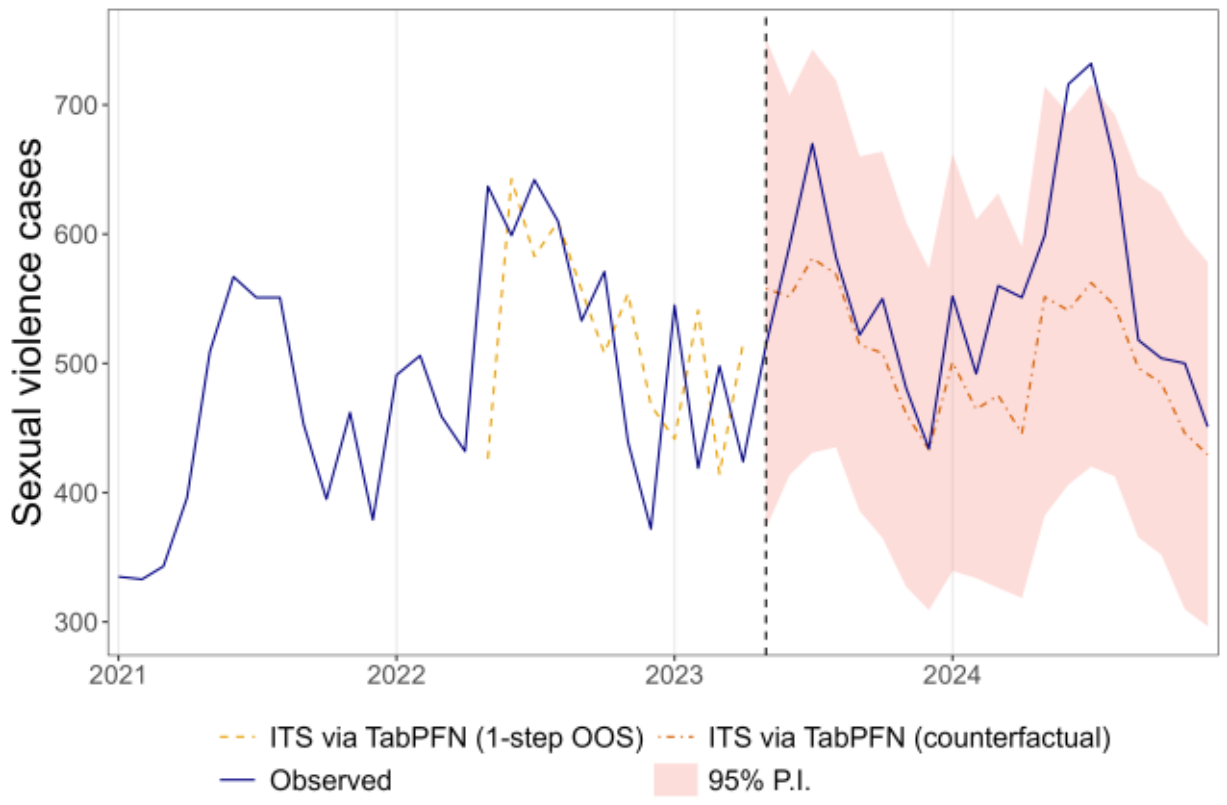

d) Sexual violence cases

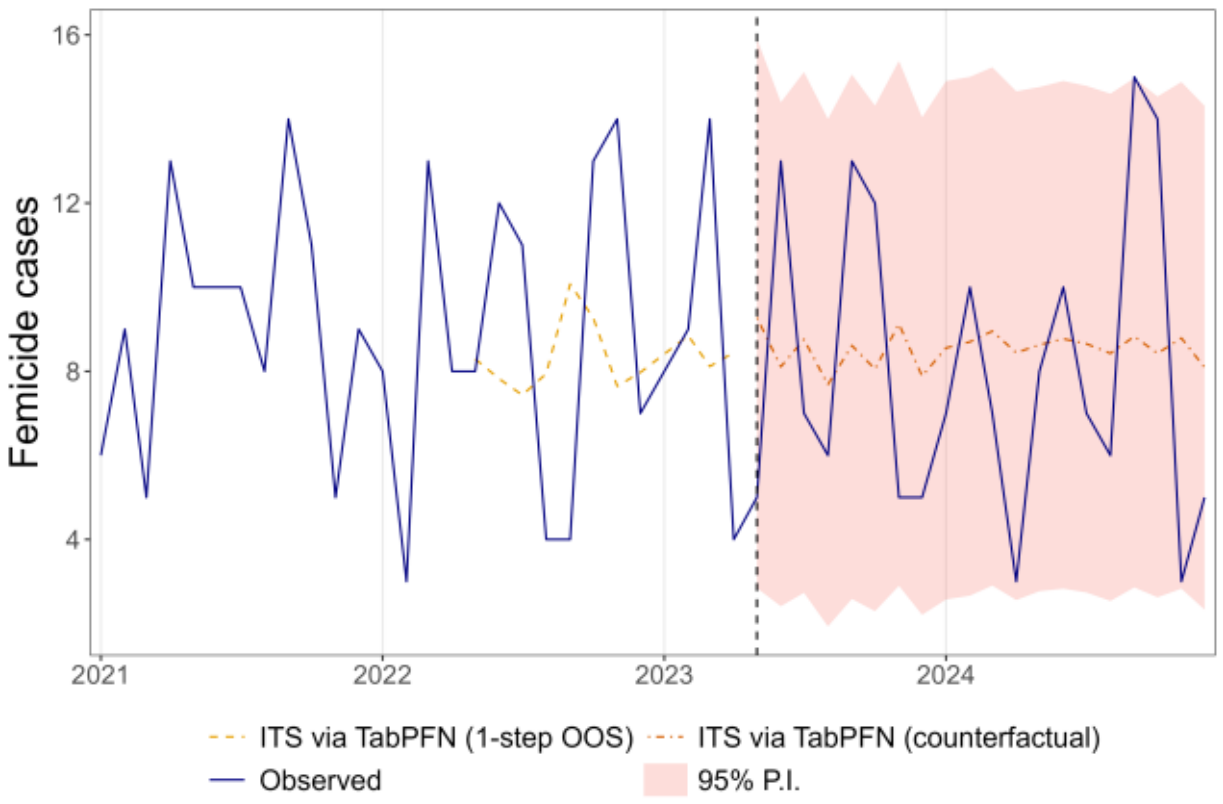

e) All femicides

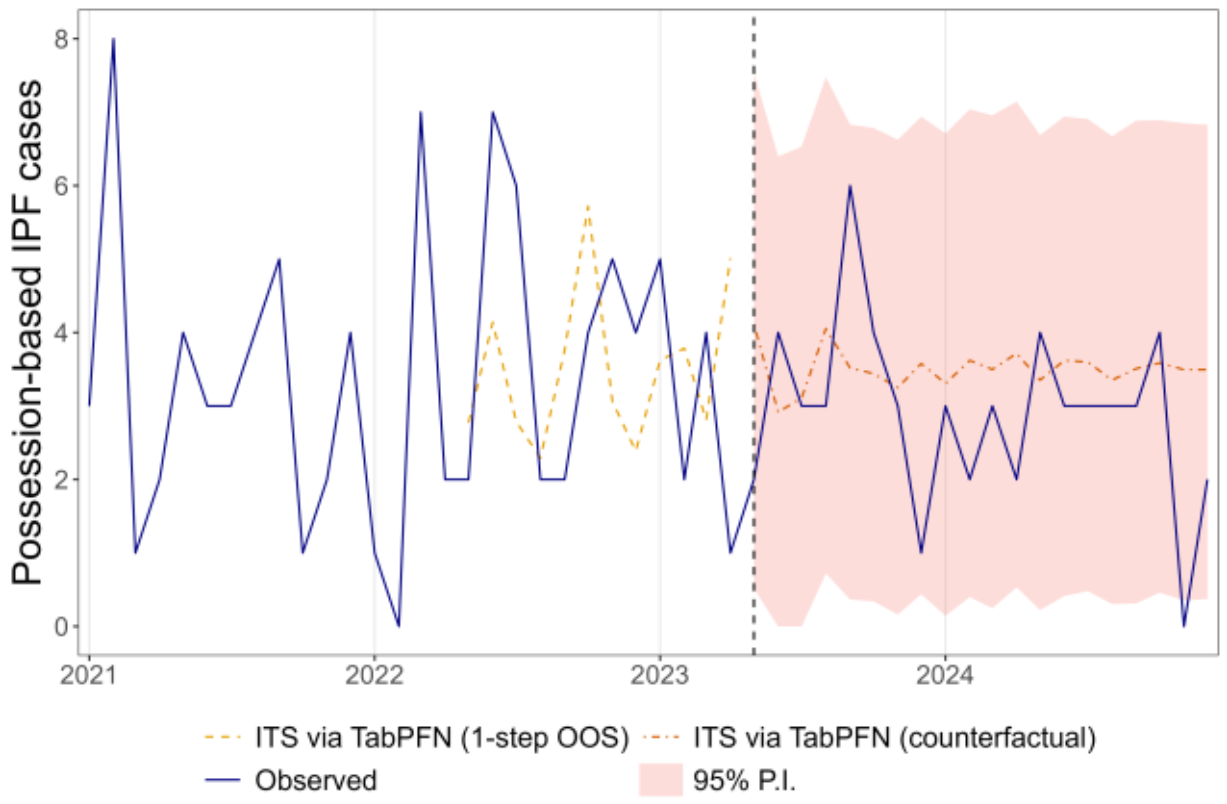

f) IPFs driven by jealousy or possessiveness

**Fig. 6 | Joint longer-horizon responses following the salient IPFs of Giulia Tramontano and Giulia Cecchettin. a,** Calls to the national 1522 helpline. **b,** Reports of abuse against women. **c,** Stalking. **d,** Sexual violence. **e,** Total femicides. **f,** IPFs associated with jealousy or possessiveness. The post-treatment period begins with Tramontano in May 2023; Cecchettin occurs in November 2023. These longer-horizon estimates capture the joint evolution following the two shocks and do not separately identify their effects after the second event. Lines and shading are defined as in Figs. 3-5. See Supplementary Information D for estimation and uncertainty details.

**Additional analyses on the main cases.** The main results are supported by several complementary analyses reported in Supplementary Information F. Estimates fitted separately for Northern, Central, and Southern Italy reproduce the positive help-seeking response to Tramontano and Cecchettin across macro-areas, showing that the national effects are not driven by the locations where the killings occurred. Disaggregating 1522 calls likewise shows mobilization among both victims and relatives or friends, with especially clear responses among the latter after the salient IPFs, consistent with public attention activating broader social ties as well as potential victims. Re-estimating the three main cases using the causal time series method Causal-ARIMA [25] yields conclusions broadly consistent with the main TabPFN analysis. Because this alternative approach relies on a different model and provides bootstrap-based counterfactual intervals, it also offers a more conventional inferential check on the statistical significance of the estimated effects.

**Comparison cases and mechanisms.** The three additional cases provide targeted tests of how salience translates into mobilization. First, we find no detectable mobilization following the lower-salience IPF case, supporting the role of salience intensity rather than intimate-partner status alone. Second, the analysis of the highly salient killing of a woman not coded as femicide in our dataset shows no comparable positive mobilization. Finally, we analyze the lower-salience non-IPF femicide involving a markedly different perpetrator configuration, with both a male and a female perpetrator and no intimate-partner relationship. This case likewise produces no coherent response and serves as an additional falsification test of whether femicide status alone is sufficient to generate mobilization. Together with the Abbas contrast, these cases indicate that the response depends not simply on the salience surrounding a violent death, but on the victim-perpetrator relationship structure and the gendered character of the violence made salient. All these estimates are reported in Supplementary Information G.

**Other outcomes.** The absence of a copycat increase in lethal violence is also robust to using administrative homicide data constructed independently of our femicide classification. The estimates are reported in Table S29 of Supplementary Information F. We find no evidence of an increase in homicides of women. If anything, the estimates point toward a sizeable decline following the Cecchettin case, although the estimated deviations generally remain within the predictive intervals, whereas male homicides show no analogous decline. This corroborates the pattern in our hand-curated femicide series and reduces concern that the result is an artifact of how femicides are classified.

## 4. Discussion

Our results show that public salience around femicides can generate substantial nationwide mobilization, but that this response is neither automatic nor generic. The increases in help-seeking and reporting following highly salient IPFs may arise because public attention makes previously hidden violence more recognizable, updates perceptions of risk, reduces stigma, and lowers informational or social barriers to seeking help. It may also mobilize relatives and friends, who can recognize warning signs and facilitate contact with support services. An alternative possibility, which our analysis investigates, is that greater reporting reflects an increase in underlying violence, including through imitation or copycat behavior [16–18]. Reporting data alone cannot distinguish these channels, making the evidence from lethal violence central to their interpretation.

Across both our hand-curated femicide series and administrative homicide data, we find no evidence of an increase in lethal violence following the salient IPFs. If anything, estimates after the Cecchettin case point toward declines in total femicides, jealousy- or possessiveness-related IPFs, and female homicides, although these rare outcomes are necessarily estimated with substantial uncertainty; male homicides show no analogous decline. We therefore do not interpret these estimates as establishing a preventive effect on lethal violence; nevertheless, they provide evidence against the prediction of a copycat response.

The dynamics also caution against characterizing the response as uniformly short-lived. The largest deviations occur around periods of peak public attention, especially after Cecchettin, but the joint Tramontano–Cecchettin analysis shows that helpline use remains elevated over a considerably longer horizon. Public salience may therefore operate through both an immediate attention shock and more persistent changes in awareness, discussion, or social coordination when successive salient events reinforce one another. This also suggests that sustained attention may be important for generating longer-lasting behavioral responses.

Taken together, these findings suggest that public salience can mobilize social and institutional responses to violence against women. Our evidence, however, concerns salience arising organically from highly visible events. Whether comparable effects can be generated through policy communication or awareness campaigns remains an open question and is likely to depend on the intensity, credibility, and persistence of the attention generated. Salience is also not a substitute for structural policy: without adequate shelters, legal protections, economic support, and effective law enforcement, greater help-seeking need not translate into lasting reductions in violence.

More broadly, our results suggest that what societies collectively see, discuss, and recognize can change whether people act. Public salience can make violence more visible, mobilize victims and their social networks, and increase contact with institutions without generating the feared increase in lethal imitation. Harnessing such attention alongside effective legal, social, and institutional support may therefore be an important component of efforts to reduce violence against women.

# Supplementary Information for

## Public salience mobilizes nationwide responses to violence against women

Augusto Cerqua, Marco Letta, Costanza Giannantoni and Gabriele Pinto

**Supplementary Information Guide**

**Sections**

**SI section A: Data**

We assembled a new dataset by combining information from multiple sources, beginning with the construction of a database of femicide cases in Italy. This step was necessary because no official administrative dataset publicly provides case-by-case information on femicides. Although the Ministry of the Interior collects homicide data, these records are not publicly accessible and do not systematically contain information on perpetrators' motives, making it difficult to distinguish femicides from other homicides of women. A complete description of the data sources and variables is provided in Table S2.

In recent years, the Italian National Institute of Statistics (ISTAT), in collaboration with the Ministry of the Interior, has produced statistical analyses of femicides within the broader system of intentional homicide statistics. ISTAT releases annual information on homicide victims—including sex, age, and the relationship between the victim and the perpetrator—and uses these variables to estimate femicides, typically defined as homicides of women committed by partners, former partners, or other family members. However, the underlying microdata are not publicly available, and the definitions currently adopted remain preliminary and may not capture all forms of femicide. This distinction is important because different types of femicide may require different policy responses. For example, the establishment of an anti-violence center is unlikely to affect femicides related to caregiver burnout.

To address these limitations, we constructed a micro-level dataset covering all femicide cases occurring in Italy between 2017 and 2024 that could be documented from the available sources, partly expanding our previous work on open access Italian femicides in 2006-2022 [1]. The dataset was assembled by digitizing information from reports produced by the non-governmental organization *Casa delle Donne per Non Subire Violenza* as well as from the journalistic source of the leading Italian newspaper *Corriere della Sera*. These are the only sources covering the whole period considered and provide detailed information on each femicide, including the name and age of the victim, the municipality where the incident occurred, the relationship between the killer and the victim, the weapon used, and a brief description of the context in which the killing took place.

However, these non-official sources are not always fully consistent in the cases they include and occasionally contain inaccuracies in some of the information reported. We therefore conducted extensive cross-checks for each femicide reported by at least one source. Following this verification process, we excluded cases that were inconsistently reported across years and cases that, after review of the available judicial evidence, were not classified as femicides (latest update: July 2026), and specific categories of incidents that, in the absence of a standardized statistical definition of femicide, were recorded only sporadically over the sample period as reporting practices evolved.[1] We also excluded the small number of cases (7 cases in total between January 2017 and December 2024) for which the perpetrator was not identified, as this prevented us from determining with certainty whether the homicide qualified as a femicide. These exclusions ensure a stable definition of femicide that is comparable across all years of the study and substantially limit the extent to which measurement error may affect our results. We have also created an expanded version of the database, which includes the cases recorded by the non-governmental organization *Non Una di Meno* as well as from the journalistic source another leading Italian national newspaper *la Repubblica* (9 additional cases between January 2020 and December 2024). This expanded version is used as the main dataset, while that restricted to the *Casa delle Donne per Non Subire Violenza* and *Corriere della Sera* will be used in a robustness exercise. We also gathered information for additional robustness checks, including indicators for multiple femicides involving both female and male victims.[2]

Based on the relationship between the perpetrator and the victim, we classified femicides into three main categories: (i) intimate-partner, (ii) family member, and (iii) other relationship. Each category was further disaggregated into the following subcategories:

i) Intimate partner — jealousy or possessiveness motives
ii) Intimate partner — caregiver burnout
iii) Intimate partner — other motives (e.g. severe psychological disorders or serious economic distress)
iv) Family member — caregiver burnout
v) Family member — other motives (e.g. severe psychological disorders or religious extremism)

---

[1] Specifically, we exclude cases involving transgender victims and suicides allegedly linked to gender-based violence. Although these cases are increasingly recognized in public discourse, their reporting in grassroots sources is not systematic over the 2017–2024 period. We also exclude killings committed in the context of robberies. In these cases, a gender-related motive cannot be clearly established, despite their classification as femicides by some sources. This exclusion concerns only eight cases over the entire period from January 2017 to December 2024. In addition, we exclude one case classified as consensual homicide, for which the circumstances and legal qualification are conceptually distinct from those underlying femicide.

[2] To construct a consistent and policy-relevant classification of femicides, we developed a structured coding protocol covering all cases reported by at least one of the aforementioned non-governmental organizations or journalistic sources. We then supplemented this information with additional details from other national and local newspapers, focusing on the victim–perpetrator relationship and the circumstances of the killing. Given the absence of standardized administrative definitions and the heterogeneity of reporting practices, this process required harmonizing partially overlapping and occasionally inconsistent narratives across sources. The classification was conducted manually following a common set of predefined coding rules. Only after this process was completed did we use a large language model (ChatGPT, GPT-5.6) as an additional validation check to assess the consistency of the manual coding. The LLM was therefore not used either as an independent source of information or as a primary classification tool. Rather, it served exclusively as an ex post robustness check on the manually assigned categories. Its use led to only very limited revisions: in fewer than 10 cases, the LLM suggested a categorization different from the original manual coding, resulting in only marginal changes to the final classification.

vi) Other relationship — friend or colleague following sentimental or sexual rejection
vii) Other relationship — victim was a sex worker
viii) Other relationship — other motives (e.g. disputes with neighbors or random attacks by perpetrators with severe mental instability)

Table S1 shows the yearly number of femicides in each of the eight categories.

| Relationship | 2017 | 2018 | 2019 | 2020 | 2021 | 2022 | 2023 | 2024 | Total |
|---|---|---|---|---|---|---|---|---|---|
| IPF - jealousy or possessiveness | 41 | 45 | 43 | 42 | 40 | 42 | 38 | 32 | 323 |
| IPF - caregiver burnout | 9 | 11 | 8 | 11 | 14 | 11 | 9 | 15 | 88 |
| IPF – other motives | 14 | 19 | 15 | 13 | 19 | 11 | 16 | 16 | 123 |
| Family Member - caregiver burnout | 2 | 5 | 1 | 2 | 1 | 7 | 3 | 4 | 25 |
| Family Member - other motives | 22 | 19 | 20 | 21 | 24 | 22 | 22 | 20 | 170 |
| Other relationship - friend or colleague following a rejection | 2 | 1 | 4 | 3 | 3 | 3 | 1 | 2 | 19 |
| Other relationship – victim was a sex worker | 5 | 4 | 2 | 1 | 1 | 4 | 1 | 3 | 21 |
| Other relationship - other motives | 11 | 5 | 5 | 5 | 8 | 5 | 11 | 3 | 53 |
| Total | 106 | 109 | 98 | 98 | 110 | 105 | 101 | 95 | 822 |

**Table S1 | Descriptive statistics**.

Moreover, we collect administrative data on all gender-related crimes reported to the police and registered in the SDI (*Sistema di Indagine*), the central database of the Italian Police. These data, available at the provincial level, include reported cases of stalking, sexual violence, and abuse against women. A key limitation of police-recorded data is the substantial underreporting of violence against women. Evidence from ISTAT's nationally representative Survey on Women's Safety shows that only a small fraction of victims report these episodes to the police. In the 2014 survey, only 11.8% of women who had experienced physical or sexual violence by a current or former partner during the previous five years reported it to the authorities (see here, in Italian); the most recent survey provides a very similar estimate of 10.5%. Reporting rates also vary considerably across types and severity of violence. For sexual violence, only 6.7% of incidents occurring in the previous 12 months were reported, whereas the reporting rate rose to 19.4% for rape or attempted rape. Stalking is similarly characterized by substantial underreporting: Istat data imply that only about 7–10% of victims formally reported the perpetrator, depending on whether the stalker was an ex partner or another person (see here, in Italian). Hence, police records capture only a limited and selected share of the underlying incidence of gender-based violence, with the extent of underreporting varying substantially across types and severity of violence. Greater salience of gender-based violence may therefore bring part of this hidden incidence to the attention of the authorities, for instance by increasing victims' awareness and willingness to report or seek help. Accordingly, increases in police reporting following a highly salient femicide are more plausibly interpreted as previously hidden violence coming to the attention of the authorities than as additional violence induced by the salience shock itself.

The SDI also records homicides, separately for male and female victims, and provides information on whether the perpetrator was an intimate partner of the victim. However, these data do not report the motive underlying the homicide and therefore do not allow us to distinguish femicides from other homicides of women.

We collect monthly data on calls to the national helpline against gender-based violence (1522) (ISTAT). These data allow us to distinguish calls made directly by victims from those made by relatives or friends, as well as the geographic origin of each call (Northern, Central, or Southern Italy).

Figure S1 shows the dynamics of total femicides and jealousy- or possessiveness-related IPFs, as well as the number of gender-related reported crimes and calls to the national helpline.

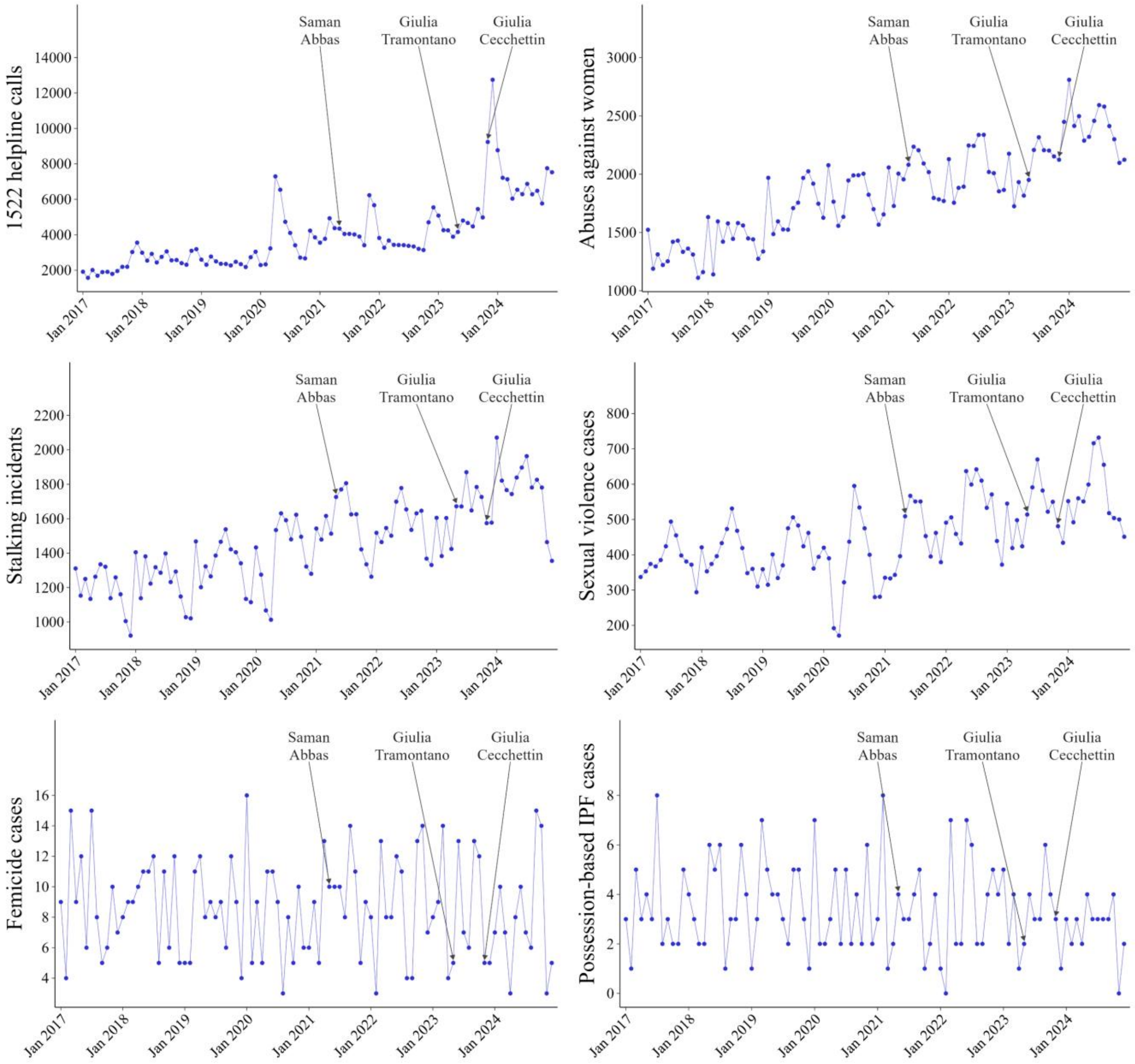

**Fig. S1 | Raw monthly outcome series.** Monthly counts of calls to the national 1522 helpline, reports of abuse against women, stalking incidents, sexual violence cases, femicides, and jealousy- or possessiveness-related intimate partner femicides (IPFs), from January 2017 onward. The timing of the three main case studies, Saman Abbas, Giulia Tramontano, and Giulia Cecchettin, is indicated in each panel.

Furthermore, we collect data on participation in women's rights demonstrations from ACLED for 2020–2026 and on online searches related to femicide from Google Trends. We also gather information on social media activity, measured by the number of posts and user reactions on Facebook and Instagram, and on the salience and media coverage of the highest-profile femicide cases in Italy in recent years using *Media Cloud*. These data are used to construct Figure 1 in the main text and Figure S2 in Supplementary Information C. Below, we provide a more detailed description of the social media and *Media Cloud* data.

**Social media data**. We collect information on social media activity using the number of Facebook posts and users' reactions to those posts. Facebook was the most widely used social networking platform in Italy throughout our study period. AGCOM–Audiweb data consistently rank Facebook first in terms of unique users, both in the earlier part of the period (see the AGCOM Observatory, 2017–2020) and in the subsequent years through 2024 (see the AGCOM Observatory, 2021–2024). This makes Facebook a particularly suitable platform for assessing the extent to which highly salient femicide cases reached and engaged a broad national audience. We retrieve Facebook activity through the Meta Content Library, considering all posts whose text contains both the first and last name of the victim. The database covers posts published by Pages with at least 15,000 likes or followers and by verified Profiles with at least 25,000 followers (i.e., those downloadable on the Meta Content Library). For each case, we record both the number of posts and the reactions generated by those posts, providing a complementary measure of social media salience alongside our news-based indicators.

**Media data**. We measure media attention using daily article counts retrieved through the Media Cloud Search API (https://www.mediacloud.org/documentation/search-api-guide). For each femicide that occurred between 1 January 2019 and 31 December 2024, we query the Italian news collection using the victim's full name. For each query, the API returns the number of articles published on each day of the period under analysis. The collection comprises 144 Italian news outlets, including national and local newspapers and online news sources; the full list is available here: https://search.mediacloud.org/collections/34412372. These data provide our main measure of media salience and allow all cases to be ranked on a common scale.

| Data Category | Variables | Level / Period | Source |
|---|---|---|---|
| Femicides dataset | Number of total femicides and IPFs, classification by motive and relationship | Municipal-level, 2017–2024 | Casa delle Donne per Non Subire Violenza; Non Una di Meno; Corriere della Sera; La Repubblica |
| Gender-related crimes | - Reported crimes: stalking, sexual violence, and abuse against women.<br>- Homicides of men, homicides of women, and homicides of women committed by intimate partners. | Provincial-level, 2017-2024 | SDI (Sistema di Indagine), Italian Police (Italian Ministry of the Interior) |
| Helpline data (1522) | Number of calls, caller type (victim vs family/friends), geographic origin (North, Center, South) | Provincial-level, 2017-2024 | ISTAT |
| Social media activity | Number of posts and user reactions (Facebook, Instagram) | Country level, 01-01-2019 to 31-12-2024 | Meta Content Library (limited to Pages with 15,000 or more likes or followers, and verified Profiles with 25,000 or more followers) |
| Online search behavior | Google search intensity for "*femminicidio*" (Google Trends index) | Country and regional level (Calabria, Lazio, and Lombardy), from 01-01-2021 to 31-12-2024 | Google Trends |
| Collective action / protests | Number of demonstrations and participants in women's rights events | Country level, (01-01-2020–to 02-2026) | ACLED |
| Media salience | Frequency ratio / salience indicators for high-profile femicide cases | 01-01-2019 to 31-12-2024 (Italian Collection) | Media Cloud |

**Table S2 | Data Sources and Variables**.

## SI section B: Background on the six cases under analysis

Our analysis considers six cases of lethal violence against women: five femicides and one homicide of a woman. In the following, we provide detailed background information for each case.

### Highly salient IPF: Giulia Tramontano

Giulia Tramontano, a 29-year-old woman who was seven months pregnant, was killed by her partner, Alessandro Impagnatiello, on 27 May 2023 in Senago, a municipality near Milan. Impagnatiello, a barman, stabbed her repeatedly and then attempted to destroy and conceal her body. In November 2024 the Milan Court of Assizes sentenced the killer to life imprisonment for the femicide, the concealment of the corpse, and the non-consensual termination of the pregnancy, and the life sentence was upheld on appeal in June 2025. In April 2026, the Court of Cassation ordered a new appeal limited to the premeditation aggravating circumstance, while leaving the life sentence unchanged. Occurring roughly six months before the Cecchettin case, the murder received extensive national media coverage, driven both by the brutality of the crime and by the fact that the victim was pregnant. In our Media Cloud salience ranking (see Figure 1 in the main text), Tramontano is the second-most salient IPF. It therefore provides the natural high-salience benchmark below Cecchettin: the two cases share the intimate-partner relationship structure but differ substantially in salience intensity, allowing us to examine whether the magnitude of mobilization varies with salience among otherwise closely related events.

### Exceptionally salient IPF: Giulia Cecchettin

The murder of Giulia Cecchettin occurred on November 11, 2023. Giulia Cecchettin, a 22-year-old university student, was brutally killed by her former partner, Filippo Turetta, whom she had met during her studies. After she was reported missing, the case received extensive national media coverage during the week-long search. Her body was found on November 18, 2023. Turetta was arrested the following morning in Germany and extradited to Italy shortly thereafter. He was later convicted and sentenced to life imprisonment; in October 2025 he withdrew his appeal, and after the Venice Prosecutor General also withdrew its appeal in November 2025, the sentence became final.

As documented in Supplementary Information C, no other femicide in our study period generated a comparable scale or intensity of public attention, and the resulting salience shock coincided with heightened public awareness of gender-based violence and lower tolerance for victim-blaming attitudes. Several factors likely contributed to the exceptional public response. The case involved a sequence of events that intensified public scrutiny, including reports of prior tensions, a warning call to the police that did not lead to intervention, and a week-long search that kept the case at the center of national attention. In addition, the timing of the event likely amplified its public resonance, as it occurred just days before the International Day for the Elimination of Violence Against Women on November 25, a date that already concentrates media attention, public demonstrations, and institutional initiatives on this issue. In 2023, the proximity between the Cecchettin case and the November 25 mobilizations may have reinforced the scale and intensity of the public response.

Following her death, Giulia Cecchettin's father and sister, Gino and Elena Cecchettin, became prominent voices in the public debate on gender-based violence. Through sustained public engagement, they contributed to keeping attention on the case and to framing it within a broader discussion of structural violence against women, thereby amplifying its societal and institutional impact.

**Highly salient non-IPF: Saman Abbas**

Unlike the previous cases, the killing of Saman Abbas did not arise from intimate-partner violence but was a femicide rooted in a family context. Saman Abbas, an 18-year-old woman of Pakistani origin living in Novellara, in the province of Reggio Emilia, was murdered on the night between April 30 and May 1, 2021, for refusing an arranged marriage with a relative in Pakistan. She was killed by members of her own family, and her body was concealed and recovered only in November 2022. Her parents, an uncle, and two cousins were identified as responsible: the parents and the two cousins received life sentences and the uncle 22 years, with the convictions becoming final in July 2026. The case drew broad national attention and became emblematic of violence linked to family control over women. Abbas is the most salient non-IPF femicide in our ranking, with salience close to that of Tramontano. This makes it the key high-salience relationship contrast: publicity is broadly comparable, while the violence is family-perpetrated rather than intimate-partner. The case therefore helps assess whether high salience is sufficient for mobilization or whether the victim–perpetrator relationship is an important part of the response.

**Highly salient killing of a woman not classified as femicide: Sharon Verzeni**

The murder of Sharon Verzeni differs from the other episodes considered in the analysis: although the victim was a woman, we do not code the case as a femicide in our dataset. Sharon Verzeni, a 33-year-old woman, was stabbed to death while walking alone at night on July 30, 2024, in Terno d'Isola, in the province of Bergamo. The man convicted at first instance, Moussa Sangare, was a stranger with no prior connection to her. On 25 February 2026, the Bergamo Court of Assizes sentenced him to life imprisonment, recognizing, among other aggravating circumstances, premeditation and futile motives.

The judicial reconstruction found no gender-related motive: Verzeni was selected as a random victim by an offender who had set out to attack someone. The reasons for judgment deposited in May 2026 describe the killing as motivated by boredom and a desire to experience strong emotions, with the victim selected during a search for someone to attack. These circumstances provide no indication that the killing was motivated by the victim's gender, sharply distinguishing the Verzeni case from the femicides considered in the analysis. We therefore do not classify the Verzeni case as a femicide. Among non-femicide killings of women considered as potential comparisons, it is the most salient. This makes the Verzeni case the strongest available test of the gender component of the response mechanism: it preserves high public salience while removing the relational structure that characterizes the femicide cases in our analysis.

**Lower-salience IPF: Alessandra Matteuzzi**

Alessandra Matteuzzi, a 56-year-old woman, was murdered on August 23, 2022, in Bologna by her former partner, also a former semi-professional footballer. After she had ended the relationship, the killer subjected her to a period of escalating stalking, and on the night of the murder he waited for her outside her home and beat her to death with a hammer and an iron bench. The case is a paradigmatic instance of intimate-partner violence: the court found that the killer treated the victim as his own property and killed her in retaliation for having been left. He was convicted of aggravated murder and stalking and sentenced to life imprisonment, a sentence that became final in September 2025. We exploit this case, which is classified as a lower-salience IPF, as a targeted comparison of salience intensity. Indeed, finding a comparable response following this femicide would weaken the interpretation that salience intensity matters for mobilization. It therefore provides a cleaner comparison for assessing how responses differ when an otherwise comparable IPF receives substantially less nationwide attention. This case is selected not simply because she received less coverage, but because she provides the closest lower-salience relational match to the main IPFs: a former intimate partner, prior stalking and controlling behavior, and a clearly timed public-information shock. She therefore provides a targeted comparison of salience intensity, asking whether an otherwise comparable IPF generates the same mobilization when nationwide salience is markedly lower. "Lower salience" is relative here: the Matteuzzi case still received substantial attention compared with most femicides in the dataset.

**Lower-salience non-IPF: Nicoletta Indelicato**

The killing of Nicoletta Indelicato is atypical relative to the other cases. Nicoletta Indelicato, a 25-year-old woman, was stabbed to death and her body partially burned on the night of 17 March 2019 in the countryside near Marsala, Sicily. She was killed by two acquaintances, Carmelo Bonetta and Margareta

Buffa. Bonetta, who confessed, was sentenced to 30 years' imprisonment, while Buffa was sentenced to life imprisonment for her role in the murder; both convictions are final. Although the courts did not establish a definitive motive, contemporaneous investigations pointed to jealousy and resentment linked to the interpersonal relationships among the victim and the two perpetrators, supporting the classification of the case within the broader category of femicide while distinguishing it from intimate-partner femicides. The absence of an intimate-partner relationship, the involvement of both a male and a female perpetrator, the unresolved motive, and the substantially lower nationwide salience distinguish the case sharply from the main IPFs. We therefore classify Indelicato as a non-IPF femicide and use it as a deliberately demanding falsification case, testing whether a lower-salience femicide with a markedly different perpetrator configuration produces the same mobilization observed after highly salient IPFs.

## SI section C: The Cecchettin salience shock

The Cecchettin femicide represented an unprecedented salience shock. To substantiate this claim, we complement the media analysis reported in Figure 1 of the main text with several additional sources capturing different dimensions of public attention and mobilization. Specifically, we combine social media activity, online search data, and indicators of public mobilization, with the resulting evidence reported in Figure S2.

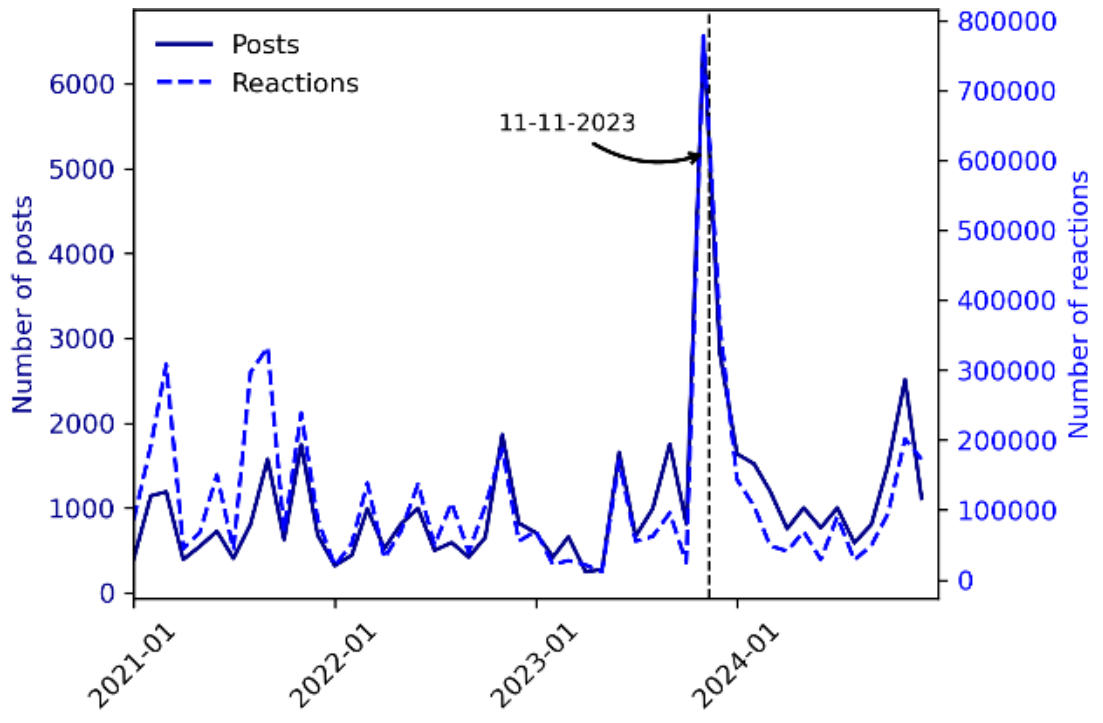

a) Facebook

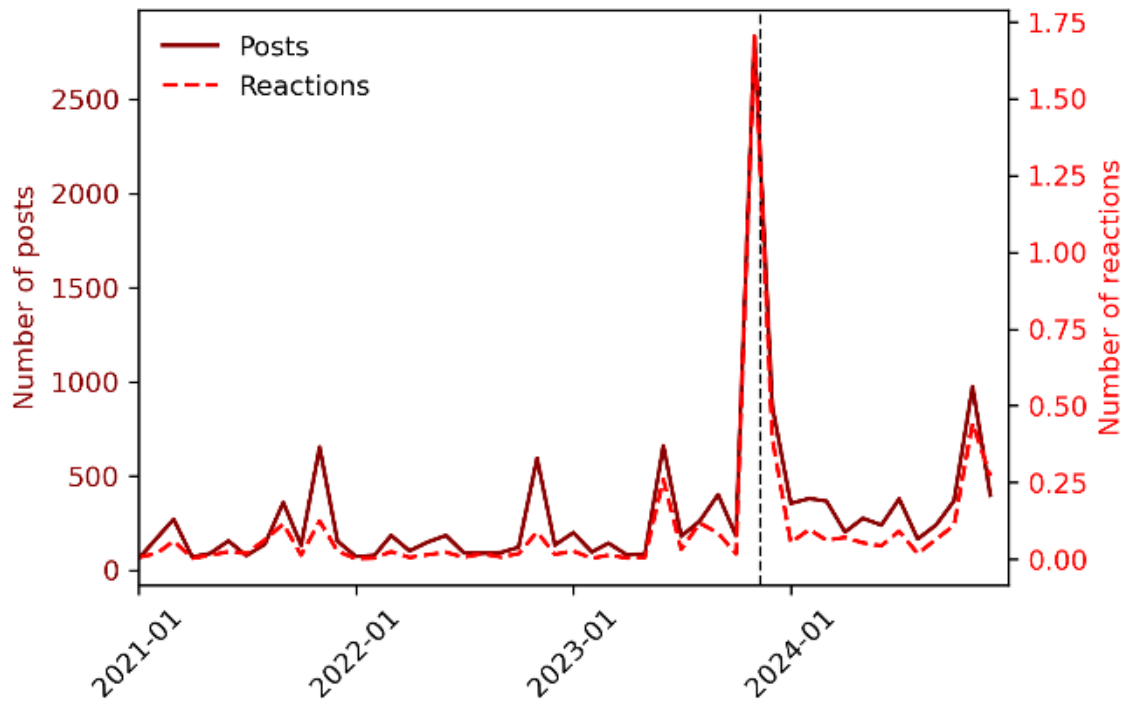

b) Instagram

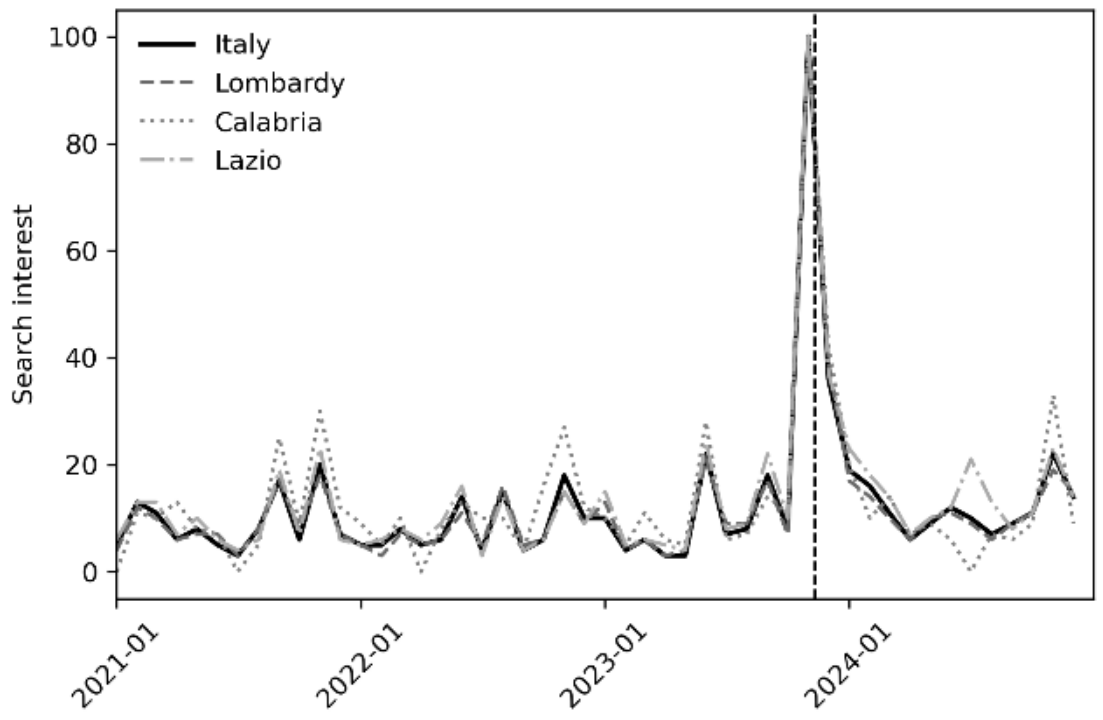

c) Google Trends

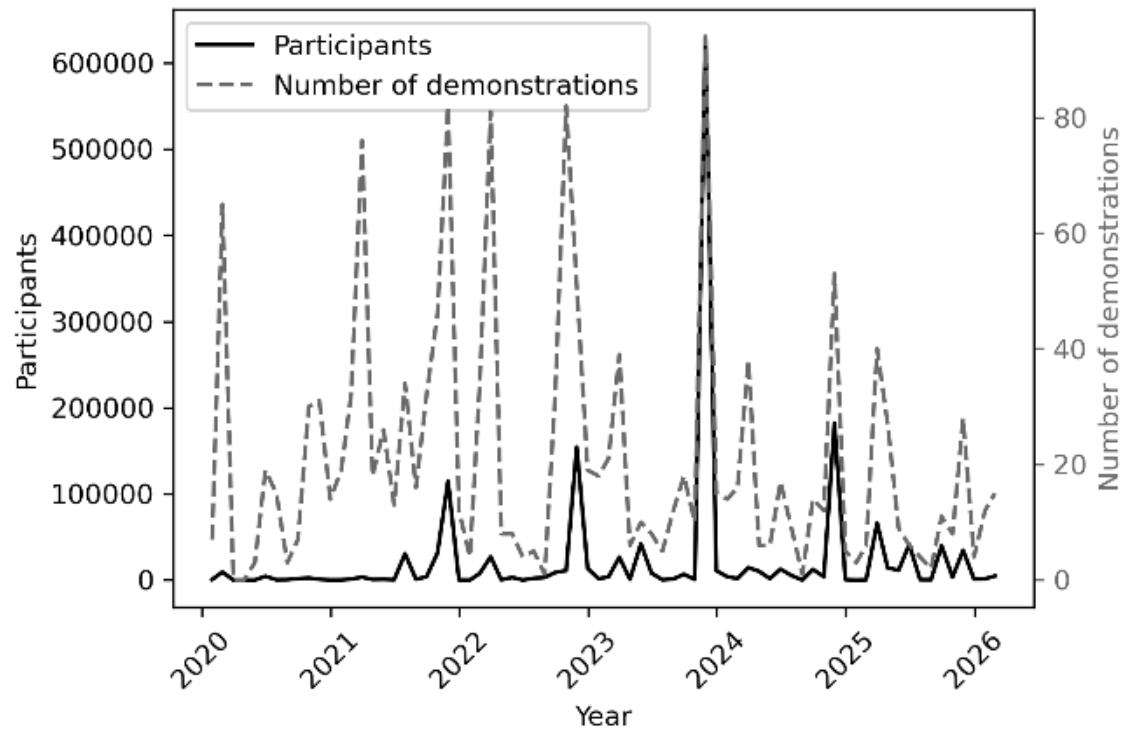

d) Women's rights demonstrations

**Fig. S2 | Descriptive evidence on the exceptional public salience of the Cecchettin case. a,** Facebook activity. **b**, Instagram activity. **c,** Google search interest. **d,** Women's-rights demonstrations. **a–c**, Frequency of the keyword "femminicidio" on Facebook, Instagram, and Google Trends; Google Trends is reported nationally and for selected regions from Northern, Central, and Southern Italy. **d**, Number of demonstrations against gender-based violence and in support of women's rights. Facebook and Instagram data are from the Meta Content Library; Google Trends data are from Google; and demonstration data are from ACLED.

The evidence in Figure S2 points to an exceptional and nationwide increase in public attention following the Cecchettin case. Facebook and Instagram activity, measured using posts containing the keyword *"femminicidio"* and the reactions generated by those posts, displays a pronounced spike in November 2023 (Fig. S2a,b), substantially exceeding the levels observed elsewhere in the series. Google Trends provides complementary evidence (Fig. S2c): weekly searches for *"femminicidio"* in Italy reached their maximum value (index 100) in the week beginning 19 November 2023, immediately following Cecchettin's murder and the arrest of her former partner. This peak far exceeds the recurrent increases in search interest observed around the International Day for the Elimination of Violence Against Women, both in previous years and in 2024. Importantly, the pattern is remarkably similar across Lombardy in Northern Italy, Lazio in Central Italy and Calabria in Southern Italy, indicating that the increase in attention was national rather than geographically concentrated.

The surge in attention was accompanied by exceptional collective mobilization. As shown in Fig. S2d, women's-rights demonstrations in late 2023 generated the largest participation peak in the period considered. Contemporary estimates for the national mobilization associated with 25 November 2023 converge on at least 500,000 participants according to organizers and multiple media reports, and the event was widely described as the most attended 25 November demonstration ever held in Italy. Taken together, the social-media, online-search and mobilization evidence documents a historically unusual surge in public attention and collective response surrounding the Cecchettin case, supporting its interpretation as an exceptionally large nationwide salience shock.

**SI section D: Methods**

**Research design and sample**. A key feature of our empirical setting is that both the outcomes and the salience shocks are national in scope. There are therefore no untreated units within Italy that can serve as contemporaneous controls. Because attention to the most salient cases spreads well beyond the location of the killing, macro-area series for the North, Center, and South are used to study geographic heterogeneity and internal consistency rather than as control groups. Treating areas farther from the event as untreated would also be difficult to justify: media coverage, public discussion, and mobilization cross regional boundaries, so outcomes in one part of the country may respond to attention generated elsewhere.

In principle, one could instead construct a synthetic control from other countries [2]. In this application, however, the key outcomes are not sufficiently harmonized internationally. Helplines differ in organization and call classification, incentives to report violence and policing practices vary across countries, legal definitions and administrative coding are not uniform, and femicide itself is measured differently across jurisdictions. A donor pool assembled from such series could reproduce the Italian pre-treatment trajectory statistically while remaining substantively incomparable, undermining the causal interpretation of the resulting estimates.

A further possibility would be to use unaffected but correlated time series as controls or auxiliary predictors [3]. This is also problematic here. Help-seeking, reports of gender-based violence, and femicides have unusually specific behavioral and institutional determinants, and plausible non-violence series are unlikely to be both strongly predictive of these outcomes and unaffected by a major nationwide salience shock. Broader violence measures do not provide an obvious solution either: general violence and gender-based violence need not move together. In Italy, for example, the long-run decline in homicide is not mirrored by an equivalent decline in femicide.

Given these constraints, we study each event using a monthly Interrupted Time Series (ITS) design, a well-established approach for evaluating interventions or shocks when a suitable contemporaneous control group is unavailable [4-6]. The basic comparison is between the outcome observed after a femicide and the path that would have been expected from its pre-event dynamics had the salience shock not occurred. This design is necessary in our setting because our focus is precisely on the most salient events that generate nationwide attention. Areas farther from the location of the killing cannot naturally be treated as unexposed controls when information and attention spread across the country, thus ruling out the use of quasi-experimental methodologies such as difference-in-differences that rely upon the availability of untreated units serving as a control group. We therefore use the North, Center, and South estimates to study geographic heterogeneity and internal consistency rather than as control groups.

The outcome series and the descriptive femicide database begin in January 2017. For event selection, however, we restrict the salience ranking to cases occurring from January 2019 onward, so that every ranked candidate has at least 24 months of raw pre-treatment history. Our forecasting approach (see below) imposes a stricter requirement. Because the specification includes outcome-level and change lags of up to 12 months and reports a 12-month expanding-window out-of-sample diagnostic, at least 37 raw pre-treatment monthly observations are required to obtain the full validation window. All main cases satisfy this requirement.[3]

The event month is treated as the first post-treatment month. Our main estimates cover six months following each event: May–October 2021 for Saman Abbas, May–October 2023 for Giulia Tramontano, and November 2023–April 2024 for Giulia Cecchettin. The supplementary comparison cases are analyzed using the same procedure.[4]

Each event is estimated using the full observed history available before it. Thus, the Cecchettin analysis conditions on the realized series through October 2023, including the months following Tramontano. The main post-event Tramontano window, by contrast, ends before Cecchettin. When the analysis is extended beyond October 2023 (see Fig. 6 in the Main Text), the two post-treatment periods overlap. We therefore interpret the longer-horizon estimates beginning in May 2023 as the joint evolution following the two salient IPFs, rather than as separately identified long-run effects of either event.

**Causal estimands and identifying assumptions**. Let $Y_t$ denote an outcome observed in month $t$, and let $T_0$ denote the final pre-treatment month. For $t > T_0$, the treated outcome is observed, whereas the outcome that would have occurred in the absence of the salience shock is not. We estimate this untreated potential outcome from the pre-treatment information set.

---

[3] The only exception is the case of Nicoletta Indelicato (March 2019): after constructing the lagged predictors, only 13 of the 26 pre-treatment monthly observations remain usable. We therefore shorten the expanding-window, one-step-ahead validation period from 12 to six months (September 2018–February 2019), with the training sample increasing from seven to 12 observations. The post-event counterfactual uses all 13 usable pre-treatment observations and covers March–August 2019. The model specification and lag structure remain unchanged.

[4] The case of Sharon Verzeni is the only exception. Although she was murdered on 30 July 2024, public and media attention to the case intensified substantially only at the end of August, when, after a month-long investigation, police identified and arrested Moussa Sangare on 30 August and he confessed to the murder. Given that this sharp increase in salience occurred at the very end of August, we treat September 2024 as the first post-treatment period in our monthly analysis. As a robustness check, we also re-estimated the models using August and July 2024 as alternative treatment start dates; the resulting estimates are qualitatively very similar.

The month-specific effect is:

$$\tau_t = Y_t(1) - \widehat{Y_t(0)} \tag{D.1}$$

For a post-treatment horizon H, the cumulative effect is:

$$\Delta_H = \sum_{h=1}^{H}\left[Y_{T_0+h}(1) - \widehat{Y_{T_0+h}(0)}\right] \tag{D.2}$$

We also report cumulative percentage effects relative to the cumulative counterfactual:

$$100 \times \frac{\sum_{h=1}^{H}\left[Y_{T_0+h}(1) - \widehat{Y_{T_0+h}(0)}\right]}{\sum_{h=1}^{H}\widehat{Y_{T_0+h}(0)}} \tag{D.3}$$

A causal interpretation rests on the following three substantive assumptions. First, the femicide and the associated increase in public attention must not be anticipated in a way that changes the outcomes before the event. Second, in the absence of the shock, the relationships governing the outcome before treatment must remain informative over the relatively short post-treatment horizon. Third, there must be no contemporaneous event that independently generates the same post-treatment deviations. A further implementation requirement is that counterfactual estimation and model selection use only pre-treatment outcomes and covariate information, together with predetermined calendar features. Any post-treatment lagged outcome values required for forecasting must be generated recursively from counterfactual predictions rather than taken from observed outcomes, thereby preventing data leakage [7].

We examine the substantive assumptions rather than relying on pre-treatment fit alone. The evidence includes rolling pre-treatment forecasts, comparisons across events and outcomes, lower-salience and falsification cases, alternative definitions of femicide, independently measured homicide outcomes, caller-type and geographic decompositions, alternative forecasting models, and an examination of potentially relevant contemporaneous events.

**Tabular foundation models and TabPFN**. Tabular foundation models (TFMs) are general-purpose prediction models pretrained on a large collection of supervised learning tasks and then applied to new datasets through in-context learning. Rather than estimating a new model from scratch for each application, they condition on the labelled observations supplied at prediction time and use the prediction rule learned during pretraining. In contrast to a conventional task-specific neural network, the pretrained model does not need to learn a new set of network weights for each outcome series. This is particularly attractive in small samples, where extensive model selection and flexible task-specific estimation can quickly exhaust the information available.

Hollmann et al. [8] introduced TabPFN as a foundation model for tabular prediction in small-data settings. TabPFN is pretrained on synthetic prediction problems and, when presented with a new dataset, jointly conditions on the training observations and the rows to be predicted. No task-specific optimization of the neural network weights is required at inference time. Hollmann et al. show that this approach can achieve strong predictive performance on small tabular datasets without the extensive tuning usually required by conventional machine learning methods.

These properties are well suited to our setting. The monthly histories available before each event are relatively short, some outcomes—notably femicides—are sparse and volatile, and the relationships among trend, seasonality, persistence, and related violence indicators need not be adequately captured by a fixed linear or autoregressive specification. We therefore use TabPFN to estimate the untreated trajectory flexibly while applying the same forecasting procedure across events and outcomes. TabPFN is not itself a causal model: causal interpretation comes from the ITS design and its identifying assumptions, while the TFM is used to construct the missing counterfactual.

Our main estimates use TabPFN 3.0 [9], the most recent version of the model available as of August 2026. Relative to the version introduced by Hollmann et al. [8], TabPFN 3.0 substantially revises the model architecture, preprocessing, and synthetic pretraining distribution, while improving predictive performance, computational scale, and inference speed. Particularly relevant for our application, the TabPFN-3 generation explicitly extends the model family to time-series forecasting and reports substantial gains on time-series tasks [9]. This development builds on earlier evidence that TabPFN can be used effectively for forecasting by recasting a time series as a tabular regression problem with appropriately constructed temporal features [10]. Our implementation follows this approach: we use the default TabPFN 3.0 regression model and represent temporal structure explicitly through lagged outcomes, changes, calendar variables, and lagged information from related series.

Each outcome is estimated separately using a 16-member ensemble. The prediction target is the ordinary month-to-month change in the outcome, expressed in its original units:

$$z_t = Y_t - Y_{t-1} \tag{D.4}$$

No logarithmic transformation is applied. Modeling monthly changes reduces sensitivity to non-stationarity in the level series, while lagged levels and the other temporal features preserve information on persistence, seasonality, and longer-run movements.

The main six-month specification contains 50 predictors, organized into four groups. First, the model uses the outcome's own history: the previous 12 monthly levels and five lags of the monthly change, at 1, 2, 3, 6, and 12 months. Second, we construct nine features that summarize recent dynamics in the outcome. These consist of the difference between the one- and three-month level lags; the difference between the one- and twelve-month level lags; the mean and standard deviation of the previous three outcome levels; the mean and standard deviation of the previous six outcome levels; the difference between the one- and two-month lags of the monthly change; and the mean and standard deviation of the previous three monthly changes. Together, these variables capture recent movement, departures from the same period of the previous year, and short-run variation without imposing a particular parametric trend.

Third, ten predictors consist of the 6- and 12-month lags of five other outcome series.[5] These variables allow information shared across the violence and help-seeking outcomes to contribute to prediction while ensuring that, throughout the six-month post-treatment horizon, their source observations remain strictly pre-treatment. Finally, the specification includes eleven standard month indicators, a separate November indicator, a linear time index, and a COVID-period indicator. The dedicated November indicator replaces the ordinary November dummy and captures the recurrent increase in attention surrounding the International Day for the Elimination of Violence against Women. The COVID indicator covers March–May and October–December 2020.

**Construction of the counterfactual and prevention of leakage**. Before the feature matrix is constructed, the outcome being forecast and the other outcome series are masked from the intervention month onward. The fitted model therefore never conditions on their realized post-treatment values.

Cross-series predictors enter the main six-month specification only at lags of 6 and 12 months. The code explicitly checks that the source date associated with every such predictor is earlier than the treatment date. In the sixth and final forecast month, a six-month lag therefore still points to the last pre-treatment observation.

[5] When total femicides is the outcome, we retain jealousy- or possessiveness-related IPFs but omit the "other femicides" series from the cross-series predictors. Because total femicides is the sum of these two components, including both components alongside lags of total femicides would create an exact linear identity among the predictors.

Shorter own-outcome lags require a different treatment because, after the first post-event month, some of them refer to months that are themselves post-treatment. These features are updated recursively. After predicting the first counterfactual outcome, that prediction is plugged into the counterfactual history and is used to construct the second forecast; the procedure is then repeated through the six-month horizon. Lagged changes and the derived rolling features are updated from the same counterfactual history. Realized post-treatment outcomes are retrieved only after the counterfactual path has been completed, when the treatment effects are calculated. Because the model predicts changes, the predicted change in each month is added to the preceding counterfactual level. Predicted levels are constrained to be non-negative. The resulting sequence is the estimated untreated path $\widehat{Y_t(0)}$.

A notable feature of the probabilistic TFMs used here is that they provide information about the predictive distribution rather than only a point forecast. For TabPFN, at each month we obtain the predictive mean together with the 2.5$^{th}$ and 97.5$^{th}$ predictive quantiles and transform them back to the level scale. These are the bands shown in the main figures.

The distinction between these predictive ranges and causal confidence intervals is important. The TabPFN bands capture uncertainty in the predicted untreated outcome conditional on the recursively constructed counterfactual history. At each step, the point counterfactual is carried forward to construct the next prediction, rather than simulating complete alternative counterfactual paths from the predictive distribution. The resulting bands therefore quantify pointwise predictive uncertainty around the estimated counterfactual and can be used to assess departures from the predicted untreated trajectory, but they should not be interpreted as causal confidence intervals. In particular, they do not fully propagate uncertainty accumulated across recursive forecasts or capture the broader uncertainty associated with causal identification. Causal-ARIMA, described below, provides a complementary inferential exercise based on simulated counterfactual paths.

**Non-stationarity and pre-treatment validation**. The main specification does not require the raw outcome levels to be stationary. TabPFN predicts the first difference rather than the level itself, while lagged levels, seasonal lags, the linear time index, and calendar indicators retain information on persistence, trends, and recurring seasonal patterns. We nevertheless examined the time-series properties of the outcomes using Augmented Dickey–Fuller (ADF) and Kwiatkowski–Phillips–Schmidt–Shin (KPSS) tests [11,12]. The two tests are useful complements because their null hypotheses run in opposite directions: the ADF test takes a unit root as the null, whereas the KPSS test takes stationarity as the null.

The tests on the raw 2017–2024 levels give a mixed picture, which is unsurprising given the very different nature of the series. Police reports and helpline calls display stronger trend and persistence, whereas the sparse lethal-violence measures are generally much closer to stationary processes in levels. For some of the reporting and help-seeking series, the level tests are consistent with trend stationarity; for others, they do not sharply discriminate between a unit-root and a trend-stationary representation. After first differencing, however, the results are much more uniform. Across the outcome families, the ADF and KPSS evidence is consistent with stationary monthly changes. This is reassuring for the transformation actually used as the TabPFN prediction target. For Causal-ARIMA (see the following subsection), non-stationarity is instead accommodated through the ordinary and seasonal integration terms in the fitted ARIMA specification.

Regarding pre-treatment validation, the more direct diagnostic for our purpose is pre-treatment forecast performance. We use an expanding-window exercise in which each forecast is generated from observations strictly preceding the month being predicted. For the three main cases, the figures display the final 12 one-step-ahead pre-treatment forecasts. This exercise asks whether the procedure can reproduce untreated observations when faced with the same informational constraint that applies at the true intervention date, rather than simply assessing in-sample fit.

**Causal-ARIMA**. We use Causal-ARIMA [13] as an alternative and more traditional estimator. Causal-ARIMA combines ARIMA forecasting with the potential-outcomes framework. It is designed for causal inference in intervention settings in which a single time series is observed before and after a treatment and an untreated contemporaneous control series is not available.

The method models the pre-intervention dynamics of each outcome and forecasts its counterfactual trajectory in the absence of the intervention. Point effects are calculated as the differences between observed outcomes and counterfactual forecasts; cumulative effects sum these differences over the six-month post-intervention horizon.

Our Causal-ARIMA specification combines the outcome's own ARIMA dynamics with deterministic calendar and time regressors: monthly indicators, including the November indicator, a linear time trend, and a COVID-19 indicator covering March–May and October–December 2020. This exercise therefore assesses whether the findings persist when the counterfactual relies on a parametric model of the outcome's own temporal dependence and predetermined calendar structure, rather than the TFM used in the baseline specification.

Bootstrap inference does not require a collection of independent time series. What is resampled is the innovation process of the fitted pre-intervention ARIMA model. In the Causal-ARIMA approach [13], estimated pre-treatment innovations are used to generate alternative realizations of the post-treatment ARIMA process under the no-treatment state. Each realization therefore gives an alternative counterfactual path for the same series. The resulting distribution reflects uncertainty about what the untreated future of that single series could have looked like.

In our implementation, the Causal-ARIMA routine returns a matrix of bootstrap counterfactual paths. We convert each path into a treatment-effect path by subtracting the simulated counterfactual from the observed post-treatment series. Pointwise intervals are obtained from the empirical quantiles of these treatment-effect draws. For cumulative inference, effects are summed within each bootstrap realization before the quantiles are calculated. This preserves the dependence across months within a simulated counterfactual path rather than treating the monthly effects as independent.

Formally, for bootstrap realization $b$, the month-specific effect draw is:

$$\hat{\tau}_t^{(b)} = Y_t(1) - \hat{Y}_t^{(b)}(0) \tag{D.5}$$

For horizon $H$, the corresponding cumulative effect draw is:

$$\hat{\Delta}_H^{(b)} = \sum_{h=1}^{H} \hat{\tau}_{T_0+h}^{(b)} \tag{D.6}$$

For each model, inference is based on 1,000 bootstrap replications. The empirical distribution of these draws provides pointwise and cumulative bootstrap inference. The code fixes the random seed for reproducibility and records the number of bootstrap draws returned by the fitted model in the run manifest rather than relying on a hard-coded number in the reporting script. This inferential exercise differs from the TabPFN uncertainty bands. TabPFN reports pointwise predictive quantiles conditional on the recursively generated counterfactual history. Causal-ARIMA instead provides simulated counterfactual paths and therefore a direct bootstrap distribution for both point and cumulative effects.

**TabICL**. As a second model-based robustness exercise, we repeat the forecasting procedure using TabICLv2 [14,15]. TabICL is an independently developed TFM based on in-context learning. The original TabICL introduced an architecture designed to scale in-context learning to larger tabular datasets. TabICLv2 extends this framework to regression and classification and introduces a new synthetic-data generator, architectural changes, and a revised pretraining procedure.

The purpose of this exercise is to change the prediction model without changing the empirical design. We therefore retain the same outcomes, intervention dates, feature construction, treatment masking, recursive procedure, and post-treatment horizon used for TabPFN.

TabICLv2 also returns predictive means and predictive quantiles. We use these in the same way as for TabPFN to construct the recursively generated counterfactual and its model-based pointwise predictive range. As in the TabPFN exercise, the predictive quantiles are conditional on the point counterfactual history rather than on a distribution of recursively simulated paths. The close correspondence of the TabPFN and TabICL procedures therefore isolates the extent to which the results depend on the particular TFM used for forecasting.

**Contemporaneous events**. In control-based designs, shocks common to treated and comparison units can, under the relevant and substantive common-shock assumption, be differenced out. This is not possible in our ITS setting because, by design, the salient events are national in scope and there is no control group. We therefore examine potentially confounding developments directly and assess whether their timing and substantive content provide a plausible alternative explanation for the estimated post-event deviations.

The most important case is Cecchettin and the legislative process that culminated in Law No. 168/2023 on violence against women and domestic violence. The bill clearly predated the femicide: it had been introduced in July 2023 and approved by the Chamber of Deputies on 26 October.[6] The Cecchettin case nevertheless has been reported to have accelerated its final parliamentary passage, with the Senate approving it unanimously on 22 November amid the nationwide mobilization that followed the murder [16]. Law No. 168/2023 strengthened the existing framework against domestic and gender-based violence rather than creating an entirely new policy regime. It reinforced preventive and precautionary measures, victim protections, restrictions on alleged perpetrators, electronic monitoring, and procedures intended to accelerate judicial intervention. Moreover, the Law appears unlikely to account for the observed pattern, as neither its approval nor its introduction is associated with a discernible spike in help-seeking behavior. This may reflect either the relatively limited salience of legislative news or the possibility that information about legislative changes does not elicit the same response among women experiencing gender-based violence as news of assaults or femicides.

[6] The law was promulgated on 24 November and entered into force on 9 December.

The timing is important for identification. The legislative process through October, including the Chamber vote, is already part of the pre-treatment history used to construct the Cecchettin counterfactual. The Senate vote occurred only after the case had become a national focal point, while the law itself entered into force in December. The salience indicators for the Cecchettin case reported in Figure S2 provide a further temporal distinction. Social media activity concerning Cecchettin, together with search activity, and public mobilization, rose sharply around the murder, the discovery of the body, the arrest, and the subsequent demonstrations. Their main surge therefore preceded the law's entry into force on 9 December. The implementation of Law No. 168/2023 cannot explain the emergence of the November salience shock or responses that begin before the law entered into force. The timing is therefore much more consistent with the Cecchettin case generating the immediate nationwide response than with the law constituting an independent contemporaneous explanation for it.

We also reviewed major national policy changes related to violence against women around the other events. We identified no comparable nationwide reform entering into force around the timing of the Abbas, Tramontano, Matteuzzi, Indelicato, or Verzeni cases. The few relevant legislative changes occurring within or close to their post-treatment windows differ substantially in timing and scope.

The main reform in the earlier part of the sample is Law No. 69/2019, commonly known as *Codice Rosso*. This was a substantial reform of criminal law and criminal procedure concerning domestic and gender-based violence. It accelerated the handling of specified offences, introduced or strengthened several criminal-law provisions, and reinforced victim protections. It entered into force on 9 August 2019, however, only in the sixth and final month of the Indelicato post-treatment window. It therefore cannot explain the absence of a response in the preceding months or generate the type of immediate mobilization observed after the salient IPFs.

For Abbas, the main contemporaneous concern is the broader COVID-19 period rather than policy change. Pandemic conditions can plausibly affect both help-seeking and police reporting. The forecasting specifications therefore include a COVID-period indicator, and the supplementary analyses also assess robustness to replacing the affected COVID-period observations with imputed values rather than relying on their realized values. These exercises reduce the concern that the Abbas estimates are driven by unusual pandemic-period observations. Moreover, despite the high salience of the Abbas case, we do not observe the positive mobilization in help-seeking and reporting found after the salient IPFs.

For Verzeni, the main concern is not a new policy intervention or another identifiable nationwide shock, but the fact that the case occurs after the exceptionally salient Cecchettin episode. We identified no new nationwide gender-violence reform entering into force during the last few months of 2024; the major reforms introduced in 2023 were already part of the pre-treatment environment. The Verzeni counterfactual is nevertheless estimated using an observed pre-treatment history that includes the unusually elevated period following Cecchettin. As the affected outcomes subsequently return toward more typical levels, this normalization likely explains the negative observed–counterfactual differences observed in the Verzeni post-treatment window.

Taken together, the timing and content of the potentially relevant contemporaneous developments provide substantial reassurance for the interpretation of the estimates. Together with the pre-treatment forecasting evidence, falsification outcomes, and robustness to alternative estimators, this institutional review makes an unrelated contemporaneous national shock an unlikely explanation for the main findings and supports interpreting the estimated deviations as responses to the events themselves.

## SI section E: Robustness

The tables in this section report robustness estimates for all outcomes across the Abbas (Tables S3-S8), Tramontano (Tables S9-S14) and Cecchettin cases (Tables S15-S20). We assess the sensitivity of our findings to five modeling choices: the treatment of pandemic-related disruptions, the exclusion of external time-varying predictors, the inclusion of additional femicide predictors, the counterfactual estimation method, and the choice of tabular foundation model. Each exercise examines whether the estimated effects depend on a particular feature of the baseline specification, while retaining the same event-specific, six-month post-treatment window.

1. Imputation of pandemic-affected observations. We replace observations for March–May and October–December 2020 in the total 1522 calls and reported sexual violence series with the corresponding calendar-month averages over 2017–2019. This modeling choice is broadly consistent with the approach used by the World Health Organization (WHO) to impute all-cause mortality in a no-pandemic counterfactual scenario [17]. The COVID-19 indicator is omitted, and the imputed series are used both as outcomes and to construct lagged predictors. This sensitivity analysis assesses whether exceptional pandemic-related movements in help-seeking and reporting disproportionately affect the estimated counterfactuals.

2. Exclusion of external time-varying predictors. We remove lagged values of the other outcome series, retaining the outcome's own lagged dynamics and deterministic calendar and time features. This specification tests whether the results depend on cross-series predictive relationships or remain similar when the counterfactual is constructed from the outcome's own history.

3. An expanded set of femicide predictors. We augment the baseline predictor set with additional femicide series differentiated by victim–perpetrator relationship, motive and victim characteristics. These include caregiver-related and other intimate-partner killings, family-related killings, killings following sentimental or sexual rejection, and killings of sex workers. All additional series enter with six- and twelve-month lags. This exercise assesses whether finer information about the composition of femicides materially changes the estimated counterfactuals and treatment effects.

4. An alternative counterfactual estimator: Causal-ARIMA. We replace the TabPFN-based forecasting model with Causal-ARIMA while retaining the interrupted time-series framework. Both approaches estimate pre-intervention relationships and project an untreated counterfactual into the post-intervention period. Causal-ARIMA instead uses a parametric time-series model with autoregressive and moving-average components, differencing where selected, and deterministic calendar and time terms. This comparison provides an alternative counterfactual specification with a different functional form and representation of temporal dependence.

5. An alternative tabular foundation model: TabICLv2. Finally, we replace TabPFN with TabICLv2 [14] while retaining the outcomes, predictor construction, event windows and recursive forecasting procedure. This isolates sensitivity to the choice of pretrained foundation model.

**Robustness tests for the case of Saman Abbas.** The robustness exercises concerning Saman Abbas (Tables S3-S8) reveal substantial sensitivity in the estimated response of 1522 calls. The baseline six-month increase of 26.8% remains broadly similar with additional femicide predictors and TabICLv2, but falls to 16.6% without external predictors and 5.6% under Causal-ARIMA, and becomes slightly negative after COVID-period imputation (−2.1%). All monthly estimated effects on calls lie within the corresponding 95% predictive ranges. Evidence for changes in recorded abuse, stalking and sexual violence is also uneven across specifications, with some monthly departures from zero but no consistently supported pattern across methods. For femicides, the six-month point estimates consistently indicate an increase in total cases and a decrease in jealousy- or possessiveness-related IPFs, although both remain within the corresponding 95% predictive ranges.

| | May 21 | Jun 21 | Jul 21 | Aug 21 | Sep 21 | Oct 21 | Cum. Effects (%) |
|---|---|---|---|---|---|---|---|
| Main estimates | 522<br>(-1808; 1844) | 663<br>(-1315; 1910) | 998<br>(-455; 1826) | 1143<br>(-305; 1879) | 1213<br>(-84; 1915) | 489<br>(-1071; 1314) | 3 mths: 21.27<br>6 mths: 26.82 |
| COVID imputed | 321<br>(-1346; 1008) | -468<br>(-4424; 1188) | -155<br>(-2116; 753) | 34<br>(-1963; 943) | 80<br>(-1971; 1011) | -318<br>(-2290; 573) | 3 mths: -2.37<br>6 mths: -2.09 |
| No external time-varying X | 530<br>(-1435; 1798) | 493<br>(-1721; 1759) | 712<br>(-1060; 1636) | 836<br>(-795; 1678) | 783<br>(-819; 1606) | 39<br>(-1571; 916) | 3 mths: 16.20<br>6 mths: 16.64 |
| Causal-ARIMA | 264<br>(-1981; 1946) | 29<br>(-1392; 1268) | 385<br>(-1260; 1832) | 283<br>(-1705; 1659) | 334<br>(-1068; 1742) | -26<br>(-1640; 1544) | 3 mths: 5.75<br>6 mths: 5.64 |
| Additional femicide predictors | 552<br>(-1360; 1794) | 582<br>(-1513; 1825) | 834<br>(-789; 1710) | 1027<br>(-405; 1799) | 1131<br>(-164; 1874) | 578<br>(-834; 1317) | 3 mths: 18.78<br>6 mths: 24.67 |
| TabICLv2 | 465<br>(-4052; 2853) | 727<br>(-4060; 3467) | 1059<br>(-2173; 2600) | 1100<br>(-1946; 2396) | 1061<br>(-1996; 2382) | 125<br>(-3472; 1573) | 3 mths: 22.08<br>6 mths: 23.58 |

**Table S3 | 1522 calls - Robustness estimates for Saman Abbas.** Entries report monthly observed-minus-counterfactual counts, with model-specific 95% uncertainty intervals below. TabPFN and TabICLv2 intervals derive from predictive quantiles; Causal-ARIMA intervals are bootstrap-based (1,000 samples). Cumulative percentage effects equal 100 times the sum of monthly effects divided by the sum of counterfactual counts over the indicated horizon. Bold indicates monthly effects whose 95% uncertainty intervals lie entirely above or below zero.

| | May 21 | Jun 21 | Jul 21 | Aug 21 | Sep 21 | Oct 21 | Cum. Effects (%) |
|---|---|---|---|---|---|---|---|
| Main estimates | 91 (-287; 383) | 236 (-106; 506) | 198 (-137; 466) | 91 (-241; 358) | 26 (-321; 322) | -170 (-503; 117) | 3 mths: 8.75<br>6 mths: 3.95 |
| COVID imputed | 91 (-271; 370) | 238 (-105; 506) | 203 (-146; 474) | 100 (-243; 362) | 35 (-319; 326) | -157 (-495; 125) | 3 mths: 8.87<br>6 mths: 4.28 |
| No external time-varying X | 102 (-251; 370) | 280 (-30; 518) | 290 (-40; 522) | 199 (-129; 433) | 156 (-178; 416) | -44 (-379; 216) | 3 mths: 11.49<br>6 mths: 8.59 |
| Causal-ARIMA | **181** (55; 290) | **213** (86; 372) | 79 (-60; 187) | **-197** (-323; -88) | **-321** (-460; -162) | **-443** (-582; -284) | 3 mths: 7.82<br>6 mths: -3.78 |
| Additional femicide predictors | 95 (-282; 386) | 231 (-116; 502) | 201 (-132; 464) | 100 (-235; 366) | 40 (-302; 330) | -149 (-476; 132) | 3 mths: 8.80<br>6 mths: 4.36 |
| TabICLv2 | 131 (-288; 468) | **373** (85; 578) | **354** (12; 627) | 205 (-205; 554) | 137 (-295; 510) | -50 (-429; 284) | 3 mths: 15.14<br>6 mths: 10.20 |

**Table S4 | Abuses against women - Robustness estimates for Saman Abbas.** See Notes of Table S3.

| | May 21 | Jun 21 | Jul 21 | Aug 21 | Sep 21 | Oct 21 | Cum. Effects (%) |
|---|---|---|---|---|---|---|---|
| Main estimates | 64 (-398; 410) | 41 (-287; 305) | 82 (-188; 295) | -16 (-272; 177) | -106 (-429; 165) | **-233** (-515; -25) | 3 mths: 3.64<br>6 mths: -1.66 |
| COVID imputed | 58 (-366; 386) | 28 (-286; 284) | 59 (-239; 283) | -54 (-344; 153) | -148 (-494; 128) | **-279** (-585; -57) | 3 mths: 2.82<br>6 mths: -3.25 |
| No external time-varying X | 88 (-334; 424) | 54 (-284; 333) | 76 (-209; 302) | -28 (-298; 172) | -128 (-460; 154) | **-264** (-554; -55) | 3 mths: 4.30<br>6 mths: -1.98 |
| Causal-ARIMA | 25 (-156; 127) | 91 (-55; 241) | 79 (-74; 270) | 29 (-120; 183) | -33 (-182; 122) | -157 (-305; 30) | 3 mths: 3.80<br>6 mths: 0.34 |
| Additional femicide predictors | 61 (-371; 396) | 29 (-274; 283) | 75 (-185; 285) | -24 (-271; 166) | -121 (-440; 152) | **-244** (-517; -38) | 3 mths: 3.22<br>6 mths: -2.19 |
| TabICLv2 | 31 (-441; 448) | 1 (-429; 350) | 76 (-354; 376) | -5 (-366; 261) | -128 (-586; 257) | -248 (-674; 38) | 3 mths: 2.08<br>6 mths: -2.66 |

**Table S5 | Stalking crimes - Robustness estimates for Saman Abbas.** See Notes of Table S3.

| | May 21 | Jun 21 | Jul 21 | Aug 21 | Sep 21 | Oct 21 | Cum. Effects (%) |
|---|---|---|---|---|---|---|---|
| Main estimates | 76 (-58; 212) | 85 (-39; 216) | 24 (-99; 156) | 37 (-79; 155) | -27 (-151; 95) | -33 (-156; 90) | 3 mths: 12.83<br>6 mths: 5.64 |
| COVID imputed | 97 (-3; 183) | 109 (-1; 205) | 4 (-125; 117) | 10 (-96; 94) | -55 (-151; 22) | -65 (-156; 10) | 3 mths: 14.77<br>6 mths: 3.40 |
| No external time-varying X | 83 (-41; 213) | 94 (-21; 217) | 36 (-83; 164) | 53 (-59; 166) | -3 (-111; 109) | -6 (-115; 105) | 3 mths: 15.09<br>6 mths: 9.30 |
| Causal-ARIMA | **128** (67; 200) | **76** (13; 145) | 39 (-25; 111) | **69** (4; 144) | 28 (-38; 99) | -22 (-85; 54) | 3 mths: 17.58<br>6 mths: 11.78 |
| Additional femicide predictors | 79 (-51; 213) | 89 (-34; 219) | 29 (-93; 160) | 39 (-77; 156) | -28 (-149; 93) | -36 (-155; 85) | 3 mths: 13.69<br>6 mths: 5.98 |
| TabICLv2 | 104 (-86; 348) | 93 (-84; 335) | 33 (-149; 287) | 62 (-104; 257) | 39 (-145; 247) | 45 (-122; 247) | 3 mths: 16.45<br>6 mths: 14.18 |

**Table S6 | Sexual violence crimes - Robustness estimates for Saman Abbas.** See Notes of Table S3.

| | May 21 | Jun 21 | Jul 21 | Aug 21 | Sep 21 | Oct 21 | Cum. Effects (%) |
|---|---|---|---|---|---|---|---|
| Main estimates | 3.01 (-2.72; 8.61) | 1.34 (-4.79; 7.02) | 2.78 (-3.25; 7.95) | 0.15 (-5.98; 5.72) | 5.89 (-0.07; 11.24) | 2.87 (-3.10; 8.17) | 3 mths: 31.22<br>6 mths: 34.16 |
| COVID imputed | 2.84 (-2.65; 8.17) | 1.45 (-4.63; 7.03) | 2.67 (-3.40; 7.92) | 0.12 (-6.03; 5.65) | 5.80 (-0.22; 11.18) | 2.93 (-3.09; 8.21) | 3 mths: 30.24<br>6 mths: 33.51 |
| No external time-varying X | 2.89 (-2.74; 8.33) | 1.58 (-4.47; 7.02) | 2.57 (-3.43; 7.78) | 0.05 (-6.10; 5.49) | 5.78 (-0.16; 11.08) | 2.89 (-3.13; 8.16) | 3 mths: 30.68<br>6 mths: 33.36 |
| Causal-ARIMA | 1.96 (-0.73; 4.53) | 0.97 (-2.22; 4.57) | 0.91 (-2.30; 4.16) | 0.31 (-2.83; 3.94) | **6.63** (3.46; 10.00) | 1.25 (-2.05; 4.84) | 3 mths: 14.70<br>6 mths: 23.61 |
| Additional femicide predictors | 2.96 (-2.71; 8.51) | 1.23 (-4.93; 6.85) | 3.03 (-2.92; 8.13) | -0.20 (-6.39; 5.43) | **6.01** (0.01; 11.23) | 2.72 (-3.26; 8.08) | 3 mths: 31.74<br>6 mths: 33.38 |
| TabICLv2 | 3.34 (-2.09; 8.78) | 0.38 (-8.54; 7.85) | 3.48 (-1.38; 8.00) | -0.55 (-8.67; 6.80) | **7.27** (1.23; 11.89) | 1.27 (-5.60; 8.24) | 3 mths: 31.56<br>6 mths: 31.76 |

**Table S7 | All femicides - Robustness estimates for Saman Abbas.** See Notes of Table S3.

| | May 21 | Jun 21 | Jul 21 | Aug 21 | Sep 21 | Oct 21 | Cum. Effects (%) |
|---|---|---|---|---|---|---|---|
| Main estimates | 0.09 (-4.13; 3.61) | -2.08 (-6.28; 1.99) | -0.50 (-4.36; 3.00) | 1.31 (-2.90; 4.00) | 1.28 (-2.81; 4.81) | -2.74 (-6.90; 0.90) | 3 mths: -19.95<br>6 mths: -11.64 |
| COVID imputed | 0.24 (-3.82; 3.68) | -2.20 (-6.39; 1.92) | -0.53 (-4.37; 3.00) | 1.33 (-2.82; 4.00) | 1.40 (-2.69; 4.91) | -2.93 (-7.11; 0.78) | 3 mths: -19.88<br>6 mths: -11.80 |
| No external time-varying X | 0.20 (-4.02; 3.72) | -1.97 (-6.05; 2.02) | -0.53 (-4.29; 2.96) | 1.36 (-2.76; 4.00) | 1.17 (-2.85; 4.67) | -2.59 (-6.56; 0.84) | 3 mths: -18.67<br>6 mths: -10.52 |
| Causal-ARIMA | -0.79 (-3.37; 2.33) | -1.83 (-4.41; 0.80) | -1.20 (-3.79; 1.43) | 2.30 (-0.28; 5.41) | 0.61 (-1.97; 3.73) | **-3.53** (-6.12; -0.90) | 3 mths: -27.63<br>6 mths: -18.16 |
| Additional femicide predictors | 0.14 (-4.03; 3.61) | -2.21 (-6.39; 1.87) | -0.40 (-4.22; 3.00) | 1.17 (-3.09; 4.00) | 1.21 (-2.93; 4.78) | -2.78 (-6.93; 0.79) | 3 mths: -19.80<br>6 mths: -12.52 |
| TabICLv2 | 0.02 (-4.18; 3.60) | -2.79 (-7.41; 1.77) | 0.26 (-4.44; 3.00) | 0.29 (-4.57; 4.00) | -0.92 (-7.46; 5.00) | -0.74 (-5.75; 1.00) | 3 mths: -20.09<br>6 mths: -16.26 |

**Table S8 | IPFs motivated by jealousy or possessiveness - Robustness estimates for Saman Abbas.** See Notes of Table S3.

**Robustness tests for the case of Giulia Tramontano.** For the Giulia Tramontano case (Tables S9-S14), the most robust finding is an increase in calls to the 1522 helpline. Six-month estimates are positive across all specifications, ranging from 18.3% to 31.0%, compared with 29.5% in the baseline, with TabICLv2 closely reproducing both the baseline magnitude and monthly pattern. The remaining outcomes show little consistent evidence of a sustained response: estimates for recorded abuse, stalking, and sexual violence are generally small relative to their uncertainty, while those for femicides are more sensitive to the forecasting method. Overall, the robustness exercises support a comparatively stable increase in the initial stage of help-seeking rather than a broader response across the other measured outcomes.

| | May 23 | Jun 23 | Jul 23 | Aug 23 | Sep 23 | Oct 23 | Cum. Effects (%) |
|---|---|---|---|---|---|---|---|
| Main estimates | 357<br>(-254; 895) | **1056**<br>(491; 1513) | **940**<br>(341; 1364) | **802**<br>(315; 1213) | **1867**<br>(1421; 2252) | **1474**<br>(1010; 1879) | 3 mths: 20.87<br>6 mths: 29.48 |
| COVID imputed | 270<br>(-775; 821) | 861<br>(-373; 1411) | 694<br>(-582; 1276) | 528<br>(-500; 1062) | **1598**<br>(744; 2074) | **1170**<br>(198; 1686) | 3 mths: 15.47<br>6 mths: 21.88 |
| No external time-varying X | 354<br>(-468; 944) | **1070**<br>(338; 1595) | **998**<br>(382; 1465) | **872**<br>(323; 1298) | **1921**<br>(1398; 2327) | **1538**<br>(983; 1959) | 3 mths: 21.62<br>6 mths: 31.01 |
| Causal-ARIMA | -39<br>(-1517; 895) | 597<br>(-1077; 1682) | 555<br>(-1150; 1705) | 368<br>(-1574; 1561) | 1548<br>(-797; 2791) | 1392<br>(-699; 2585) | 3 mths: 8.90<br>6 mths: 18.34 |
| Additional femicide predictors | 366<br>(-401; 959) | **1089**<br>(515; 1565) | **960**<br>(235; 1450) | **832**<br>(322; 1259) | **1906**<br>(1222; 2398) | **1506**<br>(1040; 1895) | 3 mths: 21.54<br>6 mths: 30.44 |
| TabICLv2 | 342<br>(-875; 1079) | **1033**<br>(424; 1510) | **899**<br>(269; 1276) | **739**<br>(308; 1151) | **1797**<br>(1457; 2112) | **1392**<br>(742; 1881) | 3 mths: 20.03<br>6 mths: 27.77 |

**Table S9 | 1522 calls - Robustness estimates for Giulia Tramontano.** See Notes of Table S3.

| | May 23 | Jun 23 | Jul 23 | Aug 23 | Sep 23 | Oct 23 | Cum. Effects (%) |
|---|---|---|---|---|---|---|---|
| Main estimates | -164<br>(-467; 117) | 0<br>(-276; 255) | 15<br>(-264; 277) | -90<br>(-356; 152) | 27<br>(-240; 280) | 67<br>(-192; 298) | 3 mths: -2.25<br>6 mths: -1.10 |
| COVID imputed | -158<br>(-477; 136) | -4<br>(-283; 257) | 25<br>(-255; 291) | -85<br>(-353; 159) | 33<br>(-236; 281) | 58<br>(-209; 297) | 3 mths: -2.06<br>6 mths: -0.99 |
| No external time-varying X | -162<br>(-458; 106) | -34<br>(-286; 211) | -4<br>(-286; 239) | -165<br>(-418; 75) | -8<br>(-256; 227) | 96<br>(-160; 309) | 3 mths: -3.00<br>6 mths: -2.08 |
| Causal-ARIMA | -78<br>(-257; 128) | 76<br>(-147; 302) | 119<br>(-116; 367) | -45<br>(-299; 199) | -9<br>(-249; 241) | 27<br>(-212; 264) | 3 mths: 1.83<br>6 mths: 0.69 |
| Additional femicide predictors | -148<br>(-492; 169) | 9<br>(-279; 272) | 10<br>(-279; 277) | -100<br>(-378; 149) | 44<br>(-222; 294) | 57<br>(-213; 300) | 3 mths: -1.96<br>6 mths: -0.97 |
| TabICLv2 | -168<br>(-537; 149) | 12<br>(-282; 268) | 12<br>(-273; 259) | -33<br>(-272; 178) | 54<br>(-208; 305) | 78<br>(-162; 305) | 3 mths: -2.17<br>6 mths: -0.35 |

**Table S10 | Abuses against women - Robustness estimates for Giulia Tramontano.** See Notes of Table S3.

| | May 23 | Jun 23 | Jul 23 | Aug 23 | Sep 23 | Oct 23 | Cum. Effects (%) |
|---|---|---|---|---|---|---|---|
| Main estimates | -9<br>(-252; 221) | -90<br>(-300; 112) | 138<br>(-82; 335) | 30<br>(-148; 188) | 140<br>(-47; 320) | 85<br>(-118; 274) | 3 mths: 0.74<br>6 mths: 2.92 |
| COVID imputed | -14<br>(-276; 230) | -90<br>(-311; 115) | 137<br>(-103; 338) | 23<br>(-161; 184) | 125<br>(-59; 304) | 66<br>(-135; 254) | 3 mths: 0.64<br>6 mths: 2.44 |
| No external time-varying X | 22<br>(-234; 259) | -37<br>(-250; 157) | 176<br>(-32; 365) | 83<br>(-88; 232) | 129<br>(-67; 321) | 116<br>(-104; 303) | 3 mths: 3.19<br>6 mths: 4.95 |
| Causal-ARIMA | -31<br>(-158; 83) | -95<br>(-263; 66) | 76<br>(-96; 249) | -16<br>(-183; 149) | 57<br>(-112; 217) | 82<br>(-73; 246) | 3 mths: -0.94<br>6 mths: 0.71 |
| Additional femicide predictors | 0<br>(-254; 250) | -74<br>(-281; 130) | 138<br>(-76; 337) | 36<br>(-150; 196) | 152<br>(-45; 340) | 79<br>(-118; 270) | 3 mths: 1.24<br>6 mths: 3.29 |
| TabICLv2 | -29<br>(-333; 220) | -120<br>(-309; 69) | 100<br>(-150; 324) | 34<br>(-173; 204) | 179<br>(-22; 373) | 91<br>(-167; 331) | 3 mths: -0.93<br>6 mths: 2.51 |

**Table S11 | Stalking crimes - Robustness estimates for Giulia Tramontano.** See Notes of Table S3.

| | May 23 | Jun 23 | Jul 23 | Aug 23 | Sep 23 | Oct 23 | Cum. Effects (%) |
|---|---|---|---|---|---|---|---|
| Main estimates | -34<br>(-197; 124) | 21<br>(-120; 154) | 77<br>(-67; 213) | 11<br>(-119; 131) | -4<br>(-134; 116) | 14<br>(-122; 143) | 3 mths: 3.75<br>6 mths: 2.56 |
| COVID imputed | -45<br>(-259; 121) | 10<br>(-142; 127) | 44<br>(-106; 167) | -22<br>(-149; 76) | -46<br>(-161; 44) | -14<br>(-149; 91) | 3 mths: 0.48<br>6 mths: -2.12 |
| No external time-varying X | -47<br>(-227; 122) | 26<br>(-117; 157) | 62<br>(-82; 199) | -3<br>(-134; 117) | -1<br>(-134; 119) | 19<br>(-126; 154) | 3 mths: 2.38<br>6 mths: 1.67 |
| Causal-ARIMA | -18<br>(-95; 74) | 17<br>(-62; 113) | 33<br>(-45; 129) | -18<br>(-103; 65) | -13<br>(-98; 74) | 23<br>(-63; 112) | 3 mths: 1.82<br>6 mths: 0.71 |
| Additional femicide predictors | -32<br>(-209; 143) | 27<br>(-116; 163) | 80<br>(-64; 220) | 9<br>(-127; 136) | -3<br>(-135; 120) | 16<br>(-123; 150) | 3 mths: 4.39<br>6 mths: 2.89 |
| TabICLv2 | -65<br>(-214; 176) | -5<br>(-197; 166) | 32<br>(-144; 219) | -19<br>(-174; 115) | -30<br>(-209; 113) | 21<br>(-196; 224) | 3 mths: -2.11<br>6 mths: -1.90 |

**Table S12 | Sexual violence crimes - Robustness estimates for Giulia Tramontano.** See Notes of Table S3.

| | May 23 | Jun 23 | Jul 23 | Aug 23 | Sep 23 | Oct 23 | Cum. Effects (%) |
|---|---|---|---|---|---|---|---|
| Main estimates | -5.26<br>(-11.77; 1.56) | 4.73<br>(-1.71; 10.49) | -2.59<br>(-8.77; 3.62) | -1.11<br>(-7.50; 4.62) | 3.67<br>(-2.43; 9.83) | 2.98<br>(-3.03; 8.84) | 3 mths: -11.12<br>6 mths: 4.49 |
| COVID imputed | -4.75<br>(-11.20; 1.77) | 4.29<br>(-2.12; 10.06) | -2.01<br>(-8.28; 4.02) | -1.57<br>(-7.90; 4.18) | 4.20<br>(-2.06; 10.20) | 3.02<br>(-3.07; 8.98) | 3 mths: -8.98<br>6 mths: 6.03 |
| No external time-varying X | -4.27<br>(-10.80; 2.00) | 4.92<br>(-1.57; 10.47) | -2.65<br>(-8.24; 3.46) | -0.69<br>(-7.66; 5.86) | 4.25<br>(-2.00; 10.35) | 3.73<br>(-2.40; 9.37) | 3 mths: -7.39<br>6 mths: 10.45 |
| Causal-ARIMA | **-9.21**<br>(-13.82; -5.47) | 2.67<br>(-1.84; 6.78) | -1.60<br>(-6.21; 2.74) | -1.85<br>(-6.39; 2.34) | 0.99<br>(-3.37; 5.45) | 3.75<br>(-0.80; 8.11) | 3 mths: -24.58<br>6 mths: -8.59 |
| Additional femicide predictors | -4.51<br>(-11.13; 2.01) | 4.74<br>(-1.61; 10.51) | -2.13<br>(-8.40; 3.89) | -1.45<br>(-7.81; 4.35) | 3.83<br>(-2.52; 10.10) | 3.01<br>(-3.12; 8.99) | 3 mths: -7.07<br>6 mths: 6.64 |
| TabICLv2 | -4.35<br>(-12.28; 3.11) | 4.06<br>(-2.32; 9.98) | -3.49<br>(-9.76; 3.12) | 0.39<br>(-7.17; 5.56) | 2.98<br>(-5.26; 10.66) | 2.30<br>(-4.18; 8.50) | 3 mths: -13.13<br>6 mths: 3.49 |

**Table S13 | All femicides - Robustness estimates for Giulia Tramontano.** See Notes of Table S3.

| | May 23 | Jun 23 | Jul 23 | Aug 23 | Sep 23 | Oct 23 | Cum. Effects (%) |
|---|---|---|---|---|---|---|---|
| Main estimates | -2.26 (-5.83; 1.50) | 0.38 (-3.26; 3.82) | -0.57 (-4.08; 2.78) | -0.67 (-4.33; 2.68) | 2.27 (-1.23; 5.59) | 0.31 (-3.15; 3.52) | 3 mths: -21.39<br>6 mths: -2.34 |
| COVID imputed | -2.10 (-5.69; 1.68) | 0.47 (-3.12; 3.89) | -0.57 (-4.09; 2.81) | -0.53 (-4.20; 2.82) | 2.31 (-1.18; 5.63) | 0.43 (-3.02; 3.63) | 3 mths: -19.60<br>6 mths: 0.11 |
| No external time-varying X | -2.17 (-5.65; 1.53) | 0.48 (-3.13; 3.87) | -0.56 (-4.01; 2.76) | -0.59 (-4.26; 2.71) | 2.14 (-1.27; 5.50) | 0.45 (-3.05; 3.55) | 3 mths: -19.98<br>6 mths: -1.14 |
| Causal-ARIMA | -2.09 (-4.41; 0.33) | 1.28 (-0.91; 3.75) | -0.05 (-2.76; 2.65) | 1.07 (-2.01; 3.72) | 1.53 (-1.26; 4.55) | 1.97 (-0.89; 4.73) | 3 mths: -8.70<br>6 mths: 20.32 |
| Additional femicide predictors | -2.16 (-5.87; 1.64) | 0.54 (-3.11; 3.98) | -0.65 (-4.17; 2.71) | -0.62 (-4.29; 2.73) | 2.31 (-1.24; 5.68) | 0.21 (-3.27; 3.48) | 3 mths: -20.18<br>6 mths: -1.67 |
| TabICLv2 | -3.23 (-6.82; 0.79) | 0.42 (-3.09; 3.41) | -0.18 (-3.62; 2.80) | -0.34 (-4.96; 3.00) | 1.71 (-1.76; 5.18) | 0.40 (-3.02; 3.22) | 3 mths: -24.93<br>6 mths: -5.25 |

**Table S14 | IPFs motivated by jealousy or possessiveness - Robustness estimates for Giulia Tramontano.** See Notes of Table S3.

**Robustness tests for the case of Giulia Cecchettin.** The Cecchettin case exhibits the strongest consistency across specifications for calls and recorded abuse and stalking. Six-month point estimates remain positive throughout, ranging from 47.5% to 63.2% for calls, 11.6% to 26.6% for abuse, and 7.9% to 12.8% for stalking. Several monthly effects fall outside the corresponding 95% predictive ranges across all methods, although their duration and magnitude vary. By contrast, all monthly effects for sexual violence remain within the corresponding 95% predictive ranges. Total-femicide point estimates consistently indicate a six-month reduction of approximately 22–27%, but monthly effects fall outside the corresponding 95% predictive ranges mainly in April and this pattern does not persist under TabICLv2. For jealousy- or possessiveness-related IPFs, the six-month estimates are negative across specifications, but their magnitude varies substantially and all monthly effects remain within the corresponding 95% predictive ranges. Thus, the robustness exercises provide stronger support for increases in help-seeking and recorded abuse and stalking than for a precisely estimated reduction in lethal violence.

| | Nov 23 | Dec 23 | Jan 24 | Feb 24 | Mar 24 | Apr 24 | Cum. Effects (%) |
|---|---|---|---|---|---|---|---|
| Main estimates | 3384 (-203; 5126) | **6659** (3985; 7969) | **3118** (1713; 4255) | **1741** (245; 2774) | 1311 (-978; 2408) | 264 (-1750; 1439) | 3 mths: 74.82<br>6 mths: 47.54 |
| COVID imputed | **3276** (60; 4890) | **6692** (4848; 7675) | **3232** (1771; 4356) | **1889** (530; 2714) | 1551 (-491; 2438) | 737 (-612; 1639) | 3 mths: 75.21<br>6 mths: 51.46 |
| No external time-varying X | 3201 (-609; 5110) | **6543** (4254; 7771) | **3192** (1838; 4324) | **1878** (541; 2878) | 1385 (-1131; 2594) | 434 (-1269; 1534) | 3 mths: 72.63<br>6 mths: 48.21 |
| Causal-ARIMA | **3249** (1743; 4191) | **7162** (5099; 8247) | **4061** (2419; 5216) | **2756** (631; 3977) | **2103** (92; 3381) | 464 (-1595; 1744) | 3 mths: 88.91<br>6 mths: 63.15 |
| Additional femicide predictors | 3026 (-1629; 5263) | **6547** (4234; 7860) | **3300** (2016; 4442) | **1914** (393; 2982) | 1673 (-195; 2693) | 461 (-2007; 1769) | 3 mths: 72.02<br>6 mths: 49.45 |
| TabICLv2 | 2944 (-99; 5189) | **6291** (2588; 7850) | **3426** (1250; 5128) | 2328 (-302; 3727) | 1853 (-1219; 3061) | 811 (-2263; 2059) | 3 mths: 69.99<br>6 mths: 52.72 |

**Table S15 | 1522 calls - Robustness estimates for Giulia Cecchettin.** See Notes of Table S3.

| | Nov 23 | Dec 23 | Jan 24 | Feb 24 | Mar 24 | Apr 24 | Cum. Effects (%) |
|---|---|---|---|---|---|---|---|
| Main estimates | 125<br>(-114; 343) | **513**<br>(271; 735) | **554**<br>(292; 802) | **556**<br>(331; 771) | **572**<br>(290; 828) | **390**<br>(134; 626) | 3 mths: 19.28<br>6 mths: 22.84 |
| COVID imputed | 121<br>(-124; 338) | **509**<br>(260; 736) | **541**<br>(269; 797) | **551**<br>(326; 769) | **556**<br>(256; 825) | **379**<br>(119; 617) | 3 mths: 18.84<br>6 mths: 22.27 |
| No external time-varying X | 126<br>(-107; 335) | **481**<br>(241; 705) | **495**<br>(220; 768) | **516**<br>(315; 723) | **548**<br>(283; 798) | **357**<br>(123; 577) | 3 mths: 17.55<br>6 mths: 20.91 |
| Causal-ARIMA | 85<br>(-97; 291) | **413**<br>(209; 625) | **344**<br>(106; 569) | **334**<br>(90; 578) | **264**<br>(29; 499) | 77<br>(-160; 329) | 3 mths: 12.88<br>6 mths: 11.62 |
| Additional femicide predictors | 132<br>(-130; 366) | **523**<br>(264; 754) | **575**<br>(317; 821) | **581**<br>(360; 793) | **578**<br>(308; 830) | **377**<br>(114; 617) | 3 mths: 20.00<br>6 mths: 23.41 |
| TabICLv2 | 190<br>(-16; 390) | **563**<br>(341; 763) | **586**<br>(323; 808) | **582**<br>(417; 755) | **647**<br>(393; 876) | **494**<br>(282; 684) | 3 mths: 22.16<br>6 mths: 26.58 |

**Table S16 | Abuses against women - Robustness estimates for Giulia Cecchettin.** See Notes of Table S3.

| | Nov 23 | Dec 23 | Jan 24 | Feb 24 | Mar 24 | Apr 24 | Cum. Effects (%) |
|---|---|---|---|---|---|---|---|
| Main estimates | -4<br>(-231; 170) | 111<br>(-61; 269) | **393**<br>(197; 582) | **316**<br>(144; 467) | 137<br>(-84; 354) | **242**<br>(60; 398) | 3 mths: 10.58<br>6 mths: 12.76 |
| COVID imputed | -7<br>(-230; 166) | 115<br>(-54; 275) | **386**<br>(177; 585) | **305**<br>(123; 462) | 127<br>(-109; 358) | **224**<br>(29; 386) | 3 mths: 10.45<br>6 mths: 12.23 |
| No external time-varying X | 14<br>(-180; 170) | 109<br>(-69; 272) | **378**<br>(167; 583) | **290**<br>(107; 446) | 146<br>(-84; 376) | **215**<br>(18; 387) | 3 mths: 10.62<br>6 mths: 12.26 |
| Causal-ARIMA | 28<br>(-104; 150) | 81<br>(-61; 243) | **286**<br>(128; 446) | **197**<br>(43; 361) | 62<br>(-95; 239) | 118<br>(-40; 271) | 3 mths: 8.16<br>6 mths: 7.88 |
| Additional femicide predictors | 1<br>(-235; 178) | 116<br>(-70; 281) | **386**<br>(180; 585) | **311**<br>(131; 466) | 144<br>(-80; 363) | **243**<br>(51; 406) | 3 mths: 10.63<br>6 mths: 12.84 |
| TabICLv2 | -17<br>(-251; 157) | 140<br>(-41; 281) | **362**<br>(171; 504) | **298**<br>(101; 440) | 94<br>(-117; 315) | **216**<br>(2; 381) | 3 mths: 10.25<br>6 mths: 11.56 |

**Table S17 | Stalking crimes - Robustness estimates for Giulia Cecchettin.** See Notes of Table S3.

| | Nov 23 | Dec 23 | Jan 24 | Feb 24 | Mar 24 | Apr 24 | Cum. Effects (%) |
|---|---|---|---|---|---|---|---|
| Main estimates | -3<br>(-125; 112) | -15<br>(-130; 92) | 35<br>(-93; 163) | 30<br>(-92; 144) | 70<br>(-58; 201) | 99<br>(-18; 211) | 3 mths: 1.17<br>6 mths: 7.56 |
| COVID imputed | 3<br>(-101; 89) | -8<br>(-113; 80) | 30<br>(-106; 149) | 23<br>(-83; 113) | 43<br>(-85; 152) | 84<br>(-20; 170) | 3 mths: 1.77<br>6 mths: 6.07 |
| No external time-varying X | -7<br>(-130; 107) | -18<br>(-134; 89) | 27<br>(-104; 161) | 20<br>(-103; 135) | 79<br>(-53; 214) | 102<br>(-19; 214) | 3 mths: 0.16<br>6 mths: 7.08 |
| Causal-ARIMA | -18<br>(-94; 69) | -49<br>(-127; 35) | 48<br>(-36; 136) | 0<br>(-80; 91) | 78<br>(-1; 165) | 71<br>(-11; 150) | 3 mths: -1.34<br>6 mths: 4.37 |
| Additional femicide predictors | -3<br>(-131; 117) | -22<br>(-137; 87) | 31<br>(-98; 161) | 26<br>(-97; 144) | 66<br>(-62; 195) | 95<br>(-25; 208) | 3 mths: 0.40<br>6 mths: 6.71 |
| TabICLv2 | 21<br>(-114; 127) | -14<br>(-163; 110) | 1<br>(-139; 158) | 10<br>(-138; 132) | 56<br>(-127; 267) | 90<br>(-62; 230) | 3 mths: 0.58<br>6 mths: 5.66 |

**Table S18 | Sexual violence crimes - Robustness estimates for Giulia Cecchettin.** See Notes of Table S3.

| | Nov 23 | Dec 23 | Jan 24 | Feb 24 | Mar 24 | Apr 24 | Cum. Effects (%) |
|---|---|---|---|---|---|---|---|
| Main estimates | -2.52 (-8.27; 2.95) | -2.87 (-9.03; 2.86) | -2.92 (-8.88; 3.02) | 1.50 (-4.46; 6.94) | -1.30 (-7.28; 4.38) | **-5.78** (-11.66; -0.01) | 3 mths: -32.84<br>6 mths: -27.29 |
| COVID imputed | -2.30 (-8.06; 3.11) | -2.75 (-9.01; 2.88) | -2.80 (-8.94; 3.05) | 1.45 (-4.47; 6.79) | -1.29 (-7.23; 4.38) | **-5.86** (-11.73; -0.06) | 3 mths: -31.58<br>6 mths: -26.81 |
| No external time-varying X | -2.34 (-8.02; 3.08) | -2.74 (-8.79; 2.90) | -3.14 (-8.89; 2.67) | 1.71 (-4.08; 6.93) | -1.18 (-7.06; 4.38) | **-6.02** (-11.72; -0.21) | 3 mths: -32.59<br>6 mths: -27.04 |
| Causal-ARIMA | -1.42 (-6.17; 2.72) | -1.07 (-5.51; 3.68) | -2.51 (-7.39; 2.04) | **4.92** (0.27; 9.86) | -1.41 (-5.78; 3.39) | **-9.01** (-13.56; -4.41) | 3 mths: -22.76<br>6 mths: -22.12 |
| Additional femicide predictors | -2.38 (-8.23; 3.23) | -3.03 (-9.14; 2.69) | -2.84 (-8.67; 2.99) | 1.34 (-4.48; 6.71) | -0.98 (-6.90; 4.61) | **-5.94** (-11.73; -0.15) | 3 mths: -32.69<br>6 mths: -27.23 |
| TabICLv2 | -1.62 (-8.18; 2.21) | -2.40 (-8.78; 4.48) | -2.79 (-9.92; 3.48) | 1.96 (-4.83; 7.47) | -2.38 (-8.83; 4.63) | -5.22 (-12.20; 0.67) | 3 mths: -28.60<br>6 mths: -25.17 |

**Table S19 | All femicides - Robustness estimates for Giulia Cecchettin.** See Notes of Table S3.

| | Nov 23 | Dec 23 | Jan 24 | Feb 24 | Mar 24 | Apr 24 | Cum. Effects (%) |
|---|---|---|---|---|---|---|---|
| Main estimates | -0.16 (-3.59; 2.98) | -2.54 (-6.22; 0.71) | -0.90 (-4.30; 2.38) | -1.28 (-4.74; 1.87) | -0.48 (-3.88; 2.80) | -1.77 (-5.19; 1.57) | 3 mths: -33.97<br>6 mths: -33.74 |
| COVID imputed | -0.20 (-3.63; 2.96) | -2.52 (-6.26; 0.75) | -0.86 (-4.28; 2.40) | -1.28 (-4.74; 1.87) | -0.46 (-3.87; 2.79) | -1.79 (-5.21; 1.55) | 3 mths: -33.79<br>6 mths: -33.65 |
| No external time-varying X | -0.20 (-3.57; 2.91) | -2.45 (-6.05; 0.70) | -0.90 (-4.26; 2.36) | -1.33 (-4.77; 1.80) | -0.49 (-3.86; 2.75) | -1.89 (-5.33; 1.45) | 3 mths: -33.63<br>6 mths: -34.16 |
| Causal-ARIMA | 0.34 (-1.77; 2.46) | 0.12 (-2.20; 2.38) | 0.83 (-2.06; 3.61) | -1.00 (-3.68; 1.83) | -0.50 (-3.27; 2.26) | -0.91 (-3.60; 1.70) | 3 mths: 22.80<br>6 mths: -7.35 |
| Additional femicide predictors | -0.16 (-3.57; 3.00) | -2.24 (-5.89; 0.98) | -1.00 (-4.37; 2.26) | -1.27 (-4.67; 1.84) | -0.33 (-3.71; 2.89) | -1.97 (-5.40; 1.39) | 3 mths: -32.73<br>6 mths: -33.24 |
| TabICLv2 | 0.00 (-3.47; 2.78) | -1.19 (-4.54; 1.00) | -1.18 (-4.73; 1.96) | -0.90 (-4.37; 1.97) | -0.34 (-3.67; 2.78) | -1.77 (-5.53; 1.49) | 3 mths: -25.29<br>6 mths: -27.77 |

**Table S20 | IPFs motivated by jealousy or possessiveness - Robustness estimates for Giulia Cecchettin.** See Notes of Table S3.

**Pre-event forecasting performance.** We complement the robustness analyses with an assessment of the pre-event forecasting performance of the alternative methods used in the analysis. Root mean squared predictive error (RMSPE) is computed over the six pre-event months to mirror the six-month post-treatment horizon used in the main analysis, so that forecasting performance is assessed over a period directly comparable to that over which the treatment effects are estimated. As shown in Table S21, no method uniformly dominates across outcomes and cases: TabPFN achieves the lowest RMSPE in nine of the eighteen comparisons, Causal-ARIMA in eight, and TabICLv2 in one. TabPFN performs particularly well for lethal-violence outcomes, whereas Causal-ARIMA consistently performs best for stalking and also for abuse and sexual violence before Cecchettin.

| Outcome | Abbas | Tramontano | Cecchettin |
|---|---|---|---|
| 1522 calls | 1,132 / **681** / 1,470 | 492 / 744 / **417** | **551** / 587 / 581 |
| Abuses against women | **154.7** / 166.9 / 209.3 | **93.2** / 119.1 / 100.2 | 109.7 / **90.0** / 102.4 |
| Stalking | 164.7 / **131.5** / 199.5 | 116.4 / **88.0** / 103.7 | 103.4 / **81.4** / 94.9 |
| Sexual violence | **51.7** / 75.7 / 71.3 | 96.9 / **66.4** / 95.0 | 42.5 / **25.7** / 43.1 |
| All femicides | **2.64** / 3.55 / 3.25 | **3.88** / 4.08 / 4.46 | **3.59** / 4.37 / 3.72 |
| Jealousy/possessiveness IPFs | **2.68** / 3.98 / 3.04 | 2.10 / **1.75** / 2.27 | **1.35** / 1.72 / 1.74 |

**Table S21 | Pre-event forecasting performance: six-month OOS RMSPE**. Entries report TabPFN / Causal-ARIMA / TabICLv2, respectively. RMSPE is the square root of the mean squared one-step-ahead forecast error over the six months immediately preceding each event: November 2020–April 2021, November 2022–April 2023, and May–October 2023. Values are expressed in the original outcome units; bold identifies the lowest RMSPE within each case–outcome comparison.

These diagnostics do not alter the key substantive conclusions, but they help assess their sensitivity to counterfactual modeling choices. For Cecchettin, the estimated increases in calls, abuse, and stalking remain positive and substantial across methods, including those with the lowest pre-event RMSPE for each outcome. For Tramontano, the increase in calls also persists under TabICLv2, the best-performing method for that series. By contrast, Causal-ARIMA's lower pre-event RMSPE for calls in the Abbas case, combined with an estimated response substantially closer to zero than the TabPFN estimates and the lack of robustness to Covid-19 imputed data, casts further doubt on the baseline increase in help-seeking for that case.

## SI section F: Additional results

This appendix presents additional estimates for the three main cases: Abbas, Tramontano and Cecchettin. The analyses distinguish between caller types, examine geographical heterogeneity, and assess effects on additional outcomes and alternative measures of femicide.

1. 1522 calls by caller type. Table S22 reports the estimated effects on calls to the 1522 helpline, distinguishing between calls made by victims and those made by family members and friends. This disaggregation helps assess whether increased public salience affects help-seeking by victims, engagement by their social networks, or both.

2. Estimates by geographical area. Tables S23–S28 report separate estimates for Northern, Central, and Southern Italy, with the islands included in the South. These analyses assess whether the national estimates mask geographical heterogeneity in responses to the three cases. Moreover, because all three cases occurred in Northern Italy, these analyses also examine whether the estimated responses extend beyond the macro-area in which the femicide occurred.

3. Additional outcomes and alternative femicide measures. Table S29 presents estimates based on two alternative versions of our femicide dataset. The first restricts coverage to the two sources available throughout 2017–2024, thereby reducing potential inconsistencies arising from changes in source coverage. The second counts each incident involving multiple femicide victims only once, rather than counting each victim separately. We also report estimates using Italian police data on homicides of women and on homicides of women committed by an intimate partner. These specifications assess whether the findings are sensitive to source coverage or to measuring femicide in terms of incidents rather than victims. Finally, we use homicides of men as a placebo outcome.

Table S22 suggests different patterns of helpline engagement across the three cases. Following Tramontano, increases in calls from family members and friends are more sustained than those from victims, with six-month cumulative point estimates of 32.5% and 18.2%, respectively. Following Cecchettin, both groups display substantial increases, with corresponding estimates of 48.2% and 46.5%. The Abbas estimates are less conclusive, with all monthly effects remaining within the corresponding 95% predictive ranges. Overall, these findings are consistent with heightened public salience around salient IPFs encouraging both direct help-seeking and engagement by victims' social networks.

a) The case of Saman Abbas

| | May 21 | Jun 21 | Jul 21 | Aug 21 | Sep 21 | Oct 21 | Cum. Effects (%) |
|---|---|---|---|---|---|---|---|
| 1522 calls | 522<br>(-1808; 1844) | 663<br>(-1315; 1910) | 998<br>(-455; 1826) | 1143<br>(-305; 1879) | 1213<br>(-84; 1915) | 489<br>(-1071; 1314) | 3 mths: 21.27<br>6 mths: 26.82 |
| 1522 calls from victims | 70<br>(-791; 605) | -22<br>(-983; 602) | -94<br>(-880; 405) | -187<br>(-1134; 376) | -149<br>(-1025; 384) | -370<br>(-1159; 117) | 3 mths: -1.08<br>6 mths: -9.11 |
| 1522 calls from family and friends | 333<br>(-1367; 1221) | 325<br>(-1234; 1214) | 472<br>(-932; 1247) | 564<br>(-795; 1219) | 532<br>(-703; 1174) | 166<br>(-1121; 782) | 3 mths: 15.65<br>6 mths: 17.23 |

b) The case of Giulia Tramontano

| | May 23 | Jun 23 | Jul 23 | Aug 23 | Sep 23 | Oct 23 | Cum. Effects (%) |
|---|---|---|---|---|---|---|---|
| 1522 calls | 357<br>(-254; 895) | **1056**<br>(491; 1513) | **940**<br>(341; 1364) | **802**<br>(315; 1213) | **1867**<br>(1421; 2252) | **1474**<br>(1010; 1879) | 3 mths: 20.87<br>6 mths: 29.48 |
| 1522 calls from victims | 43<br>(-349; 258) | 204<br>(-118; 400) | 23<br>(-429; 247) | 27<br>(-328; 231) | **418**<br>(93; 607) | **349**<br>(69; 528) | 3 mths: 9.09<br>6 mths: 18.23 |
| 1522 calls from family and friends | 278<br>(-262; 622) | **840**<br>(340; 1152) | **896**<br>(402; 1181) | **745**<br>(299; 1029) | **1433**<br>(1044; 1685) | **1108**<br>(714; 1367) | 3 mths: 24.07<br>6 mths: 32.46 |

c) The case of Giulia Cecchettin

| | Nov 23 | Dec 23 | Jan 24 | Feb 24 | Mar 24 | Apr 24 | Cum. Effects (%) |
|---|---|---|---|---|---|---|---|
| 1522 calls | 3384<br>(-203; 5126) | **6659**<br>(3985; 7969) | **3118**<br>(1713; 4255) | **1741**<br>(245; 2774) | 1311<br>(-978; 2408) | 264<br>(-1750; 1439) | 3 mths: 74.82<br>6 mths: 47.54 |
| 1522 calls from victims | **709**<br>(104; 1050) | **1327**<br>(690; 1650) | **756**<br>(224; 1052) | 360<br>(-75; 636) | 290<br>(-265; 571) | 68<br>(-435; 343) | 3 mths: 73.59<br>6 mths: 46.54 |
| 1522 calls from family and friends | 2505<br>(-428; 3978) | **5305**<br>(3399; 6157) | **2434**<br>(1310; 3260) | **1456**<br>(33; 2265) | 1110<br>(-774; 1916) | 233<br>(-1713; 1101) | 3 mths: 73.59<br>6 mths: 48.23 |

**Table S22 | 1522 calls by caller type.** Entries report monthly observed-minus-counterfactual counts, with model-specific 95% prediction intervals below. Cumulative percentage effects equal 100 times the sum of monthly effects divided by the sum of counterfactual counts over the indicated horizon. Bold indicates monthly effects whose 95% prediction intervals lie entirely above or below zero.

The subnational estimates indicate that the response to the Cecchettin case extended beyond Northern Italy. Helpline use increased across all three macro-areas, accompanied by broadly positive estimates for reported abuse and, less consistently, stalking. Following Tramontano, the clearest pattern is an increase in helpline use in Northern and Central Italy, whereas the Abbas estimates are more mixed. Estimates for sexual violence and femicides are less consistent across areas. Percentage changes in the low-count femicide series should therefore be interpreted cautiously.

a) The case of Saman Abbas

| | May 21 | Jun 21 | Jul 21 | Aug 21 | Sep 21 | Oct 21 | Cum. Effects (%) |
|---|---|---|---|---|---|---|---|
| Italy | 522<br>(-1808; 1844) | 663<br>(-1315; 1910) | 998<br>(-455; 1826) | 1143<br>(-305; 1879) | 1213<br>(-84; 1915) | 489<br>(-1071; 1314) | 3 mths: 21.27<br>6 mths: 26.82 |
| Northern regions | 186<br>(-437; 606) | 138<br>(-561; 560) | 283<br>(-358; 616) | 45<br>(-690; 353) | 179<br>(-515; 517) | 135<br>(-438; 414) | 3 mths: 17.82<br>6 mths: 14.62 |
| Central regions | 97<br>(-970; 527) | 268<br>(-424; 572) | 241<br>(-12; 417) | **292**<br>(136; 417) | 195<br>(-54; 358) | 88<br>(-116; 222) | 3 mths: 33.39<br>6 mths: 36.43 |
| Southern regions | 215<br>(-155; 511) | 294<br>(-168; 602) | 273<br>(-61; 518) | 315<br>(-13; 546) | 231<br>(-208; 485) | -67<br>(-396; 162) | 3 mths: 35.72<br>6 mths: 28.66 |

b) The case of Giulia Tramontano

| | May 23 | Jun 23 | Jul 23 | Aug 23 | Sep 23 | Oct 23 | Cum. Effects (%) |
|---|---|---|---|---|---|---|---|
| Italy | 357<br>(-254; 895) | **1056**<br>(491; 1513) | **940**<br>(341; 1364) | **802**<br>(315; 1213) | **1867**<br>(1421; 2252) | **1474**<br>(1010; 1879) | 3 mths: 20.87<br>6 mths: 29.48 |
| Northern regions | 279<br>(-151; 857) | **448**<br>(148; 642) | 375<br>(-10; 582) | **378**<br>(128; 527) | **592**<br>(351; 753) | **636**<br>(396; 799) | 3 mths: 31.31<br>6 mths: 39.89 |
| Central regions | 46<br>(-197; 299) | **306**<br>(94; 486) | **255**<br>(54; 411) | 135<br>(-18; 254) | **385**<br>(214; 512) | **342**<br>(174; 478) | 3 mths: 31.08<br>6 mths: 37.85 |
| Southern regions | 73<br>(-229; 316) | 178<br>(-115; 385) | 190<br>(-247; 427) | 134<br>(-89; 319) | 413<br>(-62; 713) | 165<br>(-70; 348) | 3 mths: 20.43<br>6 mths: 26.52 |

c) The case of Giulia Cecchettin

| | Nov 23 | Dec 23 | Jan 24 | Feb 24 | Mar 24 | Apr 24 | Cum. Effects (%) |
|---|---|---|---|---|---|---|---|
| Italy | 3384<br>(-203; 5126) | **6659**<br>(3985; 7969) | **3118**<br>(1713; 4255) | **1741**<br>(245; 2774) | 1311<br>(-978; 2408) | 264<br>(-1750; 1439) | 3 mths: 74.82<br>6 mths: 47.54 |
| Northern regions | 828<br>(-44; 1423) | **1886**<br>(1133; 2574) | **1558**<br>(430; 2797) | **1103**<br>(530; 1651) | **816**<br>(214; 1271) | 385<br>(-125; 681) | 3 mths: 76.77<br>6 mths: 64.51 |
| Central regions | 461<br>(-93; 843) | **900**<br>(401; 1263) | **610**<br>(73; 1181) | 235<br>(-302; 586) | 260<br>(-231; 589) | 60<br>(-234; 317) | 3 mths: 66.32<br>6 mths: 43.13 |
| Southern regions | **689**<br>(68; 1038) | **1233**<br>(845; 1511) | **824**<br>(428; 1377) | **582**<br>(248; 827) | **473**<br>(125; 713) | **301**<br>(30; 531) | 3 mths: 93.93<br>6 mths: 75.54 |

**Table S23 | 1522 calls – Subnational estimates.** Entries report monthly observed-minus-counterfactual counts, with model-specific 95% prediction intervals below. Cumulative percentage effects equal 100 times the sum of monthly effects divided by the sum of counterfactual counts over the indicated horizon. Bold indicates monthly effects whose 95% prediction intervals lie entirely above or below zero.

## a) The case of Saman Abbas

| | May 21 | Jun 21 | Jul 21 | Aug 21 | Sep 21 | Oct 21 | Cum. Effects (%) |
|---|---|---|---|---|---|---|---|
| Italy | 91<br>(-287; 383) | 236<br>(-106; 506) | 198<br>(-137; 466) | 91<br>(-241; 358) | 26<br>(-321; 322) | -170<br>(-503; 117) | 3 mths: 8.75<br>6 mths: 3.95 |
| Northern regions | 31<br>(-112; 162) | 95<br>(-41; 218) | 93<br>(-42; 216) | -29<br>(-161; 91) | -22<br>(-159; 106) | -79<br>(-213; 45) | 3 mths: 8.86<br>6 mths: 1.83 |
| Central regions | 63<br>(-8; 125) | 68<br>(-9; 130) | 7<br>(-63; 66) | 52<br>(-20; 111) | 33<br>(-40; 93) | -24<br>(-98; 34) | 3 mths: 11.95<br>6 mths: 8.74 |
| Southern regions | -16<br>(-216; 157) | 57<br>(-131; 211) | 67<br>(-125; 230) | 18<br>(-167; 175) | 10<br>(-170; 158) | -40<br>(-212; 103) | 3 mths: 4.49<br>6 mths: 1.97 |

## b) The case of Giulia Tramontano

| | May 23 | Jun 23 | Jul 23 | Aug 23 | Sep 23 | Oct 23 | Cum. Effects (%) |
|---|---|---|---|---|---|---|---|
| Italy | -164<br>(-467; 117) | 0<br>(-276; 255) | 15<br>(-264; 277) | -90<br>(-356; 152) | 27<br>(-240; 280) | 67<br>(-192; 298) | 3 mths: -2.25<br>6 mths: -1.10 |
| Northern regions | -102<br>(-248; 35) | -24<br>(-161; 106) | 16<br>(-127; 156) | -35<br>(-172; 91) | -2<br>(-133; 117) | 11<br>(-113; 123) | 3 mths: -4.06<br>6 mths: -2.54 |
| Central regions | -56<br>(-114; 0) | 26<br>(-31; 78) | -4<br>(-67; 53) | -10<br>(-71; 44) | 23<br>(-41; 75) | 33<br>(-29; 90) | 3 mths: -2.66<br>6 mths: 0.49 |
| Southern regions | 5<br>(-145; 139) | 6<br>(-143; 140) | -3<br>(-146; 126) | -97<br>(-250; 35) | -7<br>(-134; 105) | -14<br>(-148; 101) | 3 mths: 0.33<br>6 mths: -2.05 |

## c) The case of Giulia Cecchettin

| | Nov 23 | Dec 23 | Jan 24 | Feb 24 | Mar 24 | Apr 24 | Cum. Effects (%) |
|---|---|---|---|---|---|---|---|
| Italy | 125<br>(-114; 343) | **513**<br>(271; 735) | **554**<br>(292; 802) | **556**<br>(331; 771) | **572**<br>(290; 828) | **390**<br>(134; 626) | 3 mths: 19.28<br>6 mths: 22.84 |
| Northern regions | 65<br>(-60; 177) | **210**<br>(89; 321) | **239**<br>(121; 361) | **171**<br>(66; 277) | **217**<br>(93; 336) | **158**<br>(35; 272) | 3 mths: 19.57<br>6 mths: 20.85 |
| Central regions | 7<br>(-55; 65) | **82**<br>(19; 141) | **122**<br>(58; 188) | **105**<br>(43; 166) | **105**<br>(39; 166) | **70**<br>(8; 131) | 3 mths: 16.60<br>6 mths: 20.31 |
| Southern regions | 15<br>(-97; 118) | **171**<br>(54; 276) | **149**<br>(16; 270) | **244**<br>(148; 335) | **188**<br>(52; 308) | 117<br>(-10; 223) | 3 mths: 13.81<br>6 mths: 19.10 |

**Table S24 | Abuses against women – Subnational estimates.** See Notes of Table S23.

### a) The case of Saman Abbas

| | May 21 | Jun 21 | Jul 21 | Aug 21 | Sep 21 | Oct 21 | Cum. Effects (%) |
|---|---|---|---|---|---|---|---|
| Italy | 64<br>(-398; 410) | 41<br>(-287; 305) | 82<br>(-188; 295) | -16<br>(-272; 177) | -106<br>(-429; 165) | **-233**<br>(-515; -25) | 3 mths: 3.64<br>6 mths: -1.66 |
| Northern regions | 21<br>(-148; 137) | 98<br>(-2; 183) | 59<br>(-45; 140) | 25<br>(-63; 85) | -33<br>(-195; 106) | -53<br>(-147; 17) | 3 mths: 10.52<br>6 mths: 3.47 |
| Central regions | 45<br>(-20; 100) | 46<br>(-5; 96) | **39**<br>(1; 81) | 17<br>(-29; 60) | 44<br>(-2; 83) | -8<br>(-57; 38) | 3 mths: 13.88<br>6 mths: 10.02 |
| Southern regions | 12<br>(-238; 177) | -92<br>(-304; 46) | -41<br>(-241; 92) | -94<br>(-251; 15) | **-115**<br>(-276; -3) | **-189**<br>(-339; -82) | 3 mths: -4.87<br>6 mths: -10.36 |

### b) The case of Giulia Tramontano

| | May 23 | Jun 23 | Jul 23 | Aug 23 | Sep 23 | Oct 23 | Cum. Effects (%) |
|---|---|---|---|---|---|---|---|
| Italy | -9<br>(-252; 221) | -90<br>(-300; 112) | 138<br>(-82; 335) | 30<br>(-148; 188) | 140<br>(-47; 320) | 85<br>(-118; 274) | 3 mths: 0.74<br>6 mths: 2.92 |
| Northern regions | -31<br>(-129; 56) | -35<br>(-117; 45) | 25<br>(-63; 96) | 6<br>(-64; 62) | 74<br>(-9; 150) | -10<br>(-108; 84) | 3 mths: -2.20<br>6 mths: 0.80 |
| Central regions | -2<br>(-62; 64) | 15<br>(-42; 70) | 45<br>(-13; 102) | 5<br>(-44; 53) | 31<br>(-17; 77) | 34<br>(-15; 80) | 3 mths: 5.97<br>6 mths: 6.71 |
| Southern regions | -22<br>(-184; 114) | -90<br>(-225; 18) | 53<br>(-65; 148) | -17<br>(-118; 62) | -2<br>(-105; 79) | 71<br>(-28; 144) | 3 mths: -2.43<br>6 mths: -0.14 |

### c) The case of Giulia Cecchettin

| | Nov 23 | Dec 23 | Jan 24 | Feb 24 | Mar 24 | Apr 24 | Cum. Effects (%) |
|---|---|---|---|---|---|---|---|
| Italy | -4<br>(-231; 170) | 111<br>(-61; 269) | **393**<br>(197; 582) | **316**<br>(144; 467) | 137<br>(-84; 354) | **242**<br>(60; 398) | 3 mths: 10.58<br>6 mths: 12.76 |
| Northern regions | -7<br>(-109; 65) | 67<br>(-9; 129) | **198**<br>(100; 290) | **70**<br>(2; 129) | 68<br>(-22; 155) | 58<br>(-14; 120) | 3 mths: 15.77<br>6 mths: 13.69 |
| Central regions | 15<br>(-39; 65) | 39<br>(-20; 94) | **95**<br>(37; 153) | **98**<br>(35; 153) | **61**<br>(0; 122) | **81**<br>(22; 135) | 3 mths: 16.01<br>6 mths: 21.45 |
| Southern regions | 1<br>(-85; 66) | -9<br>(-108; 71) | 108<br>(-12; 209) | **142**<br>(48; 222) | 19<br>(-97; 111) | 90<br>(-5; 171) | 3 mths: 4.67<br>6 mths: 8.32 |

**Table S25 | Stalking crimes – Subnational estimates.** See Notes of Table S23.

## a) The case of Saman Abbas

| | May 21 | Jun 21 | Jul 21 | Aug 21 | Sep 21 | Oct 21 | Cum. Effects (%) |
|---|---|---|---|---|---|---|---|
| Italy | 76<br>(-58; 212) | 85<br>(-39; 216) | 24<br>(-99; 156) | 37<br>(-79; 155) | -27<br>(-151; 95) | -33<br>(-156; 90) | 3 mths: 12.83<br>6 mths: 5.64 |
| Northern regions | 46<br>(-22; 123) | 57<br>(-9; 132) | 50<br>(-26; 135) | 61<br>(-8; 143) | 15<br>(-52; 93) | 14<br>(-51; 90) | 3 mths: 20.51<br>6 mths: 17.18 |
| Central regions | **38**<br>(6; 71) | 24<br>(-8; 57) | 11<br>(-18; 38) | 4<br>(-24; 35) | 14<br>(-16; 44) | -9<br>(-38; 19) | 3 mths: 29.51<br>6 mths: 16.12 |
| Southern regions | 0<br>(-52; 38) | 34<br>(-17; 71) | -9<br>(-73; 44) | 3<br>(-42; 37) | -23<br>(-60; 3) | -14<br>(-47; 11) | 3 mths: 6.77<br>6 mths: -1.02 |

## b) The case of Giulia Tramontano

| | May 23 | Jun 23 | Jul 23 | Aug 23 | Sep 23 | Oct 23 | Cum. Effects (%) |
|---|---|---|---|---|---|---|---|
| Italy | -34<br>(-197; 124) | 21<br>(-120; 154) | 77<br>(-67; 213) | 11<br>(-119; 131) | -4<br>(-134; 116) | 14<br>(-122; 143) | 3 mths: 3.75<br>6 mths: 2.56 |
| Northern regions | -20<br>(-128; 84) | 5<br>(-90; 96) | 62<br>(-32; 151) | 2<br>(-78; 78) | 5<br>(-76; 83) | 3<br>(-81; 83) | 3 mths: 4.98<br>6 mths: 3.09 |
| Central regions | 23<br>(-11; 57) | **46**<br>(15; 78) | **34**<br>(3; 65) | 26<br>(-5; 57) | 7<br>(-23; 38) | -7<br>(-42; 27) | 3 mths: 35.81<br>6 mths: 21.93 |
| Southern regions | -25<br>(-71; 19) | -28<br>(-71; 11) | -16<br>(-60; 22) | -21<br>(-62; 17) | -20<br>(-56; 14) | 20<br>(-19; 55) | 3 mths: -15.59<br>6 mths: -10.25 |

## c) The case of Giulia Cecchettin

| | Nov 23 | Dec 23 | Jan 24 | Feb 24 | Mar 24 | Apr 24 | Cum. Effects (%) |
|---|---|---|---|---|---|---|---|
| Italy | -3<br>(-125; 112) | -15<br>(-130; 92) | 35<br>(-93; 163) | 30<br>(-92; 144) | 70<br>(-58; 201) | 99<br>(-18; 211) | 3 mths: 1.17<br>6 mths: 7.56 |
| Northern regions | 1<br>(-77; 77) | -24<br>(-100; 50) | -6<br>(-92; 81) | 9<br>(-71; 88) | 28<br>(-56; 115) | 52<br>(-28; 131) | 3 mths: -3.52<br>6 mths: 3.59 |
| Central regions | 9<br>(-24; 41) | 8<br>(-24; 40) | **35**<br>(2; 68) | 11<br>(-21; 42) | 24<br>(-9; 56) | 13<br>(-20; 44) | 3 mths: 18.16<br>6 mths: 17.54 |
| Southern regions | -20<br>(-54; 12) | -1<br>(-36; 32) | -2<br>(-42; 36) | 3<br>(-31; 35) | 13<br>(-26; 50) | 17<br>(-19; 49) | 3 mths: -6.62<br>6 mths: 1.61 |

**Table S26 | Sexual violence crimes – Subnational estimates.** See Notes of Table S23.

## a) The case of Saman Abbas

| | May 21 | Jun 21 | Jul 21 | Aug 21 | Sep 21 | Oct 21 | Cum. Effects (%) |
|---|---|---|---|---|---|---|---|
| Italy | 3.01<br>(-2.72; 8.61) | 1.34<br>(-4.79; 7.02) | 2.78<br>(-3.25; 7.95) | 0.15<br>(-5.98; 5.72) | 5.89<br>(-0.07; 11.24) | 2.87<br>(-3.10; 8.17) | 3 mths: 31.22<br>6 mths: 34.16 |
| Northern regions | -0.67<br>(-5.28; 3.00) | 3.37<br>(-1.80; 7.98) | -2.60<br>(-7.28; 1.83) | 1.17<br>(-3.42; 5.00) | 1.23<br>(-3.84; 5.93) | 3.29<br>(-1.23; 7.00) | 3 mths: 0.86<br>6 mths: 22.13 |
| Central regions | **2.89**<br>(0.79; 4.00) | 0.82<br>(-1.91; 2.00) | 0.99<br>(-1.24; 2.93) | 0.86<br>(-1.13; 2.00) | 0.74<br>(-2.04; 2.00) | -0.56<br>(-3.27; 1.00) | 3 mths: 109.23<br>6 mths: 69.57 |
| Southern regions | 0.71<br>(-2.83; 3.00) | -2.06<br>(-5.60; 0.00) | 2.73<br>(-0.88; 5.00) | -0.71<br>(-4.31; 1.00) | 2.55<br>(-1.15; 5.00) | 1.11<br>(-2.40; 3.00) | 3 mths: 20.95<br>6 mths: 34.20 |

## b) The case of Giulia Tramontano

| | May 23 | Jun 23 | Jul 23 | Aug 23 | Sep 23 | Oct 23 | Cum. Effects (%) |
|---|---|---|---|---|---|---|---|
| Italy | -5.26<br>(-11.77; 1.56) | 4.73<br>(-1.71; 10.49) | -2.59<br>(-8.77; 3.62) | -1.11<br>(-7.50; 4.62) | 3.67<br>(-2.43; 9.83) | 2.98<br>(-3.03; 8.84) | 3 mths: -11.12<br>6 mths: 4.49 |
| Northern regions | -2.45<br>(-7.59; 2.39) | 0.49<br>(-4.53; 5.22) | -1.52<br>(-6.25; 2.77) | -0.87<br>(-5.66; 3.50) | 0.79<br>(-3.83; 5.00) | 2.16<br>(-2.58; 6.54) | 3 mths: -22.51<br>6 mths: -4.80 |
| Central regions | -0.98<br>(-3.14; 1.00) | 2.37<br>(-0.23; 4.00) | -0.86<br>(-3.60; 1.00) | -1.60<br>(-4.51; 0.00) | 1.29<br>(-1.53; 3.00) | 0.26<br>(-2.62; 2.00) | 3 mths: 9.84<br>6 mths: 4.60 |
| Southern regions | -1.93<br>(-5.93; 1.00) | 0.56<br>(-3.10; 3.00) | -0.64<br>(-4.47; 2.96) | -0.02<br>(-3.66; 2.00) | 2.32<br>(-1.16; 5.00) | 0.23<br>(-3.31; 3.00) | 3 mths: -22.35<br>6 mths: 3.15 |

## c) The case of Giulia Cecchettin

| | Nov 23 | Dec 23 | Jan 24 | Feb 24 | Mar 24 | Apr 24 | Cum. Effects (%) |
|---|---|---|---|---|---|---|---|
| Italy | -2.52<br>(-8.27; 2.95) | -2.87<br>(-9.03; 2.86) | -2.92<br>(-8.88; 3.02) | 1.50<br>(-4.46; 6.94) | -1.30<br>(-7.28; 4.38) | **-5.78**<br>(-11.66; -0.01) | 3 mths: -32.84<br>6 mths: -27.29 |
| Northern regions | -1.96<br>(-6.04; 1.77) | -0.82<br>(-5.35; 2.00) | -1.93<br>(-6.72; 2.44) | -1.07<br>(-5.43; 3.00) | -2.70<br>(-7.20; 1.53) | -1.90<br>(-6.27; 2.00) | 3 mths: -40.23<br>6 mths: -42.57 |
| Central regions | -0.60<br>(-3.40; 1.00) | -0.76<br>(-3.59; 1.00) | -0.67<br>(-3.63; 1.00) | 1.41<br>(-1.57; 3.00) | 2.24<br>(-0.58; 4.00) | -0.70<br>(-3.66; 1.00) | 3 mths: -40.34<br>6 mths: 9.09 |
| Southern regions | -0.46<br>(-3.85; 2.00) | -0.97<br>(-4.25; 2.00) | 0.34<br>(-3.24; 3.00) | 1.35<br>(-2.10; 4.00) | -2.10<br>(-5.43; 1.00) | -1.91<br>(-5.47; 0.00) | 3 mths: -13.47<br>6 mths: -23.86 |

**Table S27 | All femicides – Subnational estimates.** See Notes of Table S23.

## a) The case of Saman Abbas

| | May 21 | Jun 21 | Jul 21 | Aug 21 | Sep 21 | Oct 21 | Cum. Effects (%) |
|---|---|---|---|---|---|---|---|
| Italy | 0.09 (-4.13; 3.61) | -2.08 (-6.28; 1.99) | -0.50 (-4.36; 3.00) | 1.31 (-2.90; 4.00) | 1.28 (-2.81; 4.81) | -2.74 (-6.90; 0.90) | 3 mths: -19.95<br>6 mths: -11.64 |
| Northern regions | -0.42 (-4.09; 2.00) | 1.17 (-2.68; 3.00) | -1.97 (-5.50; 0.00) | -0.01 (-3.60; 2.00) | -0.06 (-3.85; 2.00) | -0.90 (-4.59; 1.00) | 3 mths: -19.72<br>6 mths: -17.99 |
| Central regions | 0.86 (-0.29; 2.00) | **-0.67** (-2.21; -0.07) | 1.33 (-0.70; 2.00) | 0.42 (-1.63; 1.00) | 0.39 (-1.63; 1.00) | -0.62 (-2.64; 0.00) | 3 mths: 61.32<br>6 mths: 39.86 |
| Southern regions | -1.89 (-4.94; 0.00) | -0.88 (-3.30; 0.00) | -0.17 (-2.91; 1.00) | 0.19 (-2.32; 1.00) | 0.69 (-2.35; 2.00) | -0.93 (-3.48; 0.00) | 3 mths: -74.62<br>6 mths: -42.86 |

## b) The case of Giulia Tramontano

| | May 23 | Jun 23 | Jul 23 | Aug 23 | Sep 23 | Oct 23 | Cum. Effects (%) |
|---|---|---|---|---|---|---|---|
| Italy | -2.26 (-5.83; 1.50) | 0.38 (-3.26; 3.82) | -0.57 (-4.08; 2.78) | -0.67 (-4.33; 2.68) | 2.27 (-1.23; 5.59) | 0.31 (-3.15; 3.52) | 3 mths: -21.39<br>6 mths: -2.34 |
| Northern regions | -0.45 (-3.67; 2.00) | 0.14 (-3.46; 2.00) | -0.25 (-3.64; 1.00) | 0.30 (-3.18; 2.00) | -0.72 (-4.11; 1.00) | 0.38 (-2.98; 2.00) | 3 mths: -9.98<br>6 mths: -5.63 |
| Central regions | -0.66 (-2.04; 0.00) | 0.38 (-1.60; 1.00) | -0.65 (-2.59; 0.00) | -0.59 (-2.60; 0.00) | 1.35 (-0.62; 2.00) | 0.35 (-1.67; 1.00) | 3 mths: -48.02<br>6 mths: 5.06 |
| Southern regions | -1.19 (-3.84; 0.00) | -0.07 (-2.34; 1.00) | 0.70 (-1.30; 2.00) | 0.03 (-2.46; 1.00) | 1.84 (-0.20; 3.00) | -0.18 (-2.31; 1.00) | 3 mths: -15.80<br>6 mths: 16.32 |

## c) The case of Giulia Cecchettin

| | Nov 23 | Dec 23 | Jan 24 | Feb 24 | Mar 24 | Apr 24 | Cum. Effects (%) |
|---|---|---|---|---|---|---|---|
| Italy | -0.16 (-3.59; 2.98) | -2.54 (-6.22; 0.71) | -0.90 (-4.30; 2.38) | -1.28 (-4.74; 1.87) | -0.48 (-3.88; 2.80) | -1.77 (-5.19; 1.57) | 3 mths: -33.97<br>6 mths: -33.74 |
| Northern regions | 0.37 (-2.90; 2.00) | -0.34 (-3.57; 1.00) | -0.54 (-3.68; 1.00) | -0.60 (-3.76; 1.00) | -0.69 (-4.04; 1.00) | -0.57 (-3.75; 1.00) | 3 mths: -11.23<br>6 mths: -25.22 |
| Central regions | -0.52 (-2.05; 0.00) | -0.68 (-2.55; 0.00) | 0.33 (-1.59; 1.00) | 0.35 (-1.57; 1.00) | 0.35 (-1.62; 1.00) | 0.29 (-1.68; 1.00) | 3 mths: -46.31<br>6 mths: 3.19 |
| Southern regions | -0.15 (-2.12; 1.00) | **-1.19** (-3.31; -0.04) | -0.36 (-2.33; 0.91) | -1.18 (-3.52; 0.00) | -0.78 (-3.33; 0.88) | -0.82 (-3.89; 0.00) | 3 mths: -45.92<br>6 mths: -59.88 |

**Table S28 | IPFs motivated by jealousy or possessiveness – Subnational estimates.** See Notes of Table S23.

Table S29 shows consistently negative six-month point estimates following Cecchettin across alternative femicide measures and police-recorded female homicide outcomes, with estimated reductions of approximately 28–34%. The consistency lies in the direction and magnitude of the point estimates, rather than in uniformly strong inferential support, as most monthly effects remain within the corresponding 95% predictive ranges. Abbas and Tramontano show no comparably consistent pattern. Male homicides show no sustained decline, providing little indication of a comparable response in the placebo outcome.

a) The case of Saman Abbas

| | May 21 | Jun 21 | Jul 21 | Aug 21 | Sep 21 | Oct 21 | Cum. Effects (%) |
|---|---|---|---|---|---|---|---|
| All femicides (alt. def. 1) | 1.02 (-10.11; 9.51) | 0.04 (-9.69; 9.04) | 0.13 (-9.24; 8.00) | 0.82 (-8.71; 8.00) | 5.31 (-3.77; 13.52) | 2.33 (-6.89; 10.85) | 3 mths: 4.44<br>6 mths: 18.79 |
| All femicides (alt. def. 2) | 1.23 (-10.85; 9.55) | 0.48 (-9.60; 9.31) | 0.14 (-9.15; 8.00) | -0.29 (-9.50; 7.00) | 5.53 (-3.35; 13.70) | 2.92 (-6.15; 11.00) | 3 mths: 7.09<br>6 mths: 20.03 |
| IPF jel. poss. femicides (alt. def. 1) | 0.01 (-4.17; 3.51) | -1.99 (-6.19; 2.04) | -0.38 (-4.23; 3.00) | 1.43 (-2.82; 4.00) | 1.03 (-2.98; 4.52) | -2.76 (-6.86; 0.85) | 3 mths: -19.09<br>6 mths: -11.75 |
| IPF jel. poss. femicides (alt. def. 2) | -0.04 (-4.26; 3.55) | -1.73 (-5.97; 2.31) | -0.58 (-4.43; 3.00) | 1.27 (-2.85; 4.00) | 1.11 (-2.93; 4.70) | -2.50 (-6.56; 1.00) | 3 mths: -19.06<br>6 mths: -11.02 |
| Homicides of women | 3.11 (-3.50; 9.47) | -0.86 (-7.76; 5.51) | -0.59 (-7.51; 5.86) | -3.03 (-9.99; 3.75) | 5.25 (-1.31; 11.70) | 2.26 (-4.74; 9.02) | 3 mths: 5.62<br>6 mths: 10.07 |
| Homicides of women committed by IP | 2.73 (-2.06; 7.55) | -1.98 (-6.53; 2.60) | 1.18 (-3.38; 5.66) | 0.03 (-4.38; 4.42) | 1.12 (-3.35; 5.45) | 0.96 (-3.48; 5.35) | 3 mths: 11.29<br>6 mths: 11.54 |
| Homicides of men | 2.81 (-8.94; 13.28) | -2.37 (-14.31; 8.27) | -1.68 (-13.32; 8.67) | -2.72 (-14.24; 7.61) | -4.19 (-16.03; 6.17) | -2.62 (-14.84; 8.01) | 3 mths: -2.70<br>6 mths: -11.62 |

b) The case of Giulia Tramontano

| | May 23 | Jun 23 | Jul 23 | Aug 23 | Sep 23 | Oct 23 | Cum. Effects (%) |
|---|---|---|---|---|---|---|---|
| All femicides (alt. def. 1) | -3.96 (-12.22; 4.51) | 4.07 (-4.37; 12.44) | -1.97 (-10.37; 6.22) | -2.36 (-10.94; 5.75) | 4.09 (-4.19; 12.20) | 3.49 (-4.79; 11.39) | 3 mths: -6.91<br>6 mths: 6.39 |
| All femicides (alt. def. 2) | -3.76 (-11.82; 4.45) | 4.59 (-3.54; 12.51) | -1.21 (-9.34; 6.75) | -1.71 (-10.13; 5.93) | 3.42 (-4.60; 11.28) | 3.83 (-4.12; 11.45) | 3 mths: -1.51<br>6 mths: 10.36 |
| IPF jel. poss. femicides (alt. def. 1) | -2.26 (-5.86; 1.51) | 0.34 (-3.33; 3.81) | -0.60 (-4.15; 2.77) | -0.56 (-4.29; 2.75) | 2.20 (-1.36; 5.59) | 0.34 (-3.19; 3.56) | 3 mths: -21.87<br>6 mths: -2.40 |
| IPF jel. poss. femicides (alt. def. 2) | -2.26 (-5.80; 1.45) | 0.50 (-3.09; 3.85) | -0.51 (-4.02; 2.91) | -0.47 (-4.27; 2.93) | 2.18 (-1.39; 5.63) | 0.44 (-3.09; 3.71) | 3 mths: -20.08<br>6 mths: -0.47 |
| Homicides of women | -0.93 (-7.30; 5.84) | 1.75 (-4.71; 8.34) | 1.06 (-5.33; 7.49) | -1.96 (-8.69; 4.97) | 4.27 (-2.31; 11.06) | -0.42 (-6.80; 6.14) | 3 mths: 5.85<br>6 mths: 5.94 |
| Homicides of women committed by IP | -2.62 (-6.89; 1.93) | 0.05 (-4.25; 4.52) | 0.88 (-3.57; 5.57) | -2.25 (-6.44; 2.19) | 3.07 (-1.14; 7.44) | 0.13 (-4.02; 4.41) | 3 mths: -10.13<br>6 mths: -2.12 |
| Homicides of men | 1.76 (-8.70; 11.35) | 3.86 (-6.75; 13.81) | 5.19 (-5.41; 14.68) | 4.20 (-6.08; 13.66) | 4.43 (-5.80; 13.83) | -2.77 (-13.26; 6.77) | 3 mths: 22.00<br>6 mths: 17.31 |

## c) The case of Giulia Cecchettin

| | Nov 23 | Dec 23 | Jan 24 | Feb 24 | Mar 24 | Apr 24 | Cum. Effects (%) |
|---|---|---|---|---|---|---|---|
| All femicides (alt. def. 1) | -3.19 (-11.44; 4.56) | -3.07 (-11.60; 4.76) | -2.35 (-10.39; 5.77) | 1.45 (-6.82; 9.35) | -1.59 (-9.91; 6.51) | -5.80 (-14.00; 2.26) | 3 mths: -33.61<br>6 mths: -28.23 |
| All femicides (alt. def. 2) | -2.97 (-11.07; 4.76) | -2.79 (-10.92; 4.98) | -3.19 (-11.00; 4.68) | 0.67 (-7.24; 8.25) | -1.29 (-9.26; 6.54) | -5.38 (-13.40; 2.48) | 3 mths: -35.86<br>6 mths: -29.93 |
| IPF jel. poss. femicides (alt. def. 1) | -0.08 (-3.50; 3.00) | -2.50 (-6.18; 0.76) | -0.79 (-4.20; 2.48) | -1.27 (-4.76; 1.90) | -0.51 (-3.93; 2.78) | -1.59 (-5.03; 1.69) | 3 mths: -32.57<br>6 mths: -32.53 |
| IPF jel. poss. femicides (alt. def. 2) | -0.18 (-3.59; 3.00) | -2.32 (-5.98; 0.98) | -0.89 (-4.35; 2.48) | -1.12 (-4.59; 2.00) | -0.45 (-3.91; 2.88) | -1.72 (-5.24; 1.66) | 3 mths: -32.62<br>6 mths: -32.30 |
| Homicides of women | -3.79 (-9.83; 2.30) | -3.97 (-10.28; 2.39) | -2.24 (-8.35; 4.08) | 0.99 (-5.11; 7.29) | -5.31 (-11.39; 0.84) | **-7.07** (-13.40; -0.74) | 3 mths: -31.26<br>6 mths: -33.75 |
| Homicides of women committed by IP | -1.35 (-5.50; 2.88) | -2.90 (-7.11; 1.38) | -1.63 (-5.76; 2.66) | -1.62 (-5.76; 2.61) | -0.58 (-4.76; 3.70) | -2.99 (-7.22; 1.35) | 3 mths: -34.81<br>6 mths: -32.47 |
| Homicides of men | 1.33 (-8.55; 10.64) | 0.31 (-9.84; 9.67) | 3.53 (-6.53; 12.77) | **10.05** (0.17; 19.16) | -3.74 (-13.94; 5.57) | -4.36 (-14.44; 5.02) | 3 mths: 11.03<br>6 mths: 7.76 |

**Table S29 | Other lethal outcomes.** See Notes of Table S23.

## SI section G: Estimates on the additional cases

This section presents the estimated effects for each of the additional cases of lethal violence against women considered in the analysis: Nicoletta Indelicato (Fig. S3), Alessandra Matteuzzi (Fig. S4), and the homicide of Sharon Verzeni (Fig. S5). For each case, we report estimates for calls to the 1522 helpline, reported abuse against women, stalking, sexual violence, all femicides, and IPFs driven by jealousy or possessiveness.

Concerning Nicoletta Indelicato, the estimates do not reveal any sustained or consistent change in help-seeking or reported violence following the event. Observed helpline calls and most crime outcomes generally remain within the counterfactual prediction bands, with deviations varying across months and outcomes. Total femicides exceed the counterfactual point forecast throughout the post-event window, although uncertainty is substantial and most observations remain within the prediction bands. Jealousy- or possessiveness-related IPFs show no clear or persistent departure from the counterfactual. Overall, these results provide limited evidence of any systematic response.

The estimates for the Alessandra Matteuzzi case similarly show no pronounced or sustained divergence across outcomes. Helpline calls increase towards the end of the post-event period but remain within the prediction bands. Reported abuse is generally below the counterfactual, whereas stalking and sexual violence display mixed deviations, with no clear departures from the predictive intervals. Both femicide outcomes fluctuate around their counterfactual paths, providing little evidence of a sustained reduction.

Turning to the homicide of Sharon Verzeni, there is no evidence of increased help-seeking or reporting following the event. Several outcomes instead fall below their counterfactual forecasts, with reported abuse and stalking dropping below the prediction bands towards the end of the observation window. These negative deviations should, however, be interpreted cautiously. The estimation period includes the exceptional increase in help-seeking and reporting surrounding the Cecchettin case, and the helpline counterfactual projects a further substantial increase in late 2024. The negative estimates may therefore partly reflect the unwinding of that earlier response or the extrapolation of unusually elevated levels, rather than an independent effect of the Verzeni case. Overall, if anything, help-seeking declined following the Verzeni case, with no evidence of an increase attributable to the event.

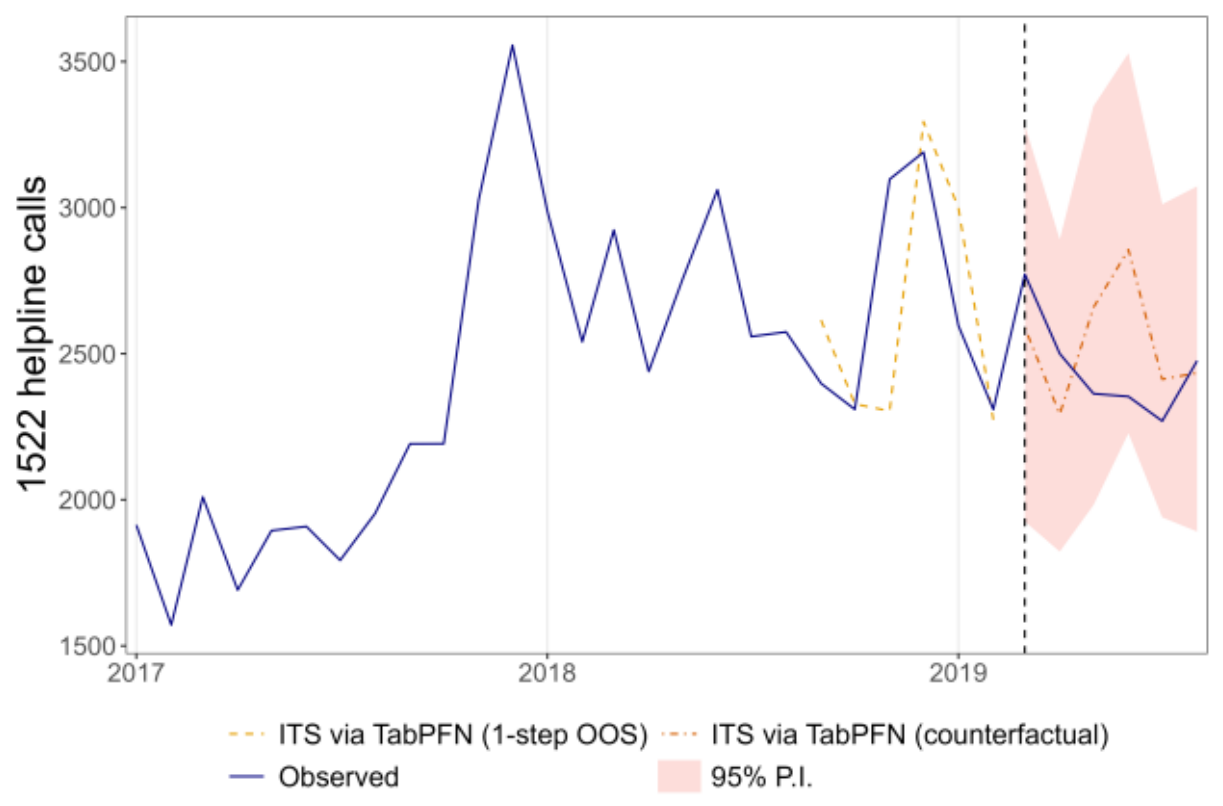

a) Calls to the helpline

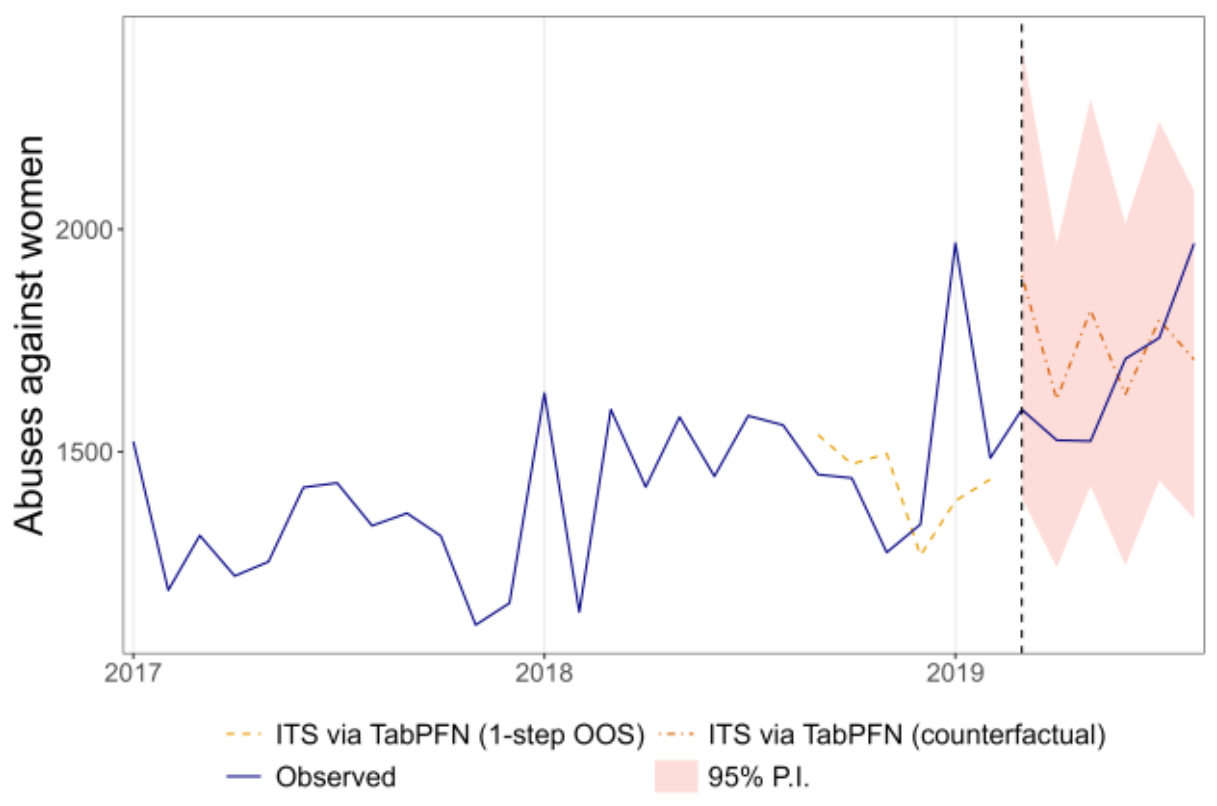

b) Abuses against women

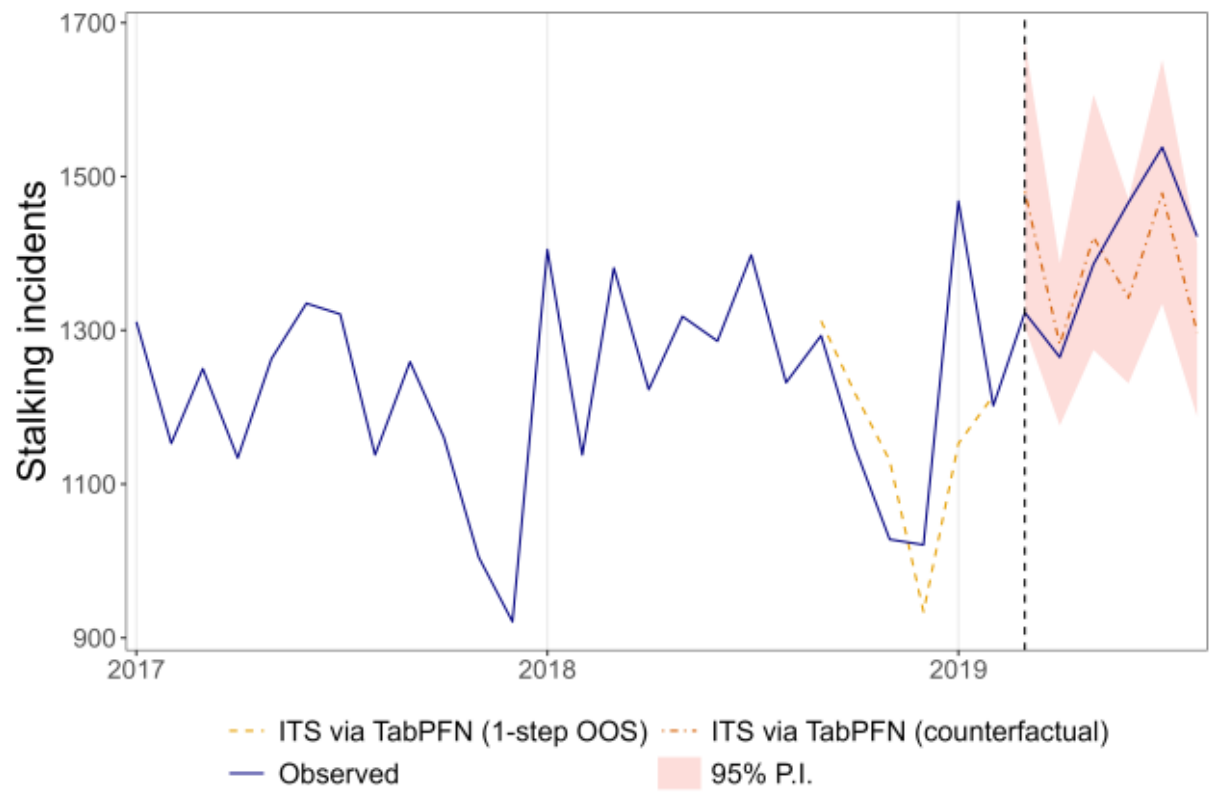

c) Stalking cases

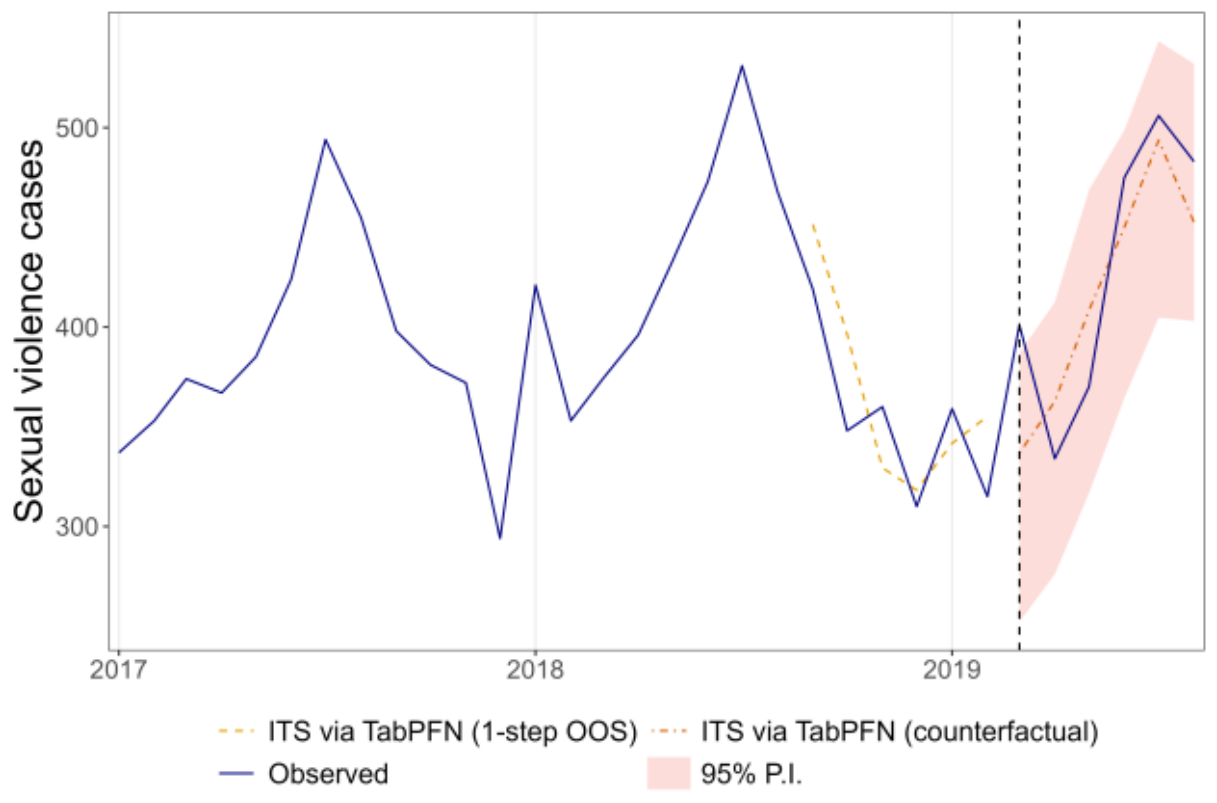

d) Sexual violence cases

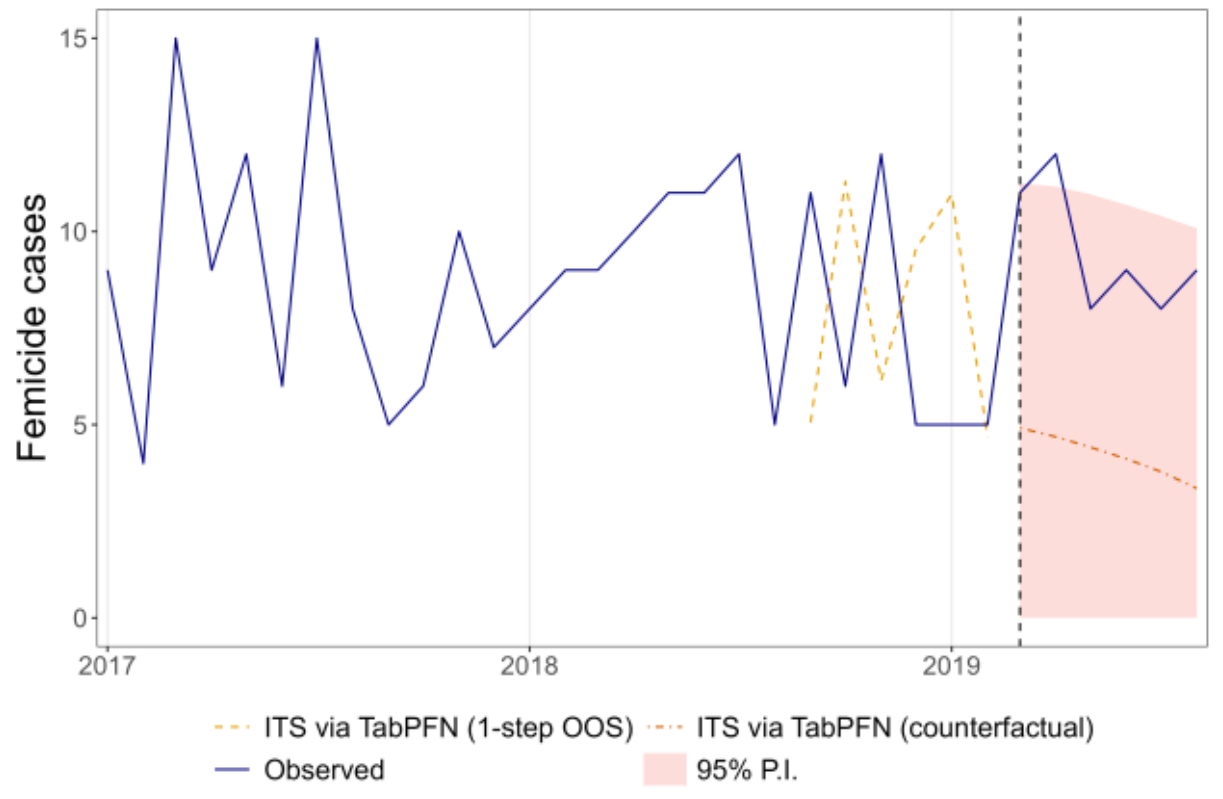

e) All femicides

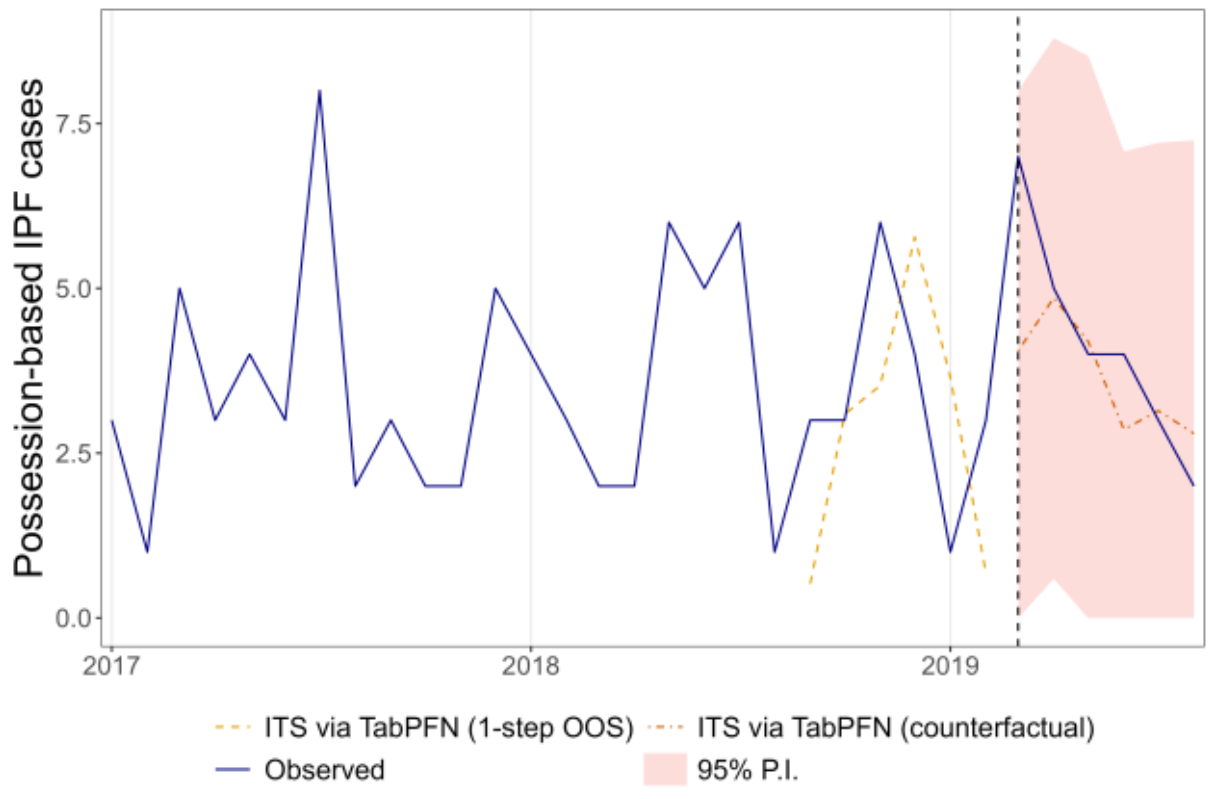

f) IPFs driven by jealousy or possessiveness

**Fig. S3 | Responses following the femicide of Nicoletta Indelicato. a,** Calls to the national 1522 helpline. **b,** Reports of abuse against women. **c,** Stalking. **d,** Sexual violence. **e,** Total femicides. **f,** IPFs associated with jealousy or possessiveness. The solid blue line shows observed monthly values; the dashed orange line shows pre-treatment one-step-ahead forecasts; the dashed red line shows the recursively generated post-treatment counterfactual. Shading denotes the 95% predictive interval around the counterfactual. The vertical dashed line marks the event month. See Supplementary Information D for estimation and uncertainty details.

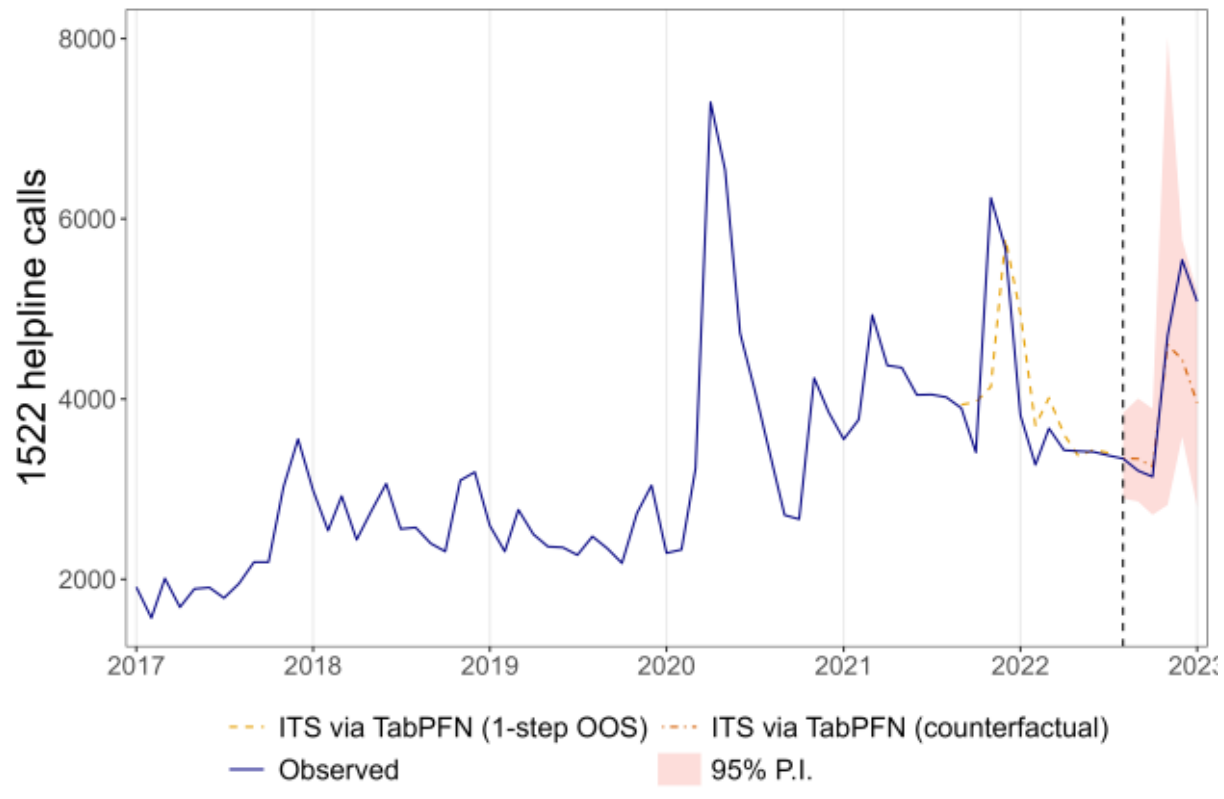

a) Calls to the helpline

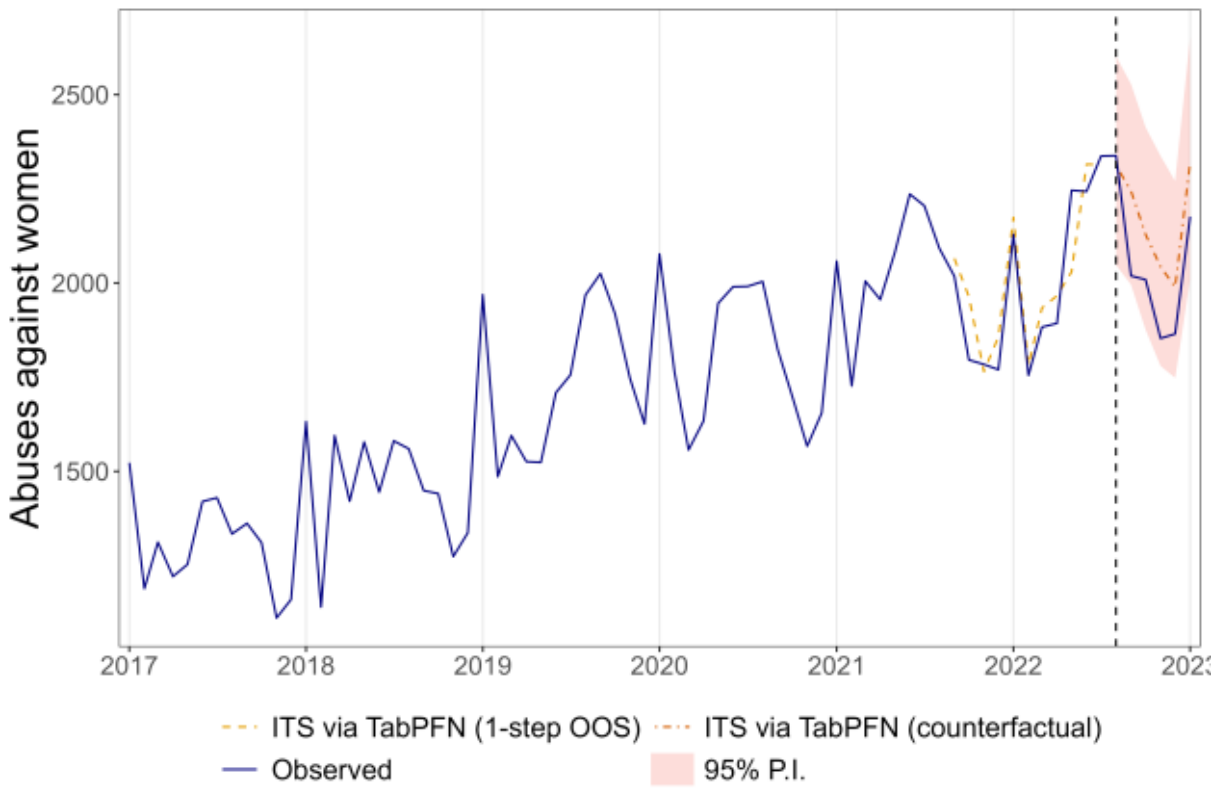

b) Abuses against women

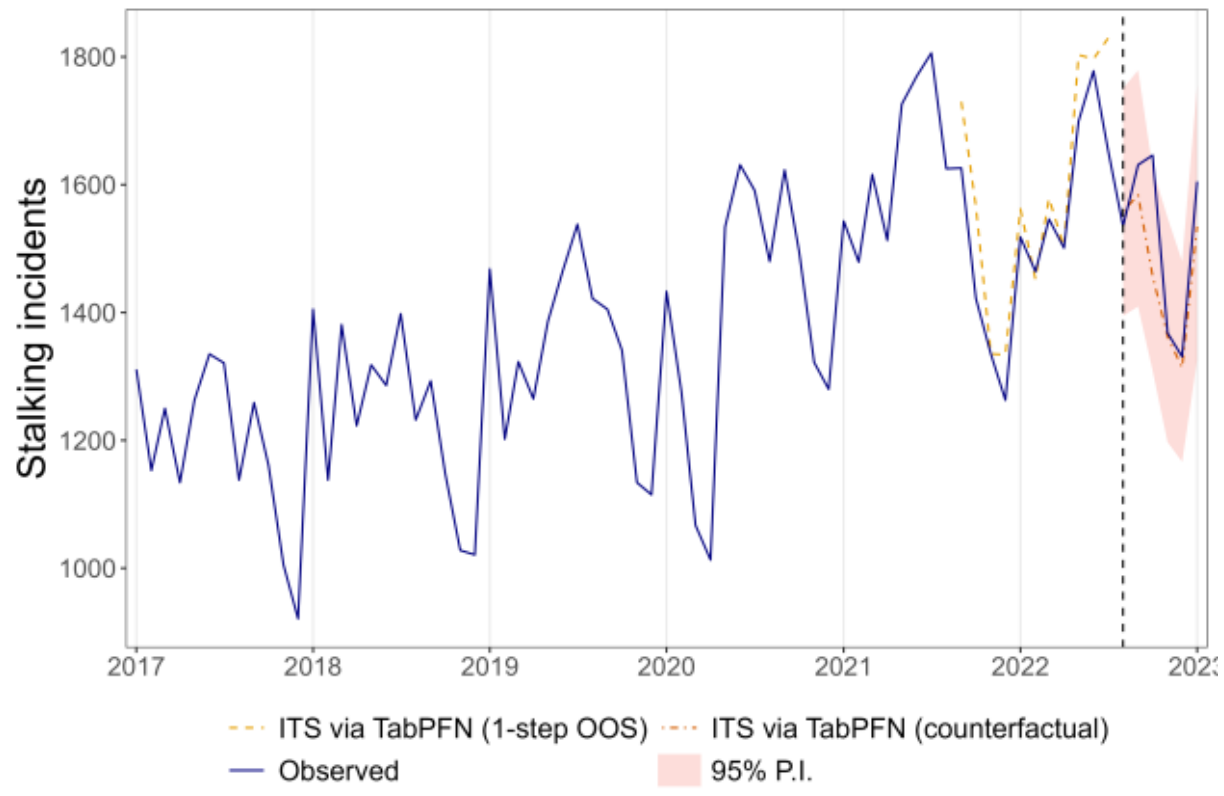

c) Stalking cases

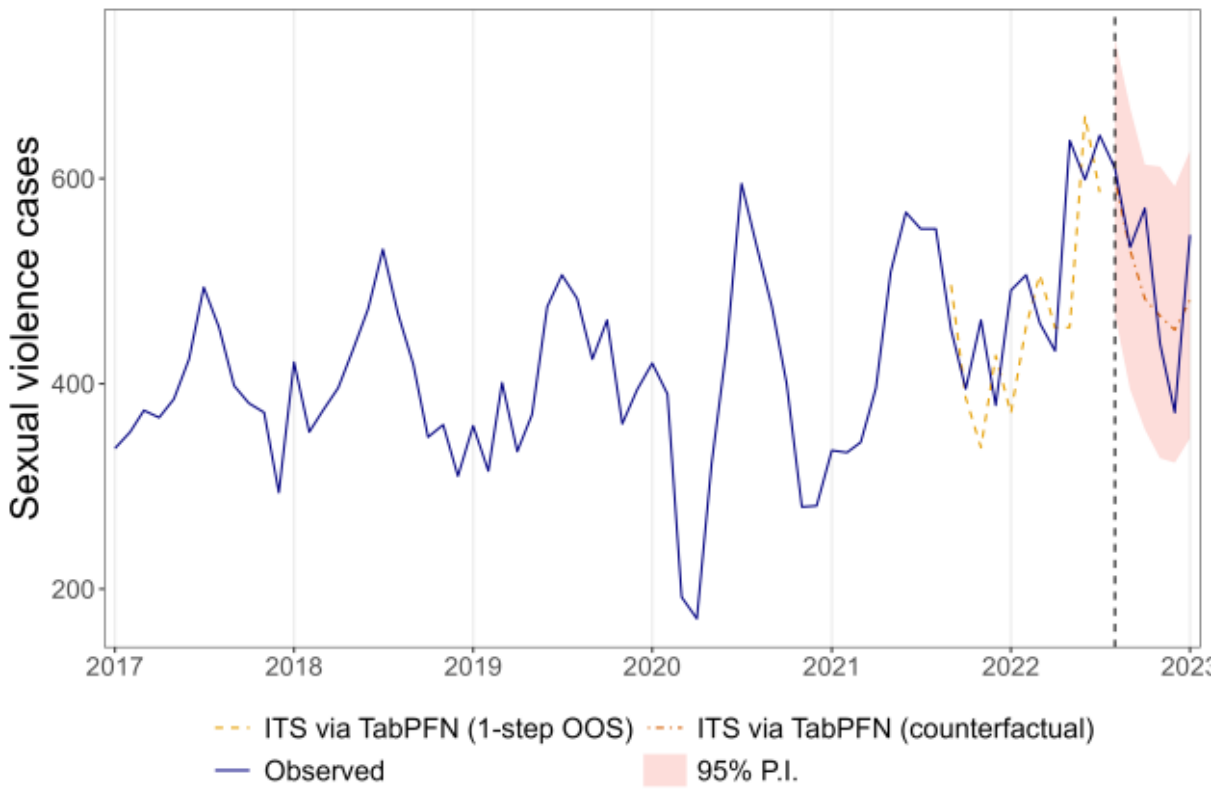

d) Sexual violence cases

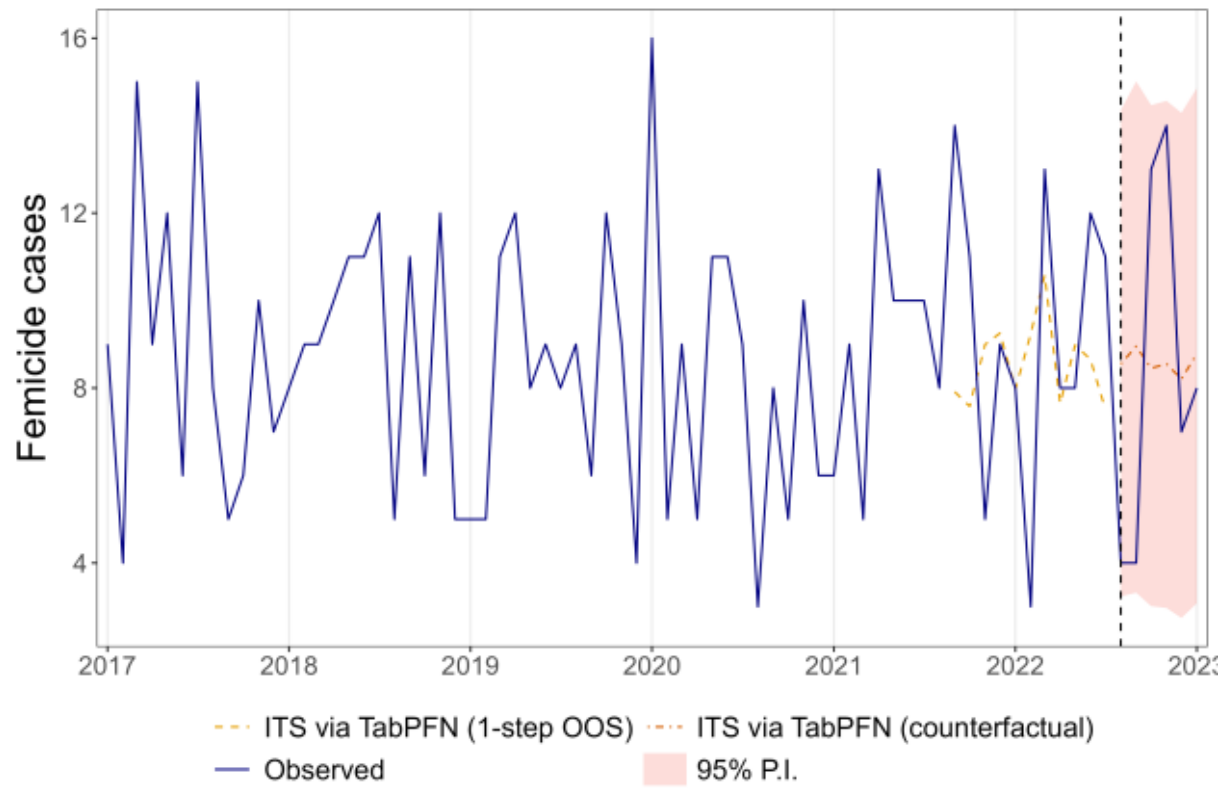

e) All femicides

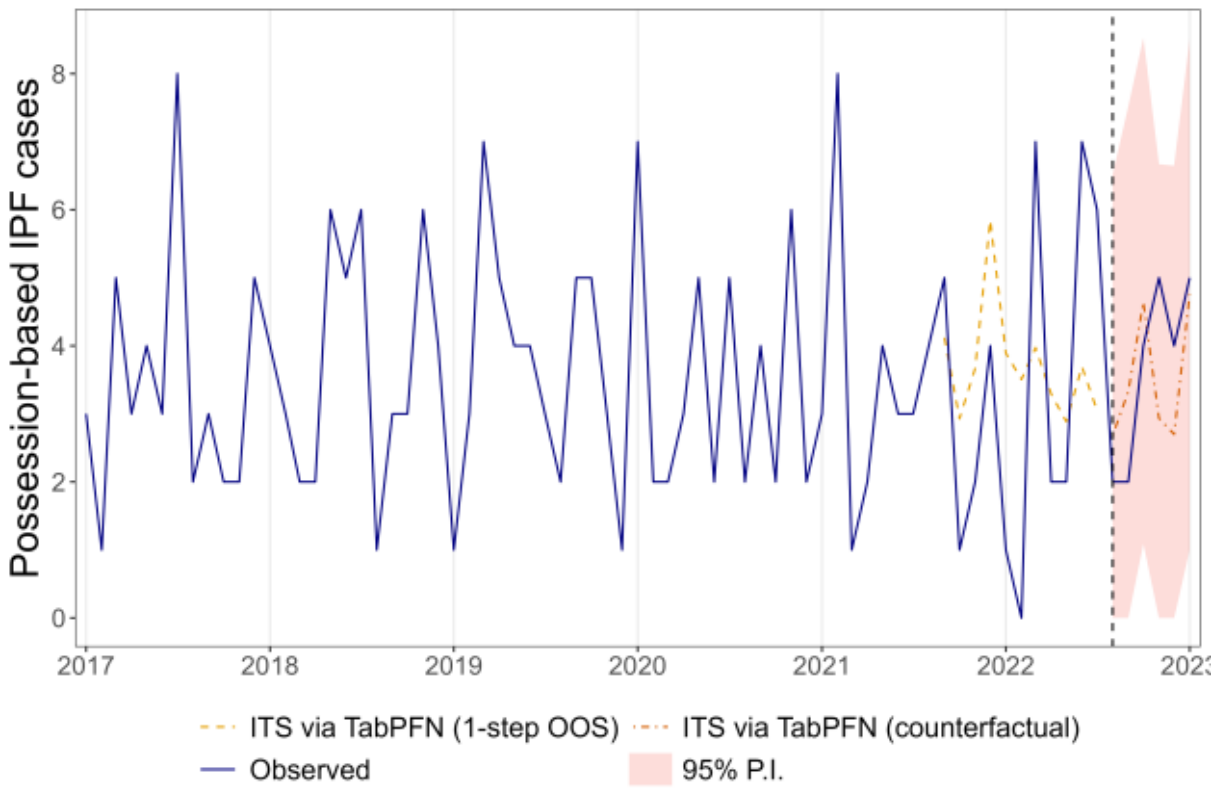

f) IPFs driven by jealousy or possessiveness

**Fig. S4 | Responses following the femicide of Alessandra Matteuzzi. a,** Calls to the national 1522 helpline. **b,** Reports of abuse against women. **c,** Stalking. **d,** Sexual violence. **e,** Total femicides. **f,** IPFs associated with jealousy or possessiveness. The solid blue line shows observed monthly values; the dashed orange line shows pre-treatment one-step-ahead forecasts; the dashed red line shows the recursively generated post-treatment counterfactual. Shading denotes the 95% predictive interval around the counterfactual. The vertical dashed line marks the event month. See Supplementary Information D for estimation and uncertainty details.

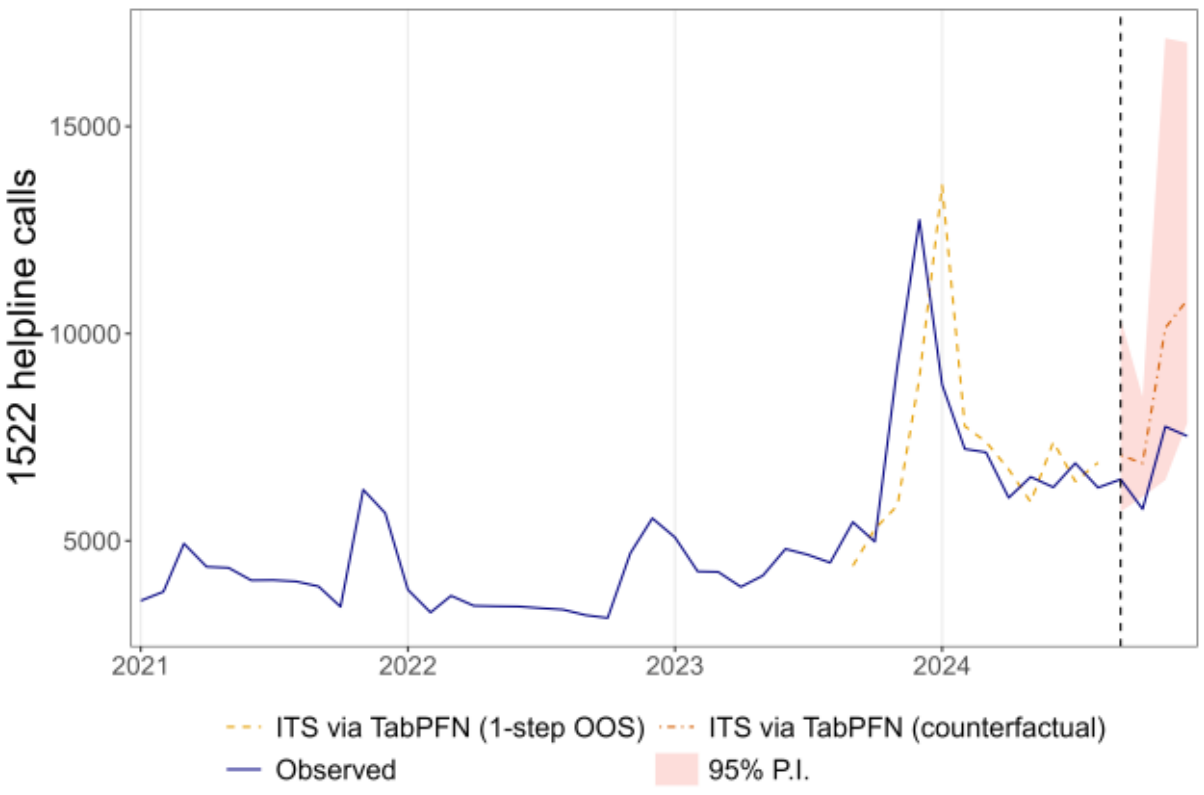

a) Calls to the helpline

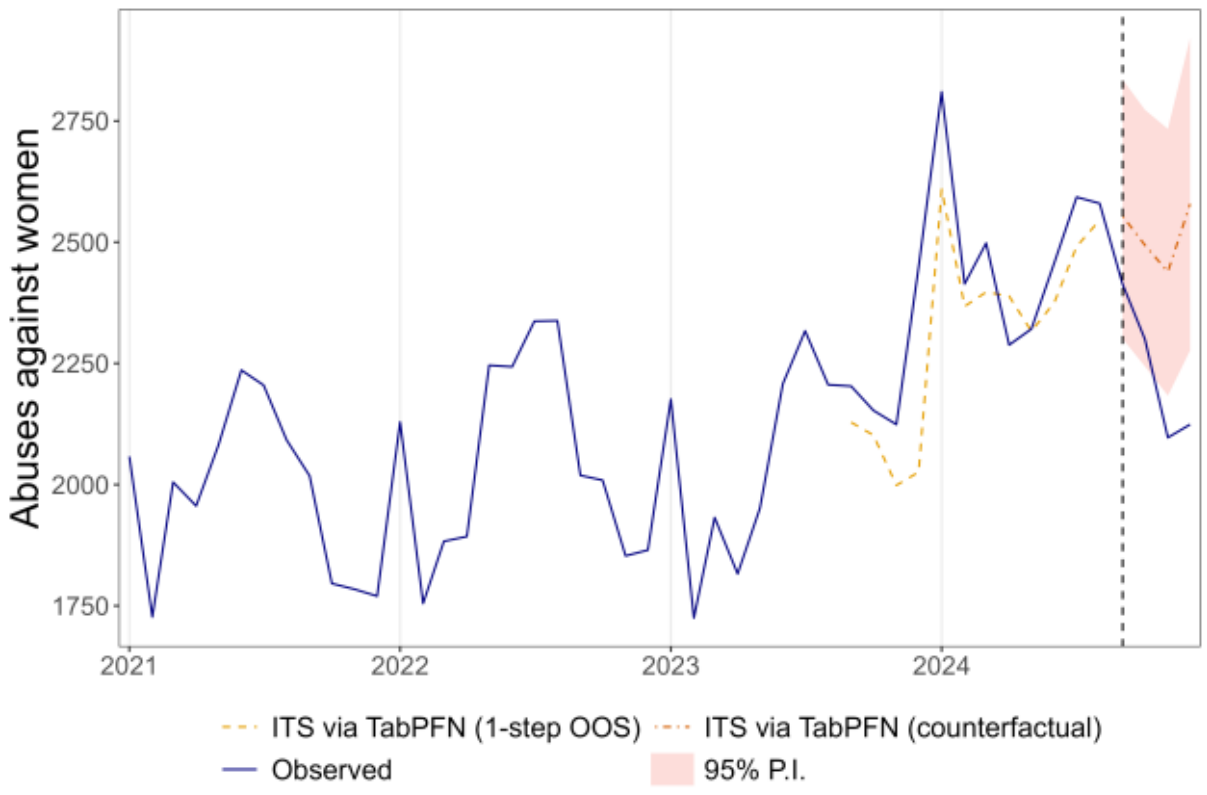

b) Abuses against women

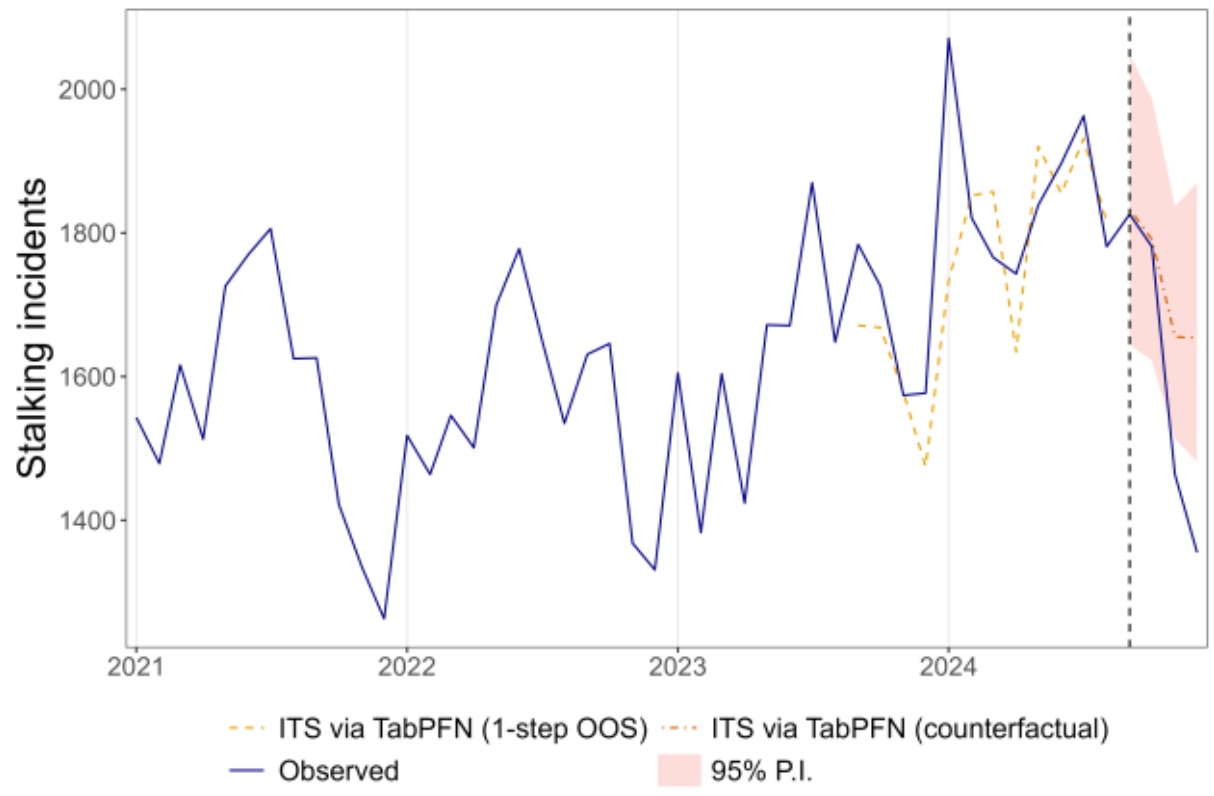

c) Stalking cases

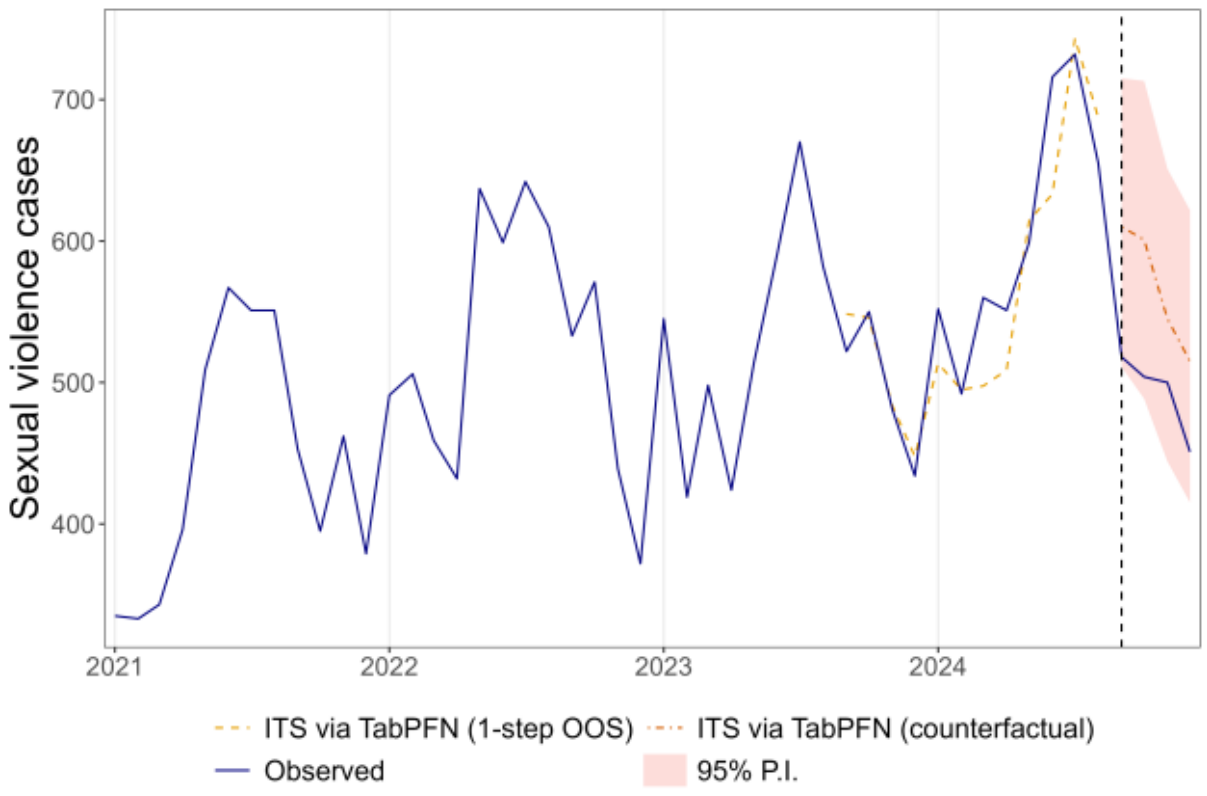

d) Sexual violence cases

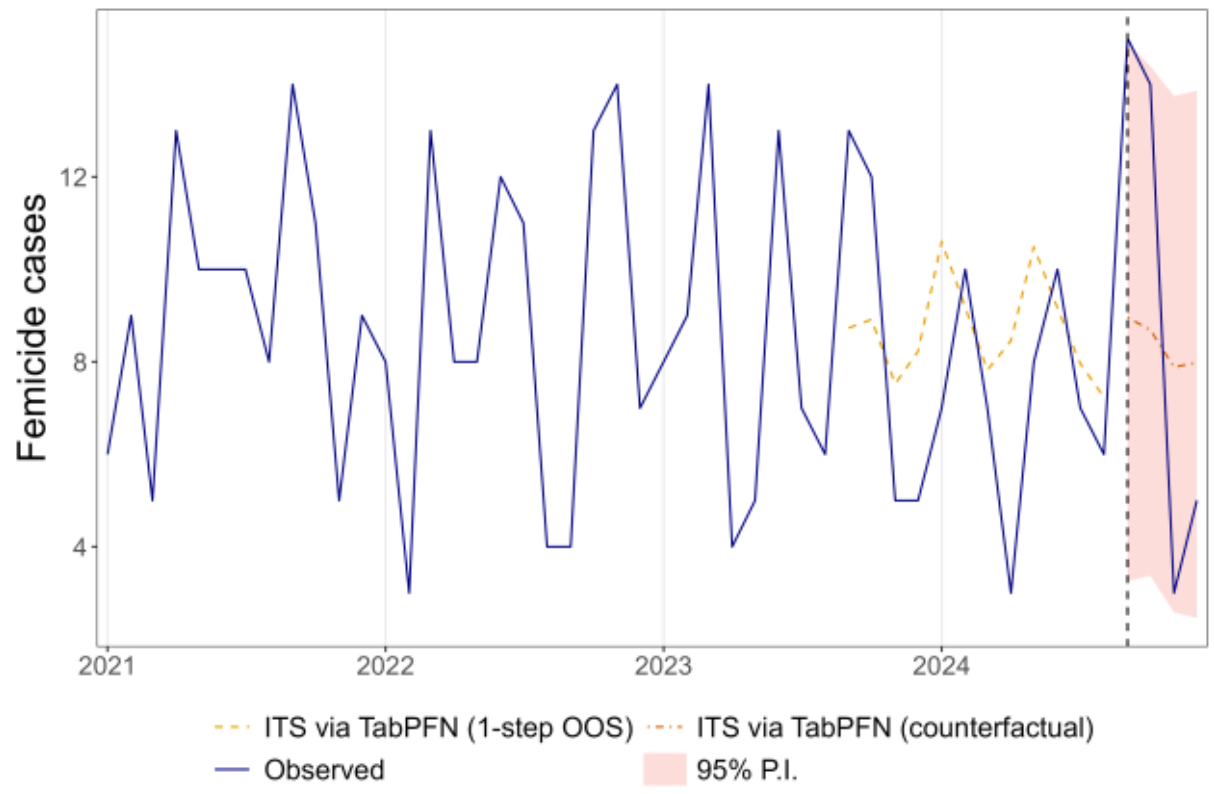

e) All femicides

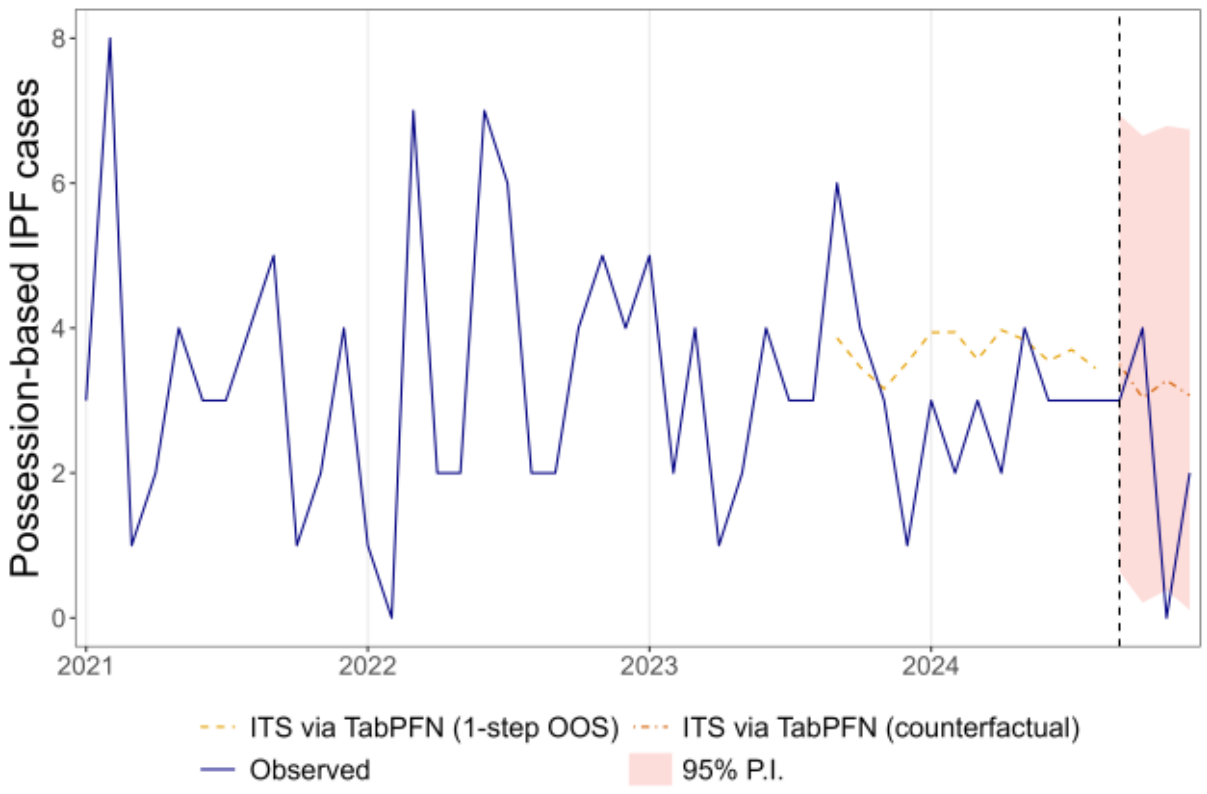

f) IPFs driven by jealousy or possessiveness

**Fig. S5 | Responses following the homicide of Sharon Verzeni. a,** Calls to the national 1522 helpline. **b,** Reports of abuse against women. **c,** Stalking. **d,** Sexual violence. **e,** Total femicides. **f,** IPFs associated with jealousy or possessiveness. The solid blue line shows observed monthly values; the dashed orange line shows pre-treatment one-step-ahead forecasts; the dashed red line shows the recursively generated post-treatment counterfactual. Shading denotes the 95% predictive interval around the counterfactual. The vertical dashed line marks the start of the post-treatment period. See Supplementary Information D for estimation and uncertainty details. Although Sharon Verzeni was murdered on 30 July 2024, public and media attention to the case intensified substantially only at the end of August, when, after a month-long investigation, police identified and arrested Moussa Sangare on 30 August and he confessed to the murder. Given that this sharp increase in salience occurred at the very end of August, we treat September 2024 as the first post-treatment period in our monthly analysis.

## References for Supplementary Information